\documentclass[aps,prl,reprint,superscriptaddress]{revtex4-2}
\usepackage{amsfonts}
\usepackage{amsmath}
\usepackage{amssymb}
\usepackage{graphicx}
\usepackage{color}
\usepackage{physics}
\usepackage{float}
\usepackage{mathtools}
\usepackage{algorithm}
\usepackage{algpseudocode}
\usepackage{wrapfig}
\usepackage{mathbbol}
\usepackage{upgreek}
\usepackage[normalem]{ulem}
\usepackage{makecell} 
\usepackage{booktabs}

\usepackage[colorinlistoftodos]{todonotes}
\usepackage[colorlinks=true, allcolors=blue]{hyperref}
\usepackage{mwe}
\usepackage{epstopdf}
\usepackage[doipre={DOI:~}]{uri}

\newcommand \be {\begin{eqnarray}}
\newcommand \ee {\end{eqnarray}}
\newcommand \ben {\begin{eqnarray}}
\newcommand \een {\end{eqnarray}}

\begin{document}
\title{Dynamics of microvesicle formation driven by glycocalyx}
\author{Ke Xiao}
\email{kexiao@health.ucsd.edu}
\affiliation{Department of Mechanical and Aerospace Engineering, University of California San Diego, La Jolla CA 92093, USA}
\affiliation{Fujian Provincial Key Lab for Soft Functional Materials Research, Research Institute for Biomimetics and Soft Matter, Department of Physics, College of Physical Science and Technology, Xiamen University, Xiamen 361005, People's Republic of China}
\author{Chen-Xu Wu}
\email{cxwu@xmu.edu.cn}
%\email{cxwu@xmu.edu.cn}
\affiliation{Fujian Provincial Key Lab for Soft Functional Materials Research, Research Institute for Biomimetics and Soft Matter, Department of Physics, College of Physical Science and Technology, Xiamen University, Xiamen 361005, People's Republic of China}
%\affiliation{Department of Mechanical and Aerospace Engineering, University of California San Diego, La Jolla CA 92093, USA}
%\affiliation{Department of Pharmacology, School of Medicine, University of California San Diego, La Jolla CA 92093, USA}

\begin{abstract} %600 characters 
Microvesicles (MVs), a type of cell-secreted extracellular vesicle (EV), are essential for various biological processes such as local and long-distance intercellular communication and the disposal of unwanted materials. 
Recent evidence suggests that overexpression of MUC1 glycoproteins facilitates MV formation on the cell membrane, but the mechanism underlying this nonequilibrium process remains unknown. 
Here, the kinetics of glycocalyx-driven MV formation is studied in the framework of Onsager's variational principle. 
Using this deterministic theory, we first show that the process of MV formation is accelerated for the denser and thicker glycocalyx-grafted cell membrane and can be improved by reducing the ionic strength or enhancing the degree of glycocalyx ionization. 
Intriguingly, we further discover that there exists an optimal glycocalyx coverage for the fastest formation of MVs. 
We then investigated the process of MV formation using a stochastic approach. Our results suggest that the transition rate is not constant, leading to the prediction of a shorter formation time compared to the deterministic approach. 
These findings provide valuable insights into the regulation of membrane morphology and MV secretion by the glycocalyx, with implications for both understanding and modulating these processes. 

\end{abstract}
\maketitle

%\iffalse
\section*{Significance Statement}
The secretion of MVs on a sugar-rich cell surface under different physiological and pathological states plays a fundamental role in many biological processes. 
Cell membranes are known to undergo significant deformation during this process. 
Although it is known that cell membranes undergo substantial deformation during MV formation, the non-equilibrium dynamics of this process, particularly as regulated by the glycocalyx, remains poorly understood. In particular, there is still a lack of a quantitative physical explanation of how glycocalyx properties and ionic strength govern the non-equilibrium dynamics of MV formation. 
Here, we first develop a deterministic model, which suggests that the interplay between driving forces generated by glycocalyx polymers and dissipative forces within the system controls the successful formation MVs. 
We then present a stochastic model that captures the essential physics of MV formation, which serves to validate the predictions obtained from the deterministic approach. 
Our results indicate that the physical properties of glycocalyx polymers and ionic strength play significant roles in regulating MV formation dynamics. 
This paper provides a framework for predicting and interpreting MV formation dynamics in small structures, with potential implications for understanding glycocalyx-mediated membrane remodeling in health and disease.
%\fi

\section*{Introduction}
\begin{figure*}[htb!]
    \centering
    \includegraphics[width=\linewidth]{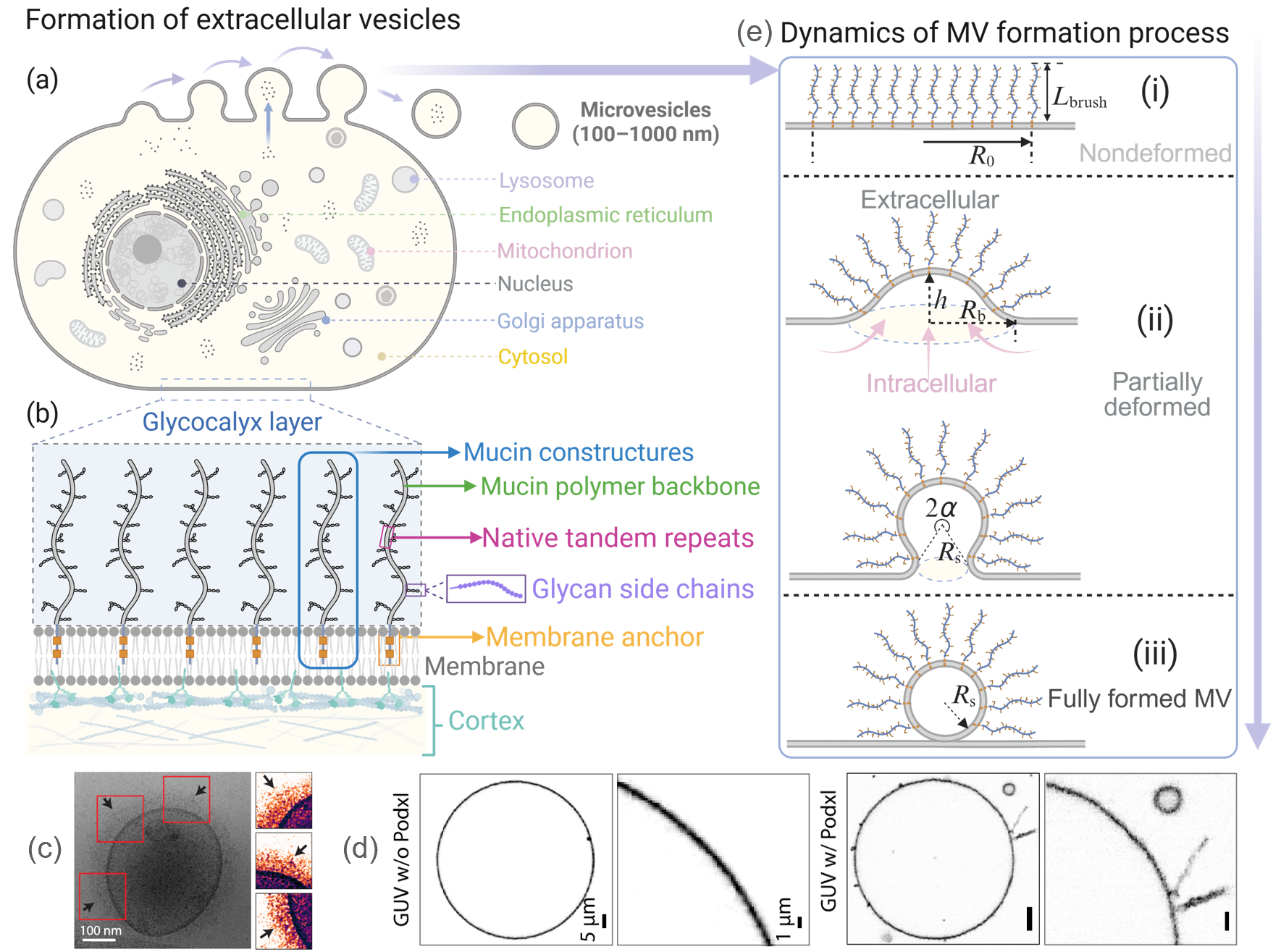}
    \caption{Schematic representation of MV formation. 
    (a) Production and release of MVs into the extracellular space. 
    (b) An enlarged view of the structure of glycocalyx layer and the membrane. 
    (c) A representative cryogenic transmission electron microscopy (cryo-TEM) image of an MV collected from cells expressing Muc1-42TR $\Delta$CT. (d) Representative confocal images of spherical membrane structures formed on giant unilamellar vesicles (GUVs) decorated with recombinant Podocalyxin. 
    %(d) Representative confocal images of daughter vesicle (middle and right) formed on giant unilamellar vesicles (GUVs, left) decorated with recombinant Podocalyxin. 
    (c) and (d) are experimental images from Ref.~\cite{SHURER20191757}, permission pending. 
    (e) Sketch of various membrane morphologies corresponds to (i) nondeformed, (ii) partially deformed, and (iii) fully formed spherical shapes, with parameters used to describe membrane geometry.
       }
    \label{fig:figure_1}
\end{figure*}

%\textit{Introduction}—
Microvesicles (MVs) are fragments of the plasma membrane secreted by all cells~\cite{JEPPESEN2023667,JEPPESEN2019428,Kumar2024,Sheta2023,AHMADZADA2020102949,BECKER2016836} into the extracellular environment under a range of conditions, from normal and stress states to physiological and pathological
states~\cite{JEPPESEN2023667,JEPPESEN2019428,Hargett2013,Katsumori2011,Kumar2024} [Fig.~\ref{fig:figure_1}(a)]. 
These EVs serve as a mechanism to modulate immune cell signaling and are also an important diagnostic marker for certain cancers~\cite{JEPPESEN2023667,JEPPESEN2019428,Kumar2024,Katherine2024,Welsh2024}. 
%These MVs are necessary for various biological processes, including long-distance intercellular communication and the disposal of unwanted materials~\cite{JEPPESEN2023667,JEPPESEN2019428,Kumar2024,Welsh2024}. 
A plethora of studies have reported that EVs also contribute to numerous pathological diseases such as cancer, heart disease, infectious diseases, and immunological and neurological diseases (Alzheimer's and Parkinson's)~\cite{Reynolds2023,GONCALVES20241,ABHANGE20213705,BECKER2016836}. 
%~\cite{Reynolds2023,Muraoka2020,Jiang2021,GONCALVES20241,ABHANGE20213705,BECKER2016836}. 
Many cell types, including cancer cells, T cells, dendritic cells, and endothelial cells, are known to commonly secrete high levels of EVs involved in various physiological and pathological conditions~\cite{Distler2005,RUHELA2020465,RUHELA2021100398}. 
These cells are often found to be densely decorated with a sugar-rich layer of brush-like mucins, hyaluronan polymers, and heavily glycosylated extramembrane compartments~\cite{J.C.H.Kuo2018,Weinbaum2007,Leonhard2020.00253,SHURER20191757}, which are collectively termed glycocalyx [Fig.~\ref{fig:figure_1}(b)]. 

It is well known that the formation of MVs involves membrane remodeling and requires disruption of the cytoskeleton. 
From a biophysical point of view, the formation of various membrane morphologies (vesicles, cellular blebs, spherical and tubular protrusions, etc.) from a plasma membrane has been mainly considered as the result of protein insertion, scaffolding, and crowding~\cite{CALLANJONES2013143,Kozlov2023,DERGANC20161152,Stachowiak2012,Busch2015,LIESE20212482,Ruhoff2022,Haleh2018}. 
In this regard, numerous modeling studies focus on inward membrane bending such as endocytosis~\cite{N.Walani2015,J.E.Hassinger2017,MA20211625,K.Xiao2023}. 
In contrast, MVs are secreted by direct outward budding or blebbing at specific sites of the plasma membrane, typically $100$ to $1000~{\rm nm}$ in size~\cite{Sheta2023,Kumar2024,AHMADZADA2020102949,JEPPESEN2023667,Matsui2021,Joanna2016,JEPPESEN2019428}. 
Recent experimental studies~\cite{SHURER20191757} have shown that bulky brush-like glycocalyx polymers on the outer membrane leaflet can regulate membrane morphologies and further revealed that glycocalyx is closely linked to the secretion of extracellular vesicles (referred to as MVs). 
Specifically, in the experiments by Shurer \textit{et al}.~\cite{SHURER20191757}, cryo-TEM confirmed that Muc1-42TR $\Delta$CT-expressing cells secrete glycocalyx-decorated membrane vesicles [Fig.~\ref{fig:figure_1}(c)], and confocal imaging showed that functionalizing GUVs with the S/T-rich domain of Podxl induces similar spherical structures [Fig.~\ref{fig:figure_1}(d)]. 
Another notable example of polymer-induced vesicle production is that polymer-like intrinsically disordered proteins are found to be capable of generating sufficient steric pressure to reshape membrane surfaces, including driving vesiculation and breaking up membrane vesicles to form smaller daughter vesicles~\cite{Wilton2017,HOUSER20223320,FengYuan2021,FengYuan2023}.  
Hence, these experiments provide evidence that the secretion of extracellular vesicles (referred to as MVs) is closely linked to mucins and other biopolymers on the cell surface. 

In addition, ionic strength is of crucial importance for a wide array of cell biological processes, such as determining the structure of intrinsic disordered proteins~\cite{konig2015single}, regulating the activity of enzymes~\cite{Jens1997}, affecting protein aggregation and phase separations~\cite{Peter2012,Shana2015}, and influencing membrane curvature generation~\cite{Wilton2017,HOUSER20223320,FengYuan2021,FengYuan2023}. 
In particular, varying environmental conditions, i.e., ionic strength, is possible to change the degree of membrane bending, the sensitivity of membrane curvature, and control the direction of membrane bending~\cite{Wilton2017,HOUSER20223320,FengYuan2021,FengYuan2023}. 
Together, these observations suggest that the glycocalyx acts as a potent driver of the membrane vesicle formation process, and this process is modulated by the physical properties of glycopolymers (e.g., grafting density, polymer chain length, and degree of ionization) and the ionic concentration of extracellular space. 
While the fundamental physical dynamics that bend sub-regions of the glycocalyx-grafted plasma membrane into MVs in the presence of distinct physiological intracellular ionic strengths remain scarcely understood. 

Considerable theoretical efforts have been devoted to understanding how polymer networks such as glycocalyx can induce membrane curvature. 
For example, elastic continuum models based on the Canham-Helfrich theory~\cite{Helfrich1973,EVANS1974923} have been widely employed to investigate the formation of membrane invaginations. 
Meanwhile, mean-field theory, analytical methods~\cite{Hiergeist1996,R.Lipowsky_1995,M.Breidenich_2000,Bickel2001}, scaling theory~\cite{Kim2001}, phase field models~\cite{Campelo2007,Campelo2008}, and Monte Carlo simulations~\cite{Werner2010,KUTTIKANDY20223674} have been used to show that anchored polymers on the membrane give rise to changes in
the shape, elastic properties, and spontaneous curvature of
the plasma membrane. 
Recently, we developed a general energetic framework that couples glycocalyx mechanics with lipid bilayer mechanics to investigate, from a thermodynamic equilibrium perspective, how the glycocalyx can generate spherical vesicles and tubular structures~\cite{K.Xiao2025PNAS,Ke.Xiao2024,K.Xiao2026}. 
However, most previous studies have focused primarily on studying the effects of the glycocalyx polymer density and chain length on the equilibrium state of membrane structures. 
A model that captures the nonequilibrium dynamics of MV formation and clarifies the physical mechanisms by which the dynamics of MV formation can be tuned by changing glycocalyx properties and ionic strength is lacking. 

%Here, we aim to quantitatively elucidate the effects of different physical properties on glycocalyx-mediated formation dynamics of MV under ion concentration. 
Here, we present a deterministic and a stochastic description of MV formation dynamics, quantitatively elucidating how polymer properties and ion strength affect vesicle formation, and compare our results with experimental observations. 
Our findings, grounded in experimental observations, may inform the design of a set of strategies to harness the dynamics of MV formation by tuning the physical properties of the glycocalyx and the ionic strength. 
%By combining deterministic and stochastic models, we predict that (i) fully formed MVs are within a range of allowable sizes, which is in agreement with experimental observations; (ii) denser and longer glycopolymers on the membrane surface promote MV formation, with formation efficiency further facilitated by elevating the degree of ionization and reducing the ionic strength; and (iii) this formation process can occur more rapidly on a looser cell membrane in the presence of spontaneous curvature. 
%Taken together, these findings, grounded in experimental observations, may inform the design of a set of strategies to harness the dynamics of MV formation by tuning the physical properties of the glycocalyx and the ionic strength. 
%Based on experimental observations, these findings may guide the design of a set of strategies for harnessing the dynamics of MV formation process by tuning the physical properties of glycocalyx and ionic strength. 

\section*{Model Development and Methods}
\subsection*{Deterministic dynamics}
The formation of an MV on the plasma membrane is modeled using the geometry depicted in Fig.~\ref{fig:figure_1}(e), where the cell surface is grafted with a thick glycocalyx layer on its outer surface and is attached to a thin cortical layer underneath through protein linkers. 
The total free energy of this composite system can be written as 
\begin{align}
\mathcal{F}_{\rm tot}=&\underbrace{\frac{N_{\rm p}}{\beta}\int_{R_{\rm s}}^{R_{\rm s}+L_{\rm brush}}\biggl[\frac{3}{2a^2c_{\rm p}(r)s^2(r)}+Wc_{\rm p}^2(r)\biggr]s(r) {\rm d} r}_{\rm Glycocalyx} \notag\\
&+\underbrace{\int_{A_{\rm mem}}\frac{\kappa}{2}(2H-c_0)^2{\rm d} A + \sigma\Delta A}_{\rm Membrane~(bending+tension)} \notag\\ 
&+ \underbrace{\int\frac{k}{2}\rho_{\rm b}z^2~{\rm d}A}_{\rm Cortex} + \underbrace{\Delta PV}_{\rm Pressure}. \label{F_tot}
\end{align}
The first term is the energy contribution associated with the glycocalyx, which includes the elastic stretching of the polymer chains, the interactions between monomers in the polymer brush, and the electrostatic contribution for ionic brushes. 
Here $N_{\rm p}$ is the number of polymer chain grafted on the membrane, $\beta=1/(k_BT)$ with $k_B$ being the Boltzmann's constant and $T$ being the temperature, $a$ is the effective monomeric segment size, $c_{\rm p}(r)$ is the local monomer density profile along the thickness with $r$ being the radial distance defined from the center of the spherical surface, $s(r)$ is the area per chain at distance from the polymer grafting surface, $W=v+\frac{(\phi_{\rm b}\cdot N_{\rm ion})^2}{2I_{\rm ion}}$ with $v$, $\phi_{\rm b}$, $N_{\rm ion}$, and $I_{\rm ion}$ being the second virial coefficient, the degree of ionization, the number of ionizable sites, and the ion concentration, respectively. 
The second and third terms appearing in Eq.~(\ref{F_tot}) represent the elastic energy of the cell membrane, accounting for bending and surface tension, respectively, where $\kappa$ is the bending modulus, $H=(c_1+c_2)/2$ is the mean curvature (defined by principal curvatures $c_1$ and $c_2$), $c_0$ is the spontaneous curvature, $\sigma$ is the membrane surface tension, and $\Delta A$ denotes the increase in membrane area relative to the flat state due to deformation. 
The fourth term describes the attachment between the cell membrane and the actin cortex and and is given by a harmonic potential where $k$ is the effective elastic stiffness (spring constant) of each membrane-cortex linker, $\rho_{\rm b}$ is the local density of bound springlike linkers, and $z$ is the stretching length of the linkers. 
The last term corresponds to the work done by the pressure difference $\Delta P$ across the membrane. 
The detailed derivation of each individual energy component in Eq.~(\ref{F_tot}) is provided in SI Appendix, section~1.1\ref{Total_free_energy}. 

Given that the MVs observed in experiments bear resemblance to spherical geometry, here we adopt a simplified geometric approach. 
Specifically, we consider a glycocalyx-anchored membrane domain with a fixed surface area $A_0=\pi R_0^2$ that is deformed into a family of spherical caps with radius $R_s$, where $R_0$ is the in-plane radius of the initially flat membrane, as shown in Fig.~\ref{fig:figure_1}(e). 
Over the evolution time, the membrane geometry is described by a shape parameter $0\leq\eta=R_0^2/(4R_{\rm s}^2)\leq1$, where $\eta=0$ corresponds to a completely flat circular domain of radius $R_0$, while $\eta=1$ represents the fully deformed state corresponding to a situation where the initial flat membrane has entirely evolved into a complete sphere, implying the formation of an MV. 

The dynamic equation governing the formation of an MV is derived using the Onsager variational principle \cite{Onsager1931_1, Onsager1931_2}, which states that the evolution of the system minimizes the Rayleighian, defined as $\mathcal{R} = \dot{\mathcal{F}}_{\rm tot} + \mathcal{D}_{\rm tot}$, where $\dot{\mathcal{F}}_{\rm tot}$ is the rate of change of the total free energy and $\mathcal{D}_{\rm tot}$ is the energy dissipation function. 
Hereafter, the dot denotes the derivative with respect to time $t$. 
The total dissipation is assumed to comprise four distinct contributions, i.e., $\mathcal{D}_{\rm tot}=\mathcal{D}_{\rm drag}+\mathcal{D}_{\rm ct}+\mathcal{D}_{\rm bulk}+\mathcal{D}_{\rm mem}$. 
The first source of energy dissipation describes the viscous drag experienced by the spherical cap as it expands in a velocity $u=\dot{h}$ where $h$ is the height of the spherical cap, which can be calculated as (see SI Appendix, section~1.2\ref{Dissipations} for the detailed derivation)
\begin{align}
\mathcal{D}_{\rm drag}=3\pi \mu_{\rm w} u^2R_{\rm s}, \label{drag dissipation}
\end{align}
where $\mu_{\rm w}$ viscosity of the fluid outside the cell (water). 
The second dissipation term originates from friction at the circle line near the detach point between the membrane and the cortex. 
In the limit of small Reynolds number, this dissipation function reads (see
the detailed derivation in SI Appendix, section1.2~\ref{Dissipations})
\begin{align}
\mathcal{D}_{\rm ct}=\mu_{\rm ct}\pi R_{\rm s}^3\sin({\alpha})\dot{\alpha}^2,
\label{friction dissipation_SI2}
\end{align}
where $\mu_{\rm ct}$ is the effective dynamic microviscosity of a typical order of $1~{\rm Pa\cdot s}$, and $\alpha$ is the opening angle of the spherical cap. 
The last two terms in $\mathcal{D}_{\rm tot}$ are bulk and membrane viscous dissipations, respectively, with the related calculation given in Refs.~\cite{P.Sens2004,ALERT20151878}
\begin{align}
\mathcal{D}_{\rm bulk}+\mathcal{D}_{\rm mem}=\frac{22}{5}\mu_{\rm cyt}\frac{\dot{V}^2}{\pi R_{\rm b}^3}+\frac{\mu_{\rm m}l_{\rm m}[{\rm d}_t(\Delta A)]^2}{2\pi R_{\rm b}^2},
\label{bulk_mem_dissipation}
\end{align}
where $\mu_{\rm cyt}$ and $\mu_{\rm m}$ are the viscosities of the cytosol and of the membrane lipid, $l_{\rm m}$ is the membrane thickness, $V$ is the volume of the spherical cap, and $R_{\rm b}$ is the base radius. 

Substituting Eqs.~(\ref{F_tot})-(\ref{bulk_mem_dissipation}) into $\mathcal{R}$, we obtain the explicit expression for the Rayleighian. 
Following the Onsager variational principle and choosing $\eta$ as the state variable, the minimization condition $\partial\mathcal{R}/\partial\dot{\eta}=0$ yields the time evolution equation for the MV
\begin{widetext}
\begin{align}
\dot{\eta}=\frac{\mathcal{M}_1-\mathcal{M}_2-8\kappa\biggl(1-\dfrac{R_0c_0}{4\sqrt{\eta}}\biggr)-R_0^2\sigma-\dfrac{\Delta PR_0^3(2\eta-1)}{4\sqrt{\eta}}}{\dfrac{3\mu_{\rm w}R_0^3}{4\eta^{3/2}}+\dfrac{\mu_{\rm ct} R_0^3}{8\eta^2\sqrt{1-\eta}}+\dfrac{\mu_{\rm m}l_{\rm m} R_0^2}{(1-\eta)}+\dfrac{11\mu_{\rm cyt} R_0^3(1-2\eta)^2}{20\eta(1-\eta)^{3/2}}} - \frac{1}{\mathcal{P}}\times\begin{cases}
\dfrac{1}{6} k\rho_0R_0^4, \quad &0<\eta\leq \Bigl(\dfrac{l_{\rm c}}{R_0}\Bigr)^2 \\
\dfrac{k\rho_0R_0l_c^3}{6\eta^{3/2}},\quad &\eta> \Bigl(\dfrac{l_{\rm c}}{R_0}\Bigr)^2
\end{cases},\label{deta_dt}
\end{align}
\end{widetext}
where
\begin{align}
\mathcal{M}_1=\frac{9R_0^3}{8\beta\xi^2\eta^{3/2}}\biggl(\frac{3\sqrt{W}}{\xi a^2}\biggr)^{\frac{2}{3}}\biggl[\biggl(1+\frac{10N\sqrt{\eta}}{3R_0}\biggl(\frac{Wa^2}{3\xi^2}\biggr)^{\frac{1}{3}}\biggr)^{\frac{1}{5}}-1\biggr],\label{M1}
\end{align}
\begin{align}
\mathcal{M}_2=\frac{3NR_0^2}{4\beta\eta\xi^2}\biggl(\frac{3W^2}{\xi^4a^2}\biggr)^{\frac{1}{3}}\biggl[1+\frac{10N\sqrt{\eta}}{3R_0}\biggl(\frac{Wa^2}{3\xi^2}\biggr)^{\frac{1}{3}}\biggr]^{-\frac{4}{5}},\label{M2}
\end{align}
with $N$ and $\xi$ being the number of segments per polymer chain and the grafting distance of polymers on the cell membrane surface, and $\mathcal{P}$ equals the denominator of the first term in Eq.~(\ref{deta_dt}). 
This dynamic equation is fairly complex and has to be solved numerically. 

\subsection*{Stochastic dynamics} 
We proceed by treating the formation of MV as a stochastic process which is assisted by thermal fluctuations or intrinsic noise within the system. 
In this framework, the shape parameter $\eta$ is treated as a random variable that evolves from $0$ to $1$ subject to a drift driven by the gradient of the free energy given in Eq.~(\ref{F_tot}). 
This calls for a stochastic description. 
In the limit of low Reynolds number, the stochastic evolution equation for $\eta(t)$ is given by 
\begin{align}
\dot{\eta}(t)=-\frac{\mathcal{F}_{\rm tot}'(\eta)}{\gamma}+\sqrt{2D}\zeta(t),\label{Stochastic equation}
\end{align} 
where $\gamma$ is an effective friction coefficient encapsulating the dissipative processes at play, $D$ is the transition rate which characterizes how fast the membrane edge can change its arch angle and $\zeta(t)$ is a Gaussian white noise with zero mean $\langle\zeta(t)\rangle=0$ and autocorrelation function given by $\langle\zeta(t)\zeta(t')\rangle=\delta(t-t')$. 
In the absence of noise and under the assumption $\gamma=\pi\mathcal{P}$, the above stochastic equation reduces to the deterministic dynamic equation~(\ref{deta_dt}). 
In the continuous limit, one can obtain the Smoluchowski equation as
%~\cite{StephenMirigian2013} 
\begin{align}
\frac{\partial \mathbb{P}(\eta,t)}{\partial t}=\frac{\partial}{\partial\eta}D\biggl[\frac{\partial \mathbb{P}(\eta,t)}{\partial\eta}+\mathbb{P}(\eta,t)\frac{\partial(\beta\mathcal{F}_{\rm tot})}{\partial\eta}\biggr],\label{Smoluchowski equation}
\end{align} 
where $\mathbb{P}(\eta,t)$ is the probability of the membrane shape with shape parameter $\eta$ at time $t$. 
The average time at which the shape parameter first reaches the value $1$ (fully formed MV) is defined as the mean first passage time. 
When $D$ is constant, it is given by 
\begin{align}
\langle\tau\rangle=\frac{1}{D}\int_0^1{\rm d}\eta \int_\eta^1 {\rm d}\nu~e^{[-\mathcal{F}_{\rm tot}(\eta)+\mathcal{F}_{\rm tot}(\nu)]/k_BT}. \label{mean_first passage time}
\end{align}
Then we can evaluate how the properties of the glycocalyx and membrane, and the ionic strength affect the MV formation rate, which is defined as the inverse of the mean first passage time, $1/(D\tau)$. 
When $D$ is not a constant, the mean first passage time becomes
\begin{align}
\langle\tau\rangle=\int_0^1{\rm d}\eta \int_\eta^1 {\rm d}\nu~\frac{e^{[-\mathcal{F}_{\rm tot}(\eta)+\mathcal{F}_{\rm tot}(\nu)]/k_BT}}{D(\nu)}.\label{mean formation time_D_main}
\end{align}
A comprehensive description of the stochastic model is provided in SI Appendix, section~2\ref{Stochastic dynamics}.  
%Full details on the Stochastic model are described in the Supplemental Material. 

\subsection*{Numerical implementation}
Based on the model developed in this section, we numerically solve the dynamic equations governing MV formation and compute the corresponding mean first passage time. 
The values of all parameters used in our simulations, along with their sources, are gathered in Table~\ref{tab_num_params} and Table~\ref{tab_sim_params} unless otherwise specified. 
We nondimensionalize the formation time via the rescaling $t\rightarrow{tM_{\rm diff}/A_{\rm g}}$, where $M_{\rm diff}=1~\mu {\rm m}^2/s$ is the polymer diffusion constant and and $A_{\rm g}=100^2~{\rm nm}^2$ is the characteristic area.

\begin{figure*}[t!]
    \centering
    \includegraphics[width=\linewidth]{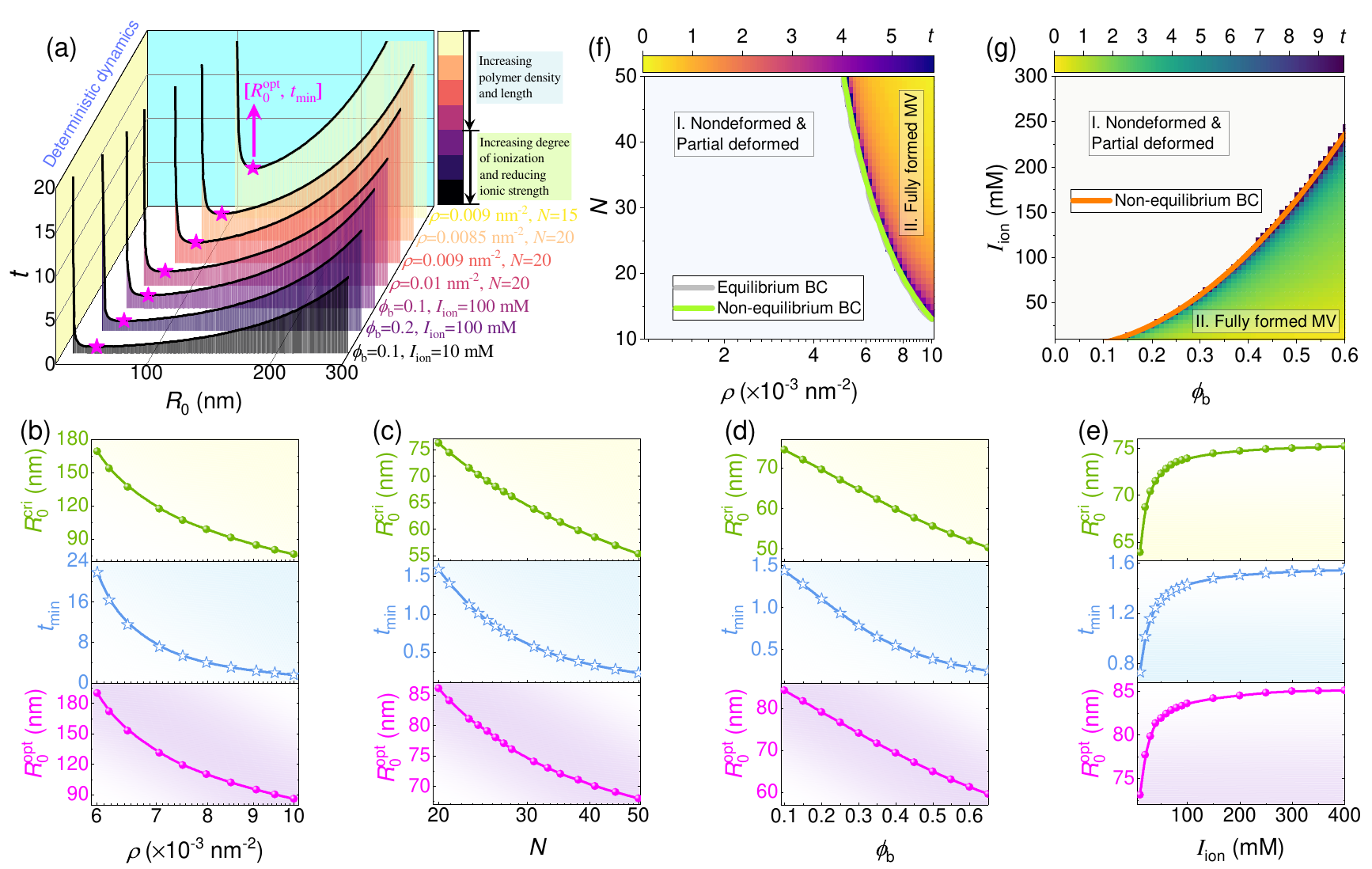}
    \caption{Deterministic MV formation dynamics. (a) Formation time as a function of glycocalyx coverage radius $R_0$ for different glycocalyx density $\rho$, length $N$, degree of ionization $\phi_{\rm b}$, and ionic strength $I_{\rm ion}$. 
    The dependence of critical radius and minimum formation time and its corresponding optimum radius on (b) $\rho$, (c) $N$, (d) $\phi_{\rm b}$, and (e) $I_{\rm ion}$. 
    The heatmaps of formation times in the parametric spaces of (f) $\rho-N$ and (g) $\phi_{\rm b}-I_{\rm ion}$. The solid lines mark the boundary condition for the successful formation of a closed MV.}
    \label{fig:figure2_2}
\end{figure*}
\section*{Results and Discussions}
\subsection*{Effects of glycocalyx properties and ionic strength on MV formation}
\textbf{Deterministic MV Formation Dynamics.} We first use the deterministic description to compare the resulting MV formation times, as shown in Fig.~\ref{fig:figure2_2}. 
Figure~\ref{fig:figure2_2}(a) shows how the MV formation time depends on the glycocalyx coverage for different glycocalyx grafting densities, polymer lengths, degrees of ionization, and ionic strengths. 
We note three main features. 
First, there exists a critical radius $R_0^{\rm cri}$ below which MV formation is impossible, corresponding to an infinite formation time. 
Hence, $R_0^{\rm cri}$ also corresponds to the certain minimum size required for the glycocalyx-coated region to generate a complete MV. 
For radii exceeding this critical value, i.e., $R_0 > R_0^{\rm cri}$, the formation time drops sharply to a minimum value $t_{\rm min}$ at the optimal radius $R_0^{\rm opt}$ (see star points). 
Intriguingly, the optimal radius lies very close to the minimum allowable size for MV formation. 
Beyond $R_0^{\rm opt}$, further increases in $R_0$ lead to a monotonic rise in formation time. 
Second, when varying the glycocalyx density (or polymer length, or degree of ionization) while keeping other parameters fixed, the simulations show that MV formation proceeds more rapidly at higher densities. 
This acceleration is attributed to the enhanced driving force provided by denser glycocalyx layers during the formation of MVs. 
In a biological context, this suggests that under physiological conditions, the overexpression of denser and longer MUC1 glycoproteins on the cell surface can accelerate MV production, in agreement with those found experimentally by Shurer \textit{et al}~\cite{SHURER20191757}. 
In contrast, increasing the ionic strength slows down MV formation. 
We can infer that, in environments with higher ionic strength, a longer time is required for glycocalyx-driven MV formation. 
Importantly, a wide range of formation times can be achieved depending on $R_0$, which helps explain the heterogeneous size distribution of MVs observed experimentally under glycocalyx regulation~\cite{SHURER20191757}. 

Furthermore, to gain more insight into the effects of glycocalyx properties and ionic strength, we examine the dependence of the critical radius $R_0^{\rm cri}$, the minimum formation time $t_{\rm min}$, and the optimal radius $R_0^{\rm opt}$ on glycocalyx properties ($\rho$, $N$, and $\phi_{\rm b}$) and ionic strength ($I_{\rm ion}$), respectively, as shown in Figs.~\ref{fig:figure2_2}(b)-\ref{fig:figure2_2}(e). 
We find that increasing glycocalyx density, polymer length, and ionization reduces $R_0^{\rm cri}$, $t_{\rm min}$, and $R_0^{\rm opt}$, shifting all these three quantities to smaller values. 
By contrast, when the ionic strength is increased, it exhibits an inverse effect on the formation of MVs, as illustrated in Fig.~\ref{fig:figure2_2}(e). 
It shows that the values of $R_0^{\rm cri}$, $t_{\rm min}$, and $R_0^{\rm opt}$ are going up with the ionic strength. 
Under physiological conditions ($I_{\rm ion} \sim 100~{\rm mM}$), Fig.~\ref{fig:figure2_2}(e) also shows that increasing the ionic strength leads to a pronounced elevation in formation time, as well as in both critical and optimal radii. 
However, above physiological levels, a further increase in ionic strength appears to have little influence on $R_0^{\rm cri}$, $t_{\rm min}$, and $R_0^{\rm opt}$. This is consistent with the fact that at high ionic strength, the electrostatic screening effect becomes significant, leading to slow formation of MVs.

Finally, to characterize the interrelated effects of glycocalyx properties and ionic strength on the formation of MVs, we constructed two heat maps of the formation time $t$ in the $\rho-N$ and $\phi_{\rm b}-I_{\rm ion}$ spaces under the condition of fixed glycocalyx coverage $R_0$, as shown in Figs.~\ref{fig:figure2_2}(f) and~\ref{fig:figure2_2}(g), respectively. 
The two-dimensional phase diagrams reveal that two regimes corresponding to a closed MV formation regime (II) and an undeformed or partially deformed regime (I) can be identified, where the color scale represents the value of the formation time. 
A detailed examination of the formation time shows that MV formation proceeds more rapidly on membranes grafted with denser and longer glycocalyx polymers [see Fig.~\ref{fig:figure2_2}(f)]. 
A similar comparison of formation times at different degrees of ionization and ionic strengths also illustrates that formation efficiency improves by reducing the ionic strength or increasing the degree of ionization [see Fig.~\ref{fig:figure2_2}(g)]. 
Here, the boundary curves are determined by $\dot{\eta}\Bigl|_{\eta=1}=0$, which is given by Eq.~(\ref{R0_cri}) in SI Appendix, section~1.3\ref{kinetic_equaiton}. 
In particular, these simulation-derived boundaries match those from thermodynamic equilibrium analysis [see Fig.~\ref{fig:Eq_phase_diagram}], confirming the consistency between the deterministic model and the equilibrium theories. 
The predicted formation timescale of the MVs agrees in order of magnitude with Refs.~\cite{FuqiangSun2023,CAVANAUGH2020152}.

\textbf{Stochastic Dynamics of MV Formation.} We now turn to the stochastic dynamics of the formation of MVs. 
Specifically, using the total free energy $\mathcal{F}_{\rm tot}$ given by Eq.~(\ref{F_tot}), we first compute the MV formation rate from Eq.~(\ref{mean_first passage time}). 
\begin{figure*}[th!]
    \centering
    \includegraphics[width=\linewidth]{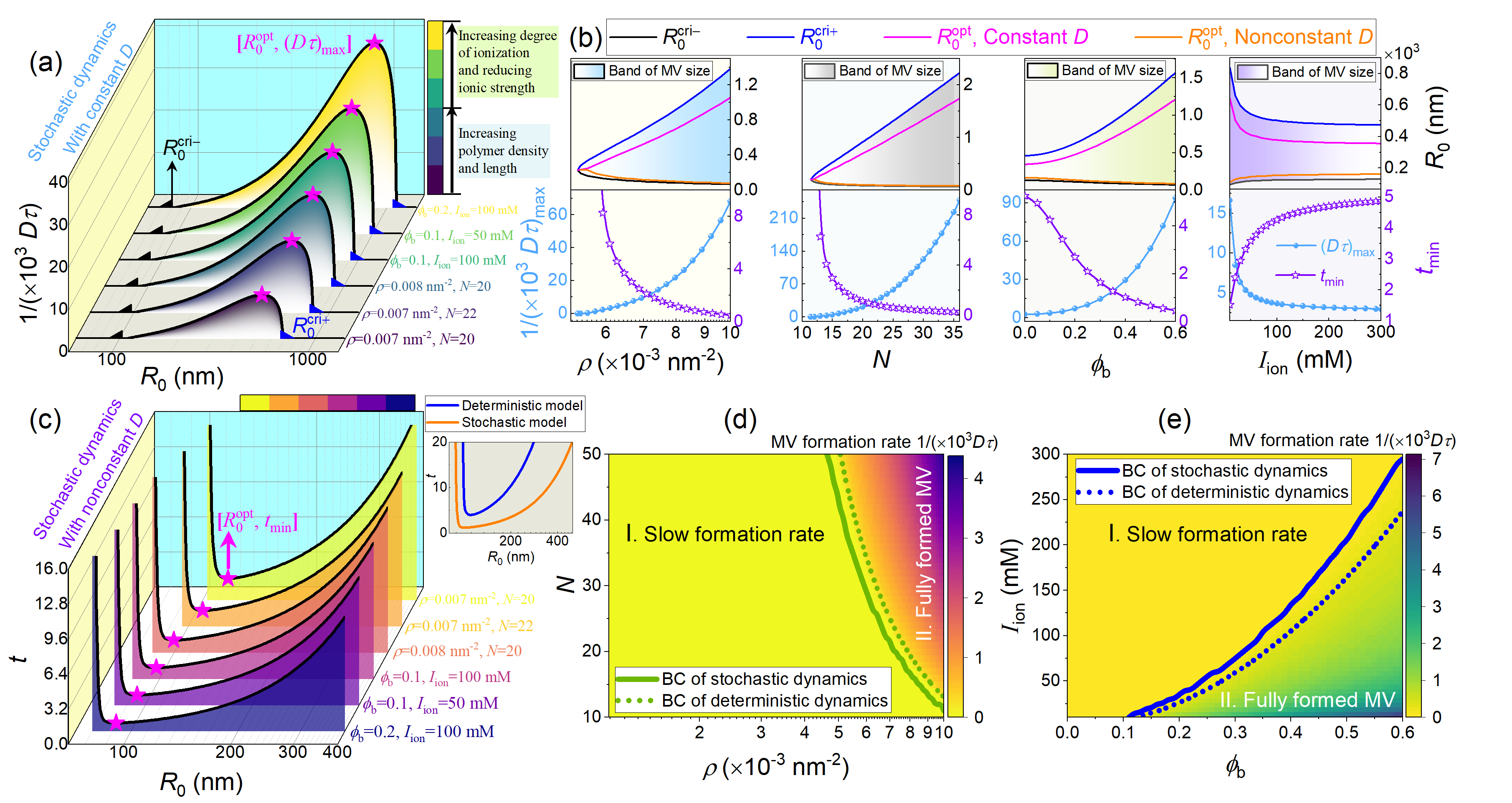}
    \caption{Stochastic dynamics of MV formation. 
    (a) Formation rate versus $R_0$ for changing $\rho$, $N$, $\phi_{\rm b}$, and $I_{\rm ion}$ with constant transition rate $D$. 
    (b) The dependence of critical radius ($R_0^{\rm cri-}$ and $R_0^{\rm cri+}$), $\tau_{\rm max}$ ($t_{\rm min}$) and $R_0^{\rm opt}$ on $\rho$, $N$, $\phi_{\rm b}$, and $I_{\rm ion}$. 
    (c) Mean formation time versus $R_0$ under different $\rho$, $N$, $\phi_{\rm b}$, and $I_{\rm ion}$ when $D$ is not constant. 
    (d) and (e) are heatmaps of formation times on the $\rho-N$ and $\phi_{\rm b}-I_{\rm ion}$ spaces, in which solid lines mark the boundary condition for the successful formation of a closed MV, and dotted lines denote the corresponding boundary curves predicted by the deterministic model.  }
    \label{fig:figure3}
\end{figure*}
Figure~\ref{fig:figure3}(a) shows the formation rate as a function of the initial glycocalyx coverage $R_0$ for a constant transition rate $D$. 
All curves exhibit a finite band of nonzero formation rates between $R_0^{\rm cri-}$ (lower limit) and $R_0^{\rm cri+}$ (upper limit), with a maximum at $R_0^{\rm opt}$. 
For values of $R_0$ outside this interval, the formation rate drops to zero, indicating an unsuccessful formation of MVs. 
The range of radii within this band defines a selectivity window for MV size, suggesting that cells may self-modulate glycocalyx properties to generate a heterogeneous collection of MV sizes under different physiological and pathological conditions. 
This prediction can be supported by previous experimental observations~\cite{SHURER20191757,HOUSER20223320}. 
For example, Shurer \textit{et al.}~\cite{SHURER20191757} observed that MUC1-42TR-expressing cells release massive concentrations of vesicles ranging from approximately $100$ to $400~\rm nm$. 
In addition, recent experimental studies using quantitative fluorescence approaches have shown that intrinsically disordered protein domains are capable of driving large membrane vesicles to fragment into smaller daughter vesicles spanning tens to hundreds of nanometers~\cite{Wilton2017, HOUSER20223320}. 
From Fig.~\ref{fig:figure3}(a), we further find that the formation rate depends on $R_0$ in a nonmonotonic manner, which increases first, then reaches a maximum at $R_0^{\rm opt}$, and is followed by a sharp drop to zero near the upper cutoff radius $R_0^{\rm cri+}$. 
In marked contrast to the deterministic model [see Fig.~\ref{fig:figure2_2}(a)], here the position of the optimal radius (corresponding to the maximum in formation rate) is quite close to the maximum allowable radius for the occurrence of MV formation.  
Moreover, Fig.~\ref{fig:figure3}(a) also demonstrates that increasing the grafting density, the thickness or the degree of ionization of glycocalyx leads to an increase in formation rate, while on the contrary to increasing ionic strength. 

Next, we plot $R_0^{\rm cri-}$, $R_0^{\rm cri+}$, $\tau_{\rm max}$, and $R_0^{\rm opt}$ against $\rho$, $N$, $\phi_{\rm b}$, and $I_{\rm ion}$ in Fig.~\ref{fig:figure3}(b), where it is noted that there exists a band of $R_0$ that allows MV formation to occur. The range of allowed MV sizes widens with increasing $\rho$, $N$, and $\phi_{\rm b}$, but narrows at higher ionic strength [top row of Fig.~\ref{fig:figure3}(b)]. 
Comparing Fig.~\ref{fig:figure3}(b) with Figs.~\ref{fig:figure2_2}(b)-\ref{fig:figure2_2}(d) reveals that while $R_0^{\rm cri-}$ and $t_{\rm min}$ follow identical trends in both stochastic (with constant $D$) and deterministic models, the behavior of $R_0^{\rm opt}$ differs markedly. 
In the stochastic model (with constant $D$), $R_0^{\rm opt}$ exhibits an inverse dependence on the control parameters [magenta curves in Fig.~\ref{fig:figure3}(b)], which scale approximately logarithmically with $\rho$ and linearly with $N$, in stark contrast to the deterministic predictions. 
At very high ionic strengths, variations in $I_{\rm ion}$ only have a weak effect on the formation rate, a result consistent with the predictions given by the deterministic model. 
These show that when $D$ is constant, stochastic and deterministic models produce contradictory predictions on the dependence of $R_0^{\rm opt}$ on parameters such as $\rho$, $N$, $\phi_{\rm b}$, and $I_{\rm ion}$, which is not reasonable. 

To test whether this prediction still holds when $D$ is not constant, we next compute the mean first passage time using Eq.~(\ref{mean formation time_D_main}). 
Figure~\ref{fig:figure3}(c) shows the mean formation time of an MV when the transition rate $D$ is not constant. Three key features emerge from the results. 
First, the overall profile resembles that obtained from the deterministic model [Fig.~\ref{fig:figure2_2}(a)]. 
Second, there exists an optimal radius at which the mean formation time reaches a minimum. 
However, when comparing Fig.~\ref{fig:figure3}(a) with Fig.~\ref{fig:figure3}(c), we surprisingly observe that the position of $R_0^{\rm opt}$ lies much closer to the lower critical radius $R_0^{\rm cri-}$, as indicated by the orange curves in Fig.~\ref{fig:figure3}(b). 
In this case, the corresponding minimum formation time follows a trend similar to that predicted by the deterministic dynamics [purple curves in Fig.~\ref{fig:figure3}(b)].  
This finding suggests that $D$ is not constant, challenging the assumption in Ref.~\cite{StephenMirigian2013} that the transition rate remains constant throughout the stochastic process. 
Third, the inset [on the right of Fig.~\ref{fig:figure3}(c)] shows that the formation time obtained from the stochastic approach ($D$ is not constant) is consistently shorter than that from the deterministic model under identical conditions. 

In the end, we scan the phase spaces of $\rho-N$ and $\phi_{\rm b}-I_{\rm ion}$ in numerical simulations to assess their influence on the MV formation rate, as shown in Figs.~\ref{fig:figure3}(d) and~\ref{fig:figure3}(e), where the color scale represents the magnitude of the formation rate $1/(D\tau)$. 
In analogy to the deterministic case, two distinct regimes are identified: a slow formation rate regime (Regime I) and a fully formed MV regime (Regime II). 
The phase diagrams also reflect that it is less time consuming to form MVs when cell surfaces are grafted with a denser and thicker glycocalyx and that the formation time can be further reduced by increasing the degree of ionization or decreasing the ionic strength.  
However, a comparison between the deterministic [Figs.~\ref{fig:figure2_2}(f) and \ref{fig:figure2_2}(g)] and stochastic [Figs.~\ref{fig:figure3}(d) and \ref{fig:figure3}(e)] phase diagrams displays notable differences. 
In the stochastic case, the fully formed MV regime is broader, and the boundary separating regime I from regime II shifts leftward (solid versus dotted lines). 
This scenario is attributed to the fact that thermal fluctuations assist the system in escaping local potential wells in the energy landscape. 
Therefore, sometimes MV formation can even occur in regions forbidden by deterministic dynamics, although formation times in these regions (regime I) can be extremely long in the stochastic model. 

\begin{figure*}[tbhp!]%[tbhp!]
    \centering
    \includegraphics[width=\linewidth]{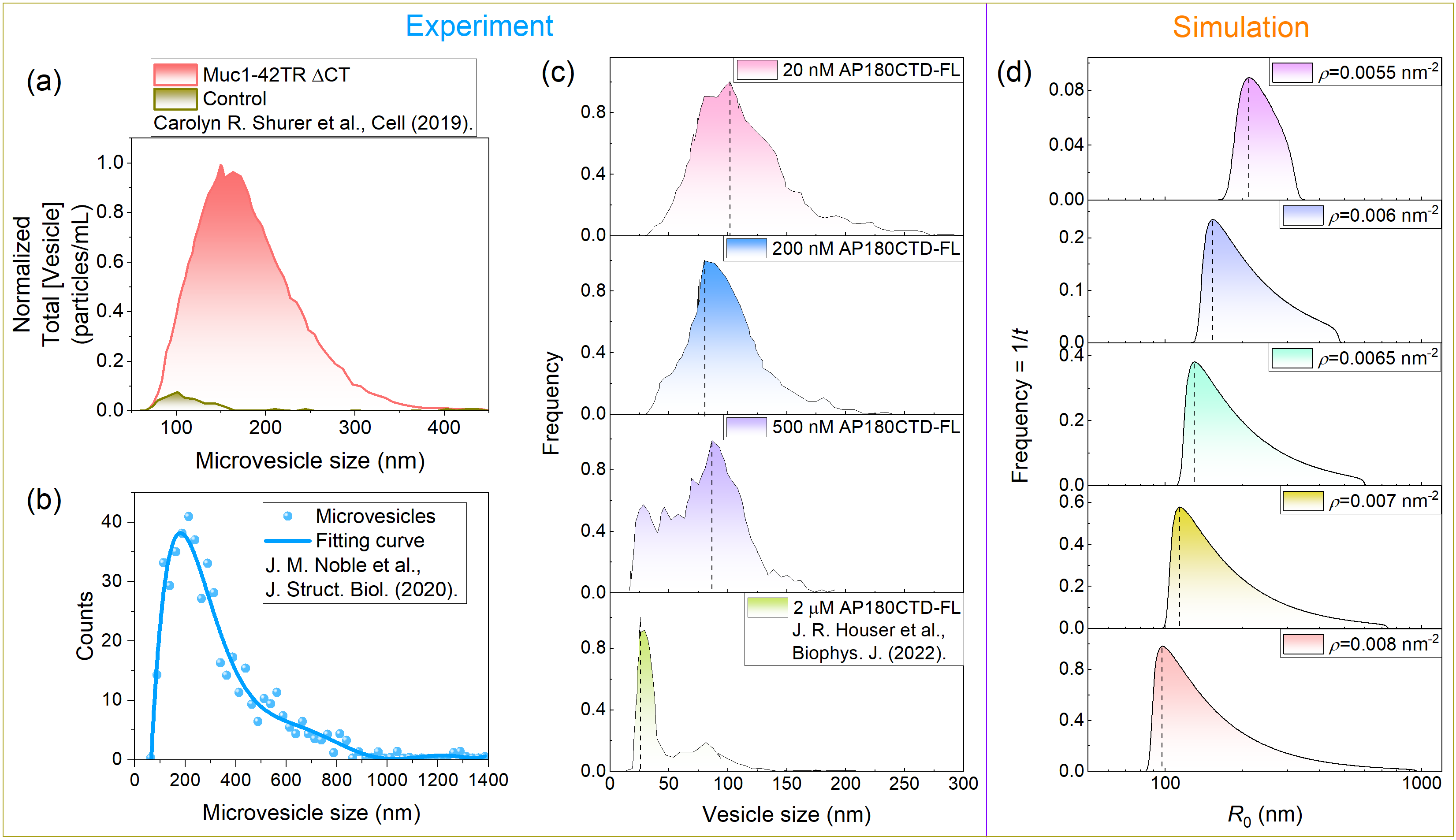}
    \caption{(a) The average concentration and size distribution of MVs released from wild-type (Control) and Muc1-42TR $\Delta$CT-expressing cells. Experimental data were extracted from Ref.~\cite{SHURER20191757}. 
    (b) The particle size distributions of MVs collected by cryo-TEM. 
    Experimental data were reproduced from Ref.~\cite{NOBLE2020107474}. 
%Data are compiled from Refs.~\cite{SHURER20191757,NOBLE2020107474}. 
    (c) Vesicle size distributions at different concentrations of AP180CTD-FL protein. Experimental data were reproduced from Ref.~\cite{HOUSER20223320}. 
    (d) MV formation frequency against $R_0$ for a sequence of glycocalyx grafting density, where the frequency is evaluated as the inverse of the mean first passage time in Eq.~(\ref{mean formation time_D_main}).  
    }
    \label{fig:figure_exp}
\end{figure*}
\textbf{Comparison with experiment.} 
Our deterministic and stochastic models suggest that a wide range of MV sizes can be produced by the regulation of glycocalyx. 
%To test this hypothesis, we resort to the experimental results from Refs.~\cite{SHURER20191757,NOBLE2020107474,HOUSER20223320}. 
To test this prediction, we compare our predictions with the experimental results of Refs.~\cite{SHURER20191757, NOBLE2020107474, HOUSER20223320}. 
In experiments~\cite{SHURER20191757}, quantitative analysis observed that Muc1-42TR $\Delta$CT-expressing cells secrete MVs in high concentration with a broad size distribution ($\sim100-400~{\rm nm}$), broader and more abundant than control cells [see Fig.~\ref{fig:figure_exp}(a)]. 
Moreover, cryo-TEM measurements from HAS3 cells also revealed heterogeneous vesicle sizes~\cite{NOBLE2020107474} [see Fig.~\ref{fig:figure_exp}(b)]. Furthermore, \textit{in vitro} experiments by Houser \textit{et al.}~\cite{HOUSER20223320} demonstrated that exposure of vesicles to the intrinsically disordered protein AP180CTD-FL can trigger membrane fission to smaller daughter vesicles with a size distribution at different protein concentrations [see Fig.~\ref{fig:figure_exp}(c)]. 
This suggests that the crowding of the intrinsically disordered protein AP180CTD-FL on membrane surfaces is sufficient to drive vesicle formation. 
The dashed lines in Fig.~\ref{fig:figure_exp}(c) exhibit a progressive decrease in vesicle size with increasing concentration of protein in solution. 
Consistent with this observation, our model also predicts that the optimal radius of the MV decreases with increasing glycocalyx density [see Fig.~\ref{fig:figure_exp}(d)]. 
Within our model, the peak frequency associated with the fastest MV formation rate implies that the presence of glycocalyx results in a larger number of MVs at the optimal radius compared to other sizes. 
An increase in glycocalyx grafting density causes the maximum frequency (or optimal radius) to shift from right to left. 
This might explain the experimental trend in the size of the vesicle demonstrated in Fig.~\ref{fig:figure_exp}(c). 
In Table~\ref{Table_Vesicle}, we summarize the median (peaks in the MV size distribution) and heterogeneous size distribution of MV derived from different cell types, where particle sizes are measured using various methodologies including cryo-TEM, cryo-TEM, EM and NTA. 
These experimental results collectively demonstrate that the vesicles are induced as a consequence of glycocalyx or biopolymers showing a heterogeneity in vesicle size. 
Our model captures this behavior, predicting a finite band of allowable MV sizes under glycocalyx regulation, in agreement with those found experimentally by Refs.~\cite{SHURER20191757, NOBLE2020107474, HOUSER20223320}. 
\begin{table*}[th!]%[!htbp]
\centering
\caption{MVs shed from different cell types were analyzed by transmission electron microscopy (TEM), electron microscopy (EM), and nanoparticle tracking analysis (NTA) to determine their size. 
The size distribution of MV derived from different cell types are shown in Fig.~\ref{fig:Exp_plots}. }
\label{Table_Vesicle}
%\begin{ruledtabular}
%\scalebox{0.9}{
\begin{tabular}{c c c c c}
MV type & Median MV size & Size distribution & Methodology & References \\
\hline
Muc1-42TR $\Delta$CT-expressing cells & $\sim150~{\rm nm}$ & $\sim50-450~{\rm nm}$ & Cryo-TEM & ~\cite{SHURER20191757} \\
\hline
HAS3-expressing cells & $\sim200~{\rm nm}$ & $\sim50-1000~{\rm nm}$ & Cryo-TEM & ~\cite{NOBLE2020107474} \\
\hline
HAS3-expressing cells & $\sim50~{\rm nm}$ & $\sim20-1000~{\rm nm}$ & NTA & ~\cite{NOBLE2020107474} \\
\hline
\makecell{Annexin V$+$ MVs in \\ human whole blood} & $\sim250~{\rm nm}$ & $\sim80-1000~{\rm nm}$ & Cryo-EM & ~\cite{W.L.Chandler2016,N.Arraud2014} \\
\hline
PC3-cell & $\sim190~{\rm nm}$ & $\sim50-500~{\rm nm}$ & NTA & ~\cite{A.Shimoda2022} \\
\hline
\makecell{GBM20/3 primary \\ human glioblastoma cell line} & $\sim83~{\rm nm}$ & $\sim20-250~{\rm nm}$ & TEM & ~\cite{shao2012protein} \\
\hline
\makecell{Primary tumour-derived (D3H1) \\ breast cancer cell lines} & $\sim110~{\rm nm}$ & $\sim25-450~{\rm nm}$ & NTA & ~\cite{Nishida-Aoki01012020} \\
\hline
\makecell{Lymph-node metastatic (D3H2LN) \\ breast cancer cell lines} & $\sim115~{\rm nm}$ & $\sim20-400~{\rm nm}$ & NTA & ~\cite{Nishida-Aoki01012020} \\
\hline
\makecell{Brain-metastatic (BMD2a) \\ breast cancer cell lines} & $\sim115~{\rm nm}$ & $\sim25-380~{\rm nm}$ & NTA & ~\cite{Nishida-Aoki01012020} \\
\hline
\makecell{MDA-231 SC human \\ breast cancer cell lines} & $\sim350~{\rm nm}$ & $\sim50-1000~{\rm nm}$ & NTA & ~\cite{WangTing2014} \\
\hline
\end{tabular}
%}
%\end{ruledtabular}
\end{table*}

\subsection*{Effects of membrane tension and spontaneous curvature on MV formation}
\begin{figure*}[tbhp!]
    \centering
    \includegraphics[width=\linewidth]{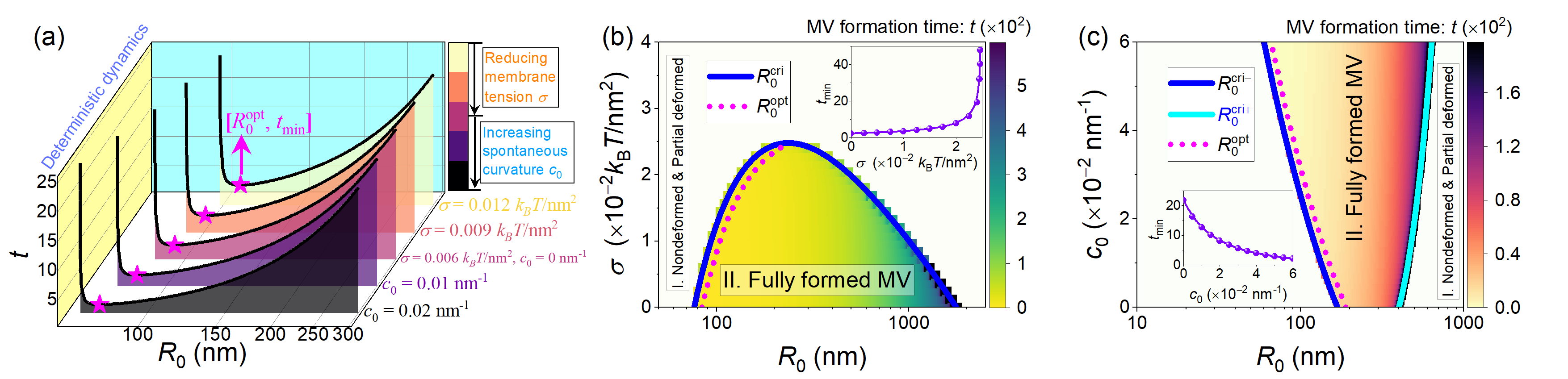}
    \caption{Deterministic model. (a) Formation time as a function of $R_0$ at different $\sigma$ and $c_0$. 
    Phase diagrams present the formation time as a function of $R_0$ and (b) membrane tension $\sigma$ and (c) spontaneous curvature $c_0$. 
    The boundary curves that separate regime I from regime II are denoted by solid lines, and the optimal radius is represented by the magenta dotted curves. 
    The inserts show the dependence of minimum formation times on membrane tension and spontaneous curvature. }
    \label{fig:figure4}
\end{figure*}

We now discuss the effects of membrane properties including membrane tension and spontaneous curvature on MV formation. 
Within the deterministic dynamics, we begin by performing the analysis of the formation time as a function of radius $R_0$ for different values of membrane tension $\sigma$ and spontaneous curvature $c_0$, as shown in Fig.~\ref{fig:figure4}(a). 
Figure.~\ref{fig:figure4}(a) exhibits several features similar to those described in the previous section, such as the existence of a minimum critical radius $R_0^{\rm cri}$ required for the formation of a closed MV, and an optimal radius $R_0^{\rm opt}$ at which the MV formation time reaches a minimum (indicated by solid star points). 
When comparing curves corresponding to various membrane tensions, we observe that an increase in the level of membrane tension will result in a longer time consuming for MV formation. 
This indicates that MV formation is faster in the low membrane tension region on cell surfaces. 
A direct comparison of the curves at distinct spontaneous curvatures illustrates that the presence of spontaneous curvature leads to an acceleration of the MV formation. 
To systematically investigate the interrelated effects of membrane tension and spontaneous curvature on the dynamics of MV formation, two phase diagrams on the $R_0-\sigma$ and $R_0-c_0$ plane are constructed [see Figs.~\ref{fig:figure4}(b) and~\ref{fig:figure4}(c)]. 
The colored contour maps of formation time $t$ show that, given a radius $R_0$, a higher formation efficiency can be achieved under a lower membrane tension or a larger spontaneous curvature, i.e., a looser and more asymmetric membrane leads to faster MV formation. 
In both phase diagrams, the solid boundary curves that separate regime I from regime II represent the lower and upper critical radii, while the magenta dotted curves denote the optimal radius with their corresponding minimum formation times presented in the insets. 
We note that the optimal radius is close to the lower critical radius. 
Intriguingly, $R_0^{\rm cri}$ and $R_0^{\rm opt}$ depend nonlinearly on membrane tension, while exhibit an almost linear dependence on spontaneous curvature.  
In Fig.~\ref{fig:figure4}(b), the solid boundary curve and the magenta dotted curve meet at a specific point where the lower and upper critical radii and the optimal radius are equal to each other. 

\begin{figure*}[tbhp!]
    \centering
    \includegraphics[width=\linewidth]{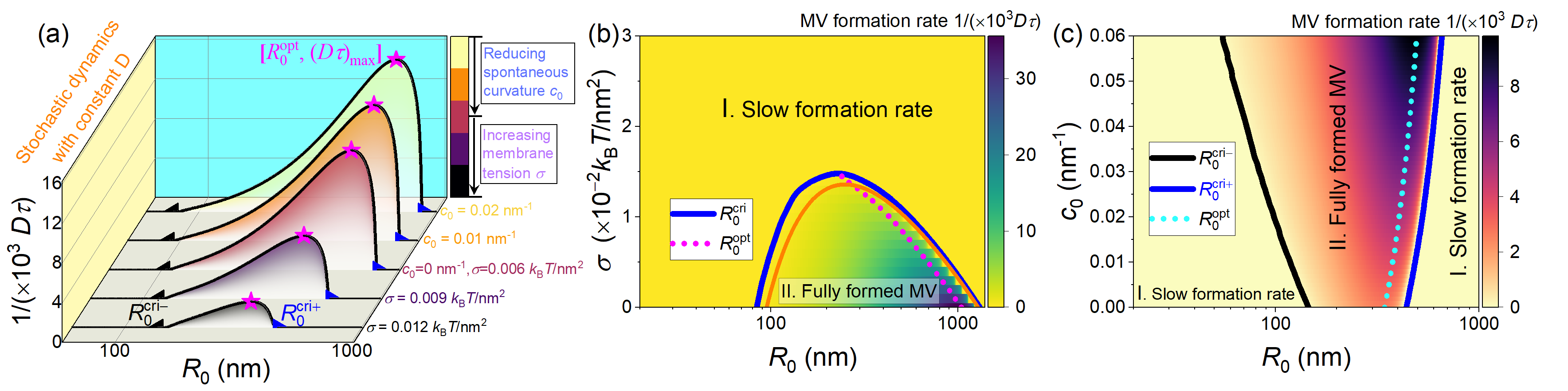}
    \caption{Stochastic model with a constant transition rate $D$. (a) Formation rate as a function of $R_0$ at different $\sigma$ and $c_0$. 
    Phase diagrams present the formation rate as a function of $R_0$ and (b) membrane tension $\sigma$ and (c) spontaneous curvature $c_0$. }
    \label{fig:figure5}
\end{figure*}

In contrast, we also plot the formation rate against $R_0$ for a
sequence of fixed membrane tension and spontaneous curvature in Fig.~\ref{fig:figure5}(a) for stochastic dynamics under a constant transition rate $D$. 
Figure.~\ref{fig:figure5}(a) shows that the range of $R_0$ that allows MV formation narrows as we increase $\sigma$ or decrease $c_0$.
This size selectivity arises from an increase in the lower critical radius $R_0^{\rm cri-}$ and a decrease in the upper critical radius $R_0^{\rm cri+}$. 
In addition to the accessible MV size range, we also find that both the maximum formation rate and the corresponding optimal radius $R_0^{\rm opt}$ are reduced under these conditions. 
Then, the interrelated effects of $\sigma-R_0$ and $c_0-R_0$ on the formation rate are summarized in the phase diagrams of Figs.~\ref{fig:figure5}(b) and~\ref{fig:figure5}(c). 
Specifically, depending on membrane tension (spontaneous curvature) and glycocalyx coverage, two regimes corresponding to MV formation and no or partial MV formation are identified, in which the color indicates the formation rate. 
Comparing Figs.~\ref{fig:figure5}(b) and~\ref{fig:figure5}(c) with their deterministic counterparts Figs.~\ref{fig:figure4}(b) and~\ref{fig:figure4}(c), we observe an analogous variation trend for the boundary curves except for the optimal radius and the maximum formation rate. 
In the stochastic model with constant $D$, the optimal radius closely follows with the upper critical radius, in contrast to the deterministic case where it is closer to the lower limit.  
%In contrast, Figs.~\ref{fig:figure5}(b) and~\ref{fig:figure5}(c) show that the optimal radius follows closely with the upper critical radius in the stochastic model. 
%Analogous analysis can be performed for the stochastic dynamics. 
In summary, varying membrane tension and spontaneous curvature not only affect the size of MV that can be produced, but also profoundly influence the formation rate. 
Importantly, both the deterministic and stochastic approaches yield qualitatively consistent predictions regarding the effects of these membrane properties on MV formation dynamics. 

\section*{Discussion and Conclusion}
In conclusion, we have studied the formation dynamics of MVs at cell membranes mediated by the glycocalyx. 
Using the Onsager variational principle, we derived a nonlinear deterministic equation for MV formation and complemented it with a stochastic description to capture the role of fluctuations. 
%Using the Onsager variational principle, we derived a nonlinear deterministic dynamical equation governing MV formation and complemented this with a stochastic description to capture the role of fluctuations. 
%The nonlinear deterministic dynamical equation for the formation of MV on cell membranes has been derived by using the Onsager variational principle, and also a stochastic description is provided for the formation dynamics. 
Given the complexity of the biological system, it is impractical to explore all possible factors involved in the formation of MVs. Therefore, we focus only on the impact of key physical factors, namely glycocalyx properties (grafting density and polymer chain length), degree of ionization, and ionic strength, on the dynamics of this formation process. 
Our findings suggest that these parameters play an essential role in regulating the formation of MVs. 
The MV formation process is largely controlled by the interplay between the energy associated with the glycocalyx, the elastic energy penalties that result from changing the membrane morphology, and different types of dissipation energy. 
Note that the deterministic approach is equivalent to a Lagrangian description in which kinetic equations can be obtained by balancing ``driving forces'' $f_{\rm driving}$ and ``dissipative forces'' $f_{\rm diss}$. 
Generally, the ``driving forces'' and the ``dissipative forces'' are calculated from the variation of the total free energy and the energy dissipated per unit time. 
In this work, the primary ``driving force'' for MV formation arises from the entropic bending force generated by the glycocalyx, whose magnitude is modulated by multiple factors, including polymer grafting density, polymer length, degree of ionization, and ionic strength.  
%In this work, the primary ``driving forces'' originate from the entropic bending force generated by the glycocalyx, with this driving force modulated by multiple factors, including polymer grafting density, polymer length, degree of ionization, and ionic strength. 

%Consequently, on the basis of these competing factors, our results predict that MV formation is accelerated on cell surfaces decorated with a layer of denser and thicker glycocalyx, with the formation efficiency further improved upon enhancing the degree of ionization or reducing the ionic strength. 
Consequently, on the basis of these competing factors, our results predict that MV formation is accelerated for cell surfaces with a layer of denser and thicker glycocalyx, and the formation efficiency can be further improved upon enhancing the degree of ionization or reducing the ionic strength. 
This finding suggests that cells may regulate glycocalyx properties to control membrane morphology, given their ability to dynamically modify glycocalyx composition and architecture to modulate cell behavior~\cite{JoeChin-HunKuo.2021}. 
In the extracellular space, changes in local ionic strength and pH provide a biologically relevant way to control MV formation, as such variations can selectively alter the charge or ionization state of glycocalyx components, thereby modulating their physical properties~\cite{J.C.H.Kuo2018,Weinbaum2007,Leonhard2020.00253}. 
%In the extracellular space, changes in local ionic strength and pH provide a biologically relevant way to control MV formation, as variations in these factors can selectively alter charge or the degree of ionization to regulate glycocalyx properties~\cite{J.C.H.Kuo2018,Weinbaum2007,Leonhard2020.00253}. 
This points to a possible mechanism by which cells endowed with dense glycocalyx can tune MV production in response to changes in the local ionic environment. 
More broadly, our findings may also have implications for other membrane trafficking processes involving large glycosylated proteins, including the formation of protrusive structures such as filopodia~\cite{RichardBennett2001,KOISTINEN2015179}, microvilli~\cite{HattrupChristineL.2008,YunminJung2016,KESAVAN2009791,KESIMER2013379,Makabe2006,EVANKO20071351} and cilia~\cite{BrianButton2012}, as well as cellular blebbing~\cite{SHURER20191757,Beyer2010-os,Ikenouchi2022-um,Charras2006-gr}. 
%In addition, these results may provide insight into budding and tubulation of lipid vesicles induced by polymer anchoring on membranes in \textit{in vitro} systems~\cite{I.Tsafrir2003,I.Tsafrir2001,Huang2022}. 
In addition, this may also have implications for understanding budding and tubulation induced in lipid vesicles by the anchoring of polymers on the membrane in \textit{in vitro} experiments~\cite{I.Tsafrir2003,I.Tsafrir2001,Huang2022}.   

%Regarding the dependence of optimal radius on glycocalyx polymer properties and solution conditions, the deterministic model and the stochastic model with a nonconstant transition rate yield consistent predictions, while the stochastic model with constant transition rates leads to opposite trends. 
Regarding the dependence of the optimal radius on glycocalyx polymer properties and solution conditions, the deterministic model and the stochastic model with a nonconstant transition rate yield consistent predictions, whereas the stochastic model with a constant transition rate predicts opposite trends. 
%The presence of noise in the stochastic model with a nonconstant transition rate gives rise to shorter formation times compared with the deterministic model, implying that fluctuations are important and expected to help the system overcome barriers. 
The presence of noise in the stochastic model with a nonconstant transition rate gives rise to shorter formation times compared with the deterministic model, implying that fluctuations are important and expected to facilitate barrier crossing and therefore accelerate MV formation. 
Giving these findings, we suggest that the transition rate is unlikely to be constant, which needs to be further tested by experiments. 
In addition to the ionic strength of the surrounding environment, stochastic fluctuations also play an important role in facilitating the formation of MV. 
Our framework could be extended to study other biologically relevant first passage problems, such as the wrapping of nanoparticles or viruses~\cite{StephenMirigian2013,FelixFrey2019}, clathrin-mediated endocytosis~\cite{FelixFrey2024}, vesicle translocation through membrane pores~\cite{Shojaei2016,BARUAH2025740}, and tissue fusion~\cite{Vincent2015}, where the coefficient $D$ is treated as a constant. 
%Moreover, our results suggest that the transition rate is not constant and the presence of noise in the stochastic model gives rise to shorter formation times compared with the deterministic model. 
%These predictions are supported by recent experimental observations. 
%Our results also reveal that this formation process may occur faster on a looser cell membrane in the presence of spontaneous curvature. 
Additionally, we find that formation proceeds more rapidly on looser membranes and in the presence of spontaneous curvature, indicating that spontaneous curvature also serves as a driving factor assisting membrane deformation. 

%\textbf{Limitations and model simplifications.} 
While our model shows agreement with experiments and provides testable predictions, we acknowledge that its simplified nature imposes certain limitations. 
For example, (i) our theoretical model prescribes the shape of the membrane with glycocalyx grafted as a simplified geometrical scheme, and hence cannot be used to simulate vesicle scission assisted by the recruitment of
specific proteins to a membrane~\cite{J.Liu2006,N.Walani2015,J.E.Hassinger2017}. 
(ii) It is assumed that the membrane is attached to the cortex by means of discrete linkers with constant density, thus neglecting the diffusion, binding, and unbinding kinetics of these protein linkers~\cite{R.Alert2016,ALERT20151878}. 
(iii) Although cytoskeletal rearrangements have been implicated in MV generation~\cite{SHURER20191757}, our current mathematical framework is not equipped to model the disassembly or remodeling of the cytoskeletal core. 
%Cytoskeletal filaments may also involved in MV generation~\cite{SHURER20191757}, but our mathematical framework is not equipped to model the disassembly of the cytoskeletal core. 

Despite there are limitations and simplifications, both deterministic and stochastic models presented here provide a useful guide for understanding the role of glycocalyx properties and ionic strength in the dynamics of MV formation. 
A few additional remarks are worth noting. 
First, the glycocalyx is typically negatively charged. 
As an example, mucins generally carry approximately five electron charges per 20 amino acids due to the ionization of acidic groups. 
Accordingly, the amount of charge can thus be modulated by the concentration of hydrogen ions in the glycocalyx, i.e., the pH value, implying that the degree of ionization of the glycocalyx is related to the pH value of the surrounding environment. 
%Second, we used the formation rate defined as $\tau_0/\tau$ to measure the formation time in the stochastic model due to the magnitude of the reduced formation time $\tau/\tau_0$ is too large, where $\tau_0=1/D$. 
Second, in the stochastic model when $D$ is constant, we characterize the formation kinetics using the normalized formation rate $\tau_0/\tau$, where $\tau_0 = 1/D$, to account for the large magnitude of the reduced formation time $\tau/\tau_0$. 
Since the transition rate $D$ is not known \textit{a priori}, Eq.~(\ref{mean_first passage time}) cannot be expected to provide the absolute formation times. 
Nevertheless, this normalization offers a convenient and meaningful way to present the results. 
%The transition rate $D$ is an unknown parameter, and therefore we use $\tau_0=1/D$ to normalize the formation rate. 
Third, the characteristic timescale associated with vesicle formation has been experimentally reported to be of the order of $10^{-2}~s$~\cite{CAVANAUGH2020152,FuqiangSun2023}, and this timescale is in good agreement with the prediction of both our deterministic model and the stochastic model when $D$ is not constant. 
This study uncovers the physical origin behind glycocalyx-regulated MV formation in nonequilibrium states. 
We expect that the findings in this work have implications for understanding how glycocalyx and different physiological conditions modulate the formation of MVs on cellular membranes and provide some guidance on how to control this process, which could be of immediate interest to promote their biomedical applications.

\textbf{Data, Materials, and Software Availability.} All study data are included in the article and/or SI Appendix. 

\section*{Acknowledgments}
The authors would like to thank Prof. Padmini Rangamani and Dr. Emmet A. Francis for their fruitful discussions and thoughtful feedback on this manuscript. 
We acknowledge the financial supports from the National Natural Science Foundation of China under Grant No.12574232 and the 111 project B16029.

\section*{Declaration of Interests}
The authors declare no competing interest.

\bibliography{apssamp}% Produces the bibliography via BibTeX.

\begin{thebibliography}{65}%
\makeatletter
\providecommand \@ifxundefined [1]{%
 \@ifx{#1\undefined}
}%
\providecommand \@ifnum [1]{%
 \ifnum #1\expandafter \@firstoftwo
 \else \expandafter \@secondoftwo
 \fi
}%
\providecommand \@ifx [1]{%
 \ifx #1\expandafter \@firstoftwo
 \else \expandafter \@secondoftwo
 \fi
}%
\providecommand \natexlab [1]{#1}%
\providecommand \enquote  [1]{``#1''}%
\providecommand \bibnamefont  [1]{#1}%
\providecommand \bibfnamefont [1]{#1}%
\providecommand \citenamefont [1]{#1}%
\providecommand \href@noop [0]{\@secondoftwo}%
\providecommand \href [0]{\begingroup \@sanitize@url \@href}%
\providecommand \@href[1]{\@@startlink{#1}\@@href}%
\providecommand \@@href[1]{\endgroup#1\@@endlink}%
\providecommand \@sanitize@url [0]{\catcode `\\12\catcode `\$12\catcode
  `\&12\catcode `\#12\catcode `\^12\catcode `\_12\catcode `\%12\relax}%
\providecommand \@@startlink[1]{}%
\providecommand \@@endlink[0]{}%
\providecommand \url  [0]{\begingroup\@sanitize@url \@url }%
\providecommand \@url [1]{\endgroup\@href {#1}{\urlprefix }}%
\providecommand \urlprefix  [0]{URL }%
\providecommand \Eprint [0]{\href }%
\providecommand \doibase [0]{https://doi.org/}%
\providecommand \selectlanguage [0]{\@gobble}%
\providecommand \bibinfo  [0]{\@secondoftwo}%
\providecommand \bibfield  [0]{\@secondoftwo}%
\providecommand \translation [1]{[#1]}%
\providecommand \BibitemOpen [0]{}%
\providecommand \bibitemStop [0]{}%
\providecommand \bibitemNoStop [0]{.\EOS\space}%
\providecommand \EOS [0]{\spacefactor3000\relax}%
\providecommand \BibitemShut  [1]{\csname bibitem#1\endcsname}%
\let\auto@bib@innerbib\@empty
%</preamble>
\bibitem [{\citenamefont {Levine}\ and\ \citenamefont
  {Steinhardt}(1984)}]{Levine1984quasicrystals}%
  \BibitemOpen
  \bibfield  {author} {\bibinfo {author} {\bibfnamefont {D.}~\bibnamefont
  {Levine}}\ and\ \bibinfo {author} {\bibfnamefont {P.~J.}\ \bibnamefont
  {Steinhardt}},\ }\bibfield  {title} {\bibinfo {title} {Quasicrystals: A new
  class of ordered structures},\ }\href
  {https://doi.org/10.1103/PhysRevLett.53.2477} {\bibfield  {journal} {\bibinfo
   {journal} {Phys. Rev. Lett.}\ }\textbf {\bibinfo {volume} {53}},\ \bibinfo
  {pages} {2477} (\bibinfo {year} {1984})}\BibitemShut {NoStop}%
\bibitem [{\citenamefont {Shechtman}\ \emph {et~al.}(1984)\citenamefont
  {Shechtman}, \citenamefont {Blech}, \citenamefont {Gratias},\ and\
  \citenamefont {Cahn}}]{Shechtman1984metallic}%
  \BibitemOpen
  \bibfield  {author} {\bibinfo {author} {\bibfnamefont {D.}~\bibnamefont
  {Shechtman}}, \bibinfo {author} {\bibfnamefont {I.}~\bibnamefont {Blech}},
  \bibinfo {author} {\bibfnamefont {D.}~\bibnamefont {Gratias}},\ and\ \bibinfo
  {author} {\bibfnamefont {J.~W.}\ \bibnamefont {Cahn}},\ }\bibfield  {title}
  {\bibinfo {title} {Metallic phase with long-range orientational order and no
  translational symmetry},\ }\href
  {https://doi.org/10.1103/PhysRevLett.53.1951} {\bibfield  {journal} {\bibinfo
   {journal} {Phys. Rev. Lett.}\ }\textbf {\bibinfo {volume} {53}},\ \bibinfo
  {pages} {1951} (\bibinfo {year} {1984})}\BibitemShut {NoStop}%
\bibitem [{\citenamefont {Ishimasa}\ \emph {et~al.}(1985)\citenamefont
  {Ishimasa}, \citenamefont {Nissen},\ and\ \citenamefont
  {Fukano}}]{Ishimasa1985new}%
  \BibitemOpen
  \bibfield  {author} {\bibinfo {author} {\bibfnamefont {T.}~\bibnamefont
  {Ishimasa}}, \bibinfo {author} {\bibfnamefont {H.-U.}\ \bibnamefont
  {Nissen}},\ and\ \bibinfo {author} {\bibfnamefont {Y.}~\bibnamefont
  {Fukano}},\ }\bibfield  {title} {\bibinfo {title} {New ordered state between
  crystalline and amorphous in ni-cr particles},\ }\href
  {https://doi.org/10.1103/PhysRevLett.55.511} {\bibfield  {journal} {\bibinfo
  {journal} {Phys. Rev. Lett.}\ }\textbf {\bibinfo {volume} {55}},\ \bibinfo
  {pages} {511} (\bibinfo {year} {1985})}\BibitemShut {NoStop}%
\bibitem [{\citenamefont {P}\ and\ \citenamefont
  {Steinhardt}(1991)}]{DiVincenzo1991quasicrystals}%
  \BibitemOpen
  \bibfield  {author} {\bibinfo {author} {\bibfnamefont {D.~D.}\ \bibnamefont
  {P}}\ and\ \bibinfo {author} {\bibfnamefont {P.~J.}\ \bibnamefont
  {Steinhardt}},\ }\href@noop {} {\emph {\bibinfo {title}
  {\href{https://search.worldcat.org/title/quasicrystals-the-state-of-the-art/oclc/24431056}{Quasicrystals:
  The State of the Art}}}}\ (\bibinfo  {publisher} {World Scientific,
  Singapore},\ \bibinfo {year} {1991})\BibitemShut {NoStop}%
\bibitem [{\citenamefont {Janot}(1994)}]{janot1994quasicrystals}%
  \BibitemOpen
  \bibfield  {author} {\bibinfo {author} {\bibfnamefont {C.}~\bibnamefont
  {Janot}},\ }\href@noop {} {\emph {\bibinfo {title}
  {\href{https://global.oup.com/academic/product/quasicrystals-a-primer-9780198513896?lang=en&cc=nl}{Quasicrystals:
  A primer}}}}\ (\bibinfo  {publisher} {Oxford University Press, UK},\ \bibinfo
  {year} {1994})\BibitemShut {NoStop}%
\bibitem [{\citenamefont {Bindi}\ \emph {et~al.}(2009)\citenamefont {Bindi},
  \citenamefont {Steinhardt}, \citenamefont {Yao},\ and\ \citenamefont
  {Lu}}]{Bindi2009natural}%
  \BibitemOpen
  \bibfield  {author} {\bibinfo {author} {\bibfnamefont {L.}~\bibnamefont
  {Bindi}}, \bibinfo {author} {\bibfnamefont {P.~J.}\ \bibnamefont
  {Steinhardt}}, \bibinfo {author} {\bibfnamefont {N.}~\bibnamefont {Yao}},\
  and\ \bibinfo {author} {\bibfnamefont {P.~J.}\ \bibnamefont {Lu}},\
  }\bibfield  {title} {\bibinfo {title} {Natural quasicrystals},\ }\href
  {https://doi.org/10.1126/science.1170827} {\bibfield  {journal} {\bibinfo
  {journal} {Science}\ }\textbf {\bibinfo {volume} {324}},\ \bibinfo {pages}
  {1306} (\bibinfo {year} {2009})}\BibitemShut {NoStop}%
\bibitem [{\citenamefont {Maciá}(2005)}]{Macia2006}%
  \BibitemOpen
  \bibfield  {author} {\bibinfo {author} {\bibfnamefont {E.}~\bibnamefont
  {Maciá}},\ }\bibfield  {title} {\bibinfo {title} {The role of aperiodic
  order in science and technology},\ }\href
  {https://doi.org/10.1088/0034-4885/69/2/R03} {\bibfield  {journal} {\bibinfo
  {journal} {Rep. Prog. Phys.}\ }\textbf {\bibinfo {volume} {69}},\ \bibinfo
  {pages} {397} (\bibinfo {year} {2005})}\BibitemShut {NoStop}%
\bibitem [{\citenamefont {Dubois}(2012)}]{dubois2012properties}%
  \BibitemOpen
  \bibfield  {author} {\bibinfo {author} {\bibfnamefont {J.-M.}\ \bibnamefont
  {Dubois}},\ }\bibfield  {title} {\bibinfo {title} {Properties and
  applications of quasicrystals and complex metallic alloys},\ }\href
  {https://doi.org/10.1039/C2CS35110B} {\bibfield  {journal} {\bibinfo
  {journal} {Chemical Society Reviews}\ }\textbf {\bibinfo {volume} {41}},\
  \bibinfo {pages} {6760} (\bibinfo {year} {2012})}\BibitemShut {NoStop}%
\bibitem [{\citenamefont {Zeng}\ \emph {et~al.}(2004)\citenamefont {Zeng},
  \citenamefont {Ungar}, \citenamefont {Liu}, \citenamefont {Percec},
  \citenamefont {Dulcey},\ and\ \citenamefont {Hobbs}}]{xianbing2004}%
  \BibitemOpen
  \bibfield  {author} {\bibinfo {author} {\bibfnamefont {X.}~\bibnamefont
  {Zeng}}, \bibinfo {author} {\bibfnamefont {G.}~\bibnamefont {Ungar}},
  \bibinfo {author} {\bibfnamefont {Y.}~\bibnamefont {Liu}}, \bibinfo {author}
  {\bibfnamefont {V.}~\bibnamefont {Percec}}, \bibinfo {author} {\bibfnamefont
  {A.~E.}\ \bibnamefont {Dulcey}},\ and\ \bibinfo {author} {\bibfnamefont
  {J.~K.}\ \bibnamefont {Hobbs}},\ }\bibfield  {title} {\bibinfo {title}
  {Supramolecular dendritic liquid quasicrystals},\ }\href
  {https://doi.org/10.1038/nature02368} {\bibfield  {journal} {\bibinfo
  {journal} {Nature}\ }\textbf {\bibinfo {volume} {428}},\ \bibinfo {pages}
  {157} (\bibinfo {year} {2004})}\BibitemShut {NoStop}%
\bibitem [{\citenamefont {Hayashida}\ \emph {et~al.}(2007)\citenamefont
  {Hayashida}, \citenamefont {Dotera}, \citenamefont {Takano},\ and\
  \citenamefont {Matsushita}}]{Hayashida2007}%
  \BibitemOpen
  \bibfield  {author} {\bibinfo {author} {\bibfnamefont {K.}~\bibnamefont
  {Hayashida}}, \bibinfo {author} {\bibfnamefont {T.}~\bibnamefont {Dotera}},
  \bibinfo {author} {\bibfnamefont {A.}~\bibnamefont {Takano}},\ and\ \bibinfo
  {author} {\bibfnamefont {Y.}~\bibnamefont {Matsushita}},\ }\bibfield  {title}
  {\bibinfo {title} {Polymeric quasicrystal: Mesoscopic quasicrystalline tiling
  in $abc$ star polymers},\ }\href
  {https://doi.org/10.1103/PhysRevLett.98.195502} {\bibfield  {journal}
  {\bibinfo  {journal} {Phys. Rev. Lett.}\ }\textbf {\bibinfo {volume} {98}},\
  \bibinfo {pages} {195502} (\bibinfo {year} {2007})}\BibitemShut {NoStop}%
\bibitem [{\citenamefont {Talapin}\ \emph {et~al.}(2009)\citenamefont
  {Talapin}, \citenamefont {Shevchenko}, \citenamefont {Bodnarchuk},
  \citenamefont {Ye}, \citenamefont {Chen},\ and\ \citenamefont
  {Murray}}]{talapin2009qc}%
  \BibitemOpen
  \bibfield  {author} {\bibinfo {author} {\bibfnamefont {D.~V.}\ \bibnamefont
  {Talapin}}, \bibinfo {author} {\bibfnamefont {E.~V.}\ \bibnamefont
  {Shevchenko}}, \bibinfo {author} {\bibfnamefont {M.~I.}\ \bibnamefont
  {Bodnarchuk}}, \bibinfo {author} {\bibfnamefont {X.}~\bibnamefont {Ye}},
  \bibinfo {author} {\bibfnamefont {J.}~\bibnamefont {Chen}},\ and\ \bibinfo
  {author} {\bibfnamefont {C.~B.}\ \bibnamefont {Murray}},\ }\bibfield  {title}
  {\bibinfo {title} {Quasicrystalline order in self-assembled binary
  nanoparticle superlattices},\ }\href {https://doi.org/10.1038/nature08439}
  {\bibfield  {journal} {\bibinfo  {journal} {Nature}\ }\textbf {\bibinfo
  {volume} {461}},\ \bibinfo {pages} {964} (\bibinfo {year}
  {2009})}\BibitemShut {NoStop}%
\bibitem [{\citenamefont {Steurer}(2012)}]{C2CS35063G}%
  \BibitemOpen
  \bibfield  {author} {\bibinfo {author} {\bibfnamefont {W.}~\bibnamefont
  {Steurer}},\ }\bibfield  {title} {\bibinfo {title} {Why are quasicrystals
  quasiperiodic?},\ }\href {https://doi.org/10.1039/C2CS35063G} {\bibfield
  {journal} {\bibinfo  {journal} {Chem. Soc. Rev.}\ }\textbf {\bibinfo {volume}
  {41}},\ \bibinfo {pages} {6719} (\bibinfo {year} {2012})}\BibitemShut
  {NoStop}%
\bibitem [{\citenamefont {Dotera}(2011)}]{Tomonari1197QC}%
  \BibitemOpen
  \bibfield  {author} {\bibinfo {author} {\bibfnamefont {T.}~\bibnamefont
  {Dotera}},\ }\bibfield  {title} {\bibinfo {title} {Quasicrystals in soft
  matter},\ }\href {https://doi.org/https://doi.org/10.1002/ijch.201100146}
  {\bibfield  {journal} {\bibinfo  {journal} {Isr. J. Chem.}\ }\textbf
  {\bibinfo {volume} {51}},\ \bibinfo {pages} {1197} (\bibinfo {year}
  {2011})}\BibitemShut {NoStop}%
\bibitem [{\citenamefont {Mikhael}\ \emph {et~al.}(2008)\citenamefont
  {Mikhael}, \citenamefont {Roth}, \citenamefont {Helden},\ and\ \citenamefont
  {Bechinger}}]{Jules2008archimedean}%
  \BibitemOpen
  \bibfield  {author} {\bibinfo {author} {\bibfnamefont {J.}~\bibnamefont
  {Mikhael}}, \bibinfo {author} {\bibfnamefont {J.}~\bibnamefont {Roth}},
  \bibinfo {author} {\bibfnamefont {L.}~\bibnamefont {Helden}},\ and\ \bibinfo
  {author} {\bibfnamefont {C.}~\bibnamefont {Bechinger}},\ }\bibfield  {title}
  {\bibinfo {title} {Archimedean-like tiling on decagonal quasicrystalline
  surfaces},\ }\href {https://doi.org/10.1038/nature07074} {\bibfield
  {journal} {\bibinfo  {journal} {Nature}\ }\textbf {\bibinfo {volume} {454}},\
  \bibinfo {pages} {501} (\bibinfo {year} {2008})}\BibitemShut {NoStop}%
\bibitem [{\citenamefont {F{\"o}rster}\ \emph {et~al.}(2013)\citenamefont
  {F{\"o}rster}, \citenamefont {Meinel}, \citenamefont {Hammer}, \citenamefont
  {Trautmann},\ and\ \citenamefont {Widdra}}]{forster2013quasicrystalline}%
  \BibitemOpen
  \bibfield  {author} {\bibinfo {author} {\bibfnamefont {S.}~\bibnamefont
  {F{\"o}rster}}, \bibinfo {author} {\bibfnamefont {K.}~\bibnamefont {Meinel}},
  \bibinfo {author} {\bibfnamefont {R.}~\bibnamefont {Hammer}}, \bibinfo
  {author} {\bibfnamefont {M.}~\bibnamefont {Trautmann}},\ and\ \bibinfo
  {author} {\bibfnamefont {W.}~\bibnamefont {Widdra}},\ }\bibfield  {title}
  {\bibinfo {title} {Quasicrystalline structure formation in a classical
  crystalline thin-film system},\ }\href {https://doi.org/10.1038/nature12514}
  {\bibfield  {journal} {\bibinfo  {journal} {Nature}\ }\textbf {\bibinfo
  {volume} {502}},\ \bibinfo {pages} {215} (\bibinfo {year}
  {2013})}\BibitemShut {NoStop}%
\bibitem [{\citenamefont {Penrose}(1974)}]{penrose1974role}%
  \BibitemOpen
  \bibfield  {author} {\bibinfo {author} {\bibfnamefont {R.}~\bibnamefont
  {Penrose}},\ }\bibfield  {title} {\bibinfo {title} {The role of aesthetics in
  pure and applied mathematical research},\ }\href
  {https://cir.nii.ac.jp/crid/1572543024090822912} {\bibfield  {journal}
  {\bibinfo  {journal} {Bull. Inst. Math. Appl.}\ }\textbf {\bibinfo {volume}
  {10}},\ \bibinfo {pages} {266} (\bibinfo {year} {1974})}\BibitemShut
  {NoStop}%
\bibitem [{\citenamefont {Lu}\ and\ \citenamefont
  {Steinhardt}(2007)}]{Lu2007medieval}%
  \BibitemOpen
  \bibfield  {author} {\bibinfo {author} {\bibfnamefont {P.~J.}\ \bibnamefont
  {Lu}}\ and\ \bibinfo {author} {\bibfnamefont {P.~J.}\ \bibnamefont
  {Steinhardt}},\ }\bibfield  {title} {\bibinfo {title} {Decagonal and
  quasi-crystalline tilings in medieval islamic architecture},\ }\href
  {https://doi.org/10.1126/science.1135491} {\bibfield  {journal} {\bibinfo
  {journal} {Science}\ }\textbf {\bibinfo {volume} {315}},\ \bibinfo {pages}
  {1106} (\bibinfo {year} {2007})}\BibitemShut {NoStop}%
\bibitem [{\citenamefont {Plati}\ \emph {et~al.}(2024)\citenamefont {Plati},
  \citenamefont {Maire}, \citenamefont {Fayen}, \citenamefont {Boulogne},
  \citenamefont {Restagno}, \citenamefont {Smallenburg},\ and\ \citenamefont
  {Foffi}}]{plati2023quasicrystalline}%
  \BibitemOpen
  \bibfield  {author} {\bibinfo {author} {\bibfnamefont {A.}~\bibnamefont
  {Plati}}, \bibinfo {author} {\bibfnamefont {R.}~\bibnamefont {Maire}},
  \bibinfo {author} {\bibfnamefont {E.}~\bibnamefont {Fayen}}, \bibinfo
  {author} {\bibfnamefont {F.}~\bibnamefont {Boulogne}}, \bibinfo {author}
  {\bibfnamefont {F.}~\bibnamefont {Restagno}}, \bibinfo {author}
  {\bibfnamefont {F.}~\bibnamefont {Smallenburg}},\ and\ \bibinfo {author}
  {\bibfnamefont {G.}~\bibnamefont {Foffi}},\ }\bibfield  {title} {\bibinfo
  {title} {Quasi-crystalline order in vibrating granular matter},\ }\href
  {https://doi.org/10.1038/s41567-023-02364-1} {\bibfield  {journal} {\bibinfo
  {journal} {Nat. Phys.}\ }\textbf {\bibinfo {volume} {20}},\ \bibinfo {pages}
  {465} (\bibinfo {year} {2024})}\BibitemShut {NoStop}%
\bibitem [{\citenamefont {Mivehvar}\ \emph {et~al.}(2019)\citenamefont
  {Mivehvar}, \citenamefont {Ritsch},\ and\ \citenamefont
  {Piazza}}]{Mivehvar2019}%
  \BibitemOpen
  \bibfield  {author} {\bibinfo {author} {\bibfnamefont {F.}~\bibnamefont
  {Mivehvar}}, \bibinfo {author} {\bibfnamefont {H.}~\bibnamefont {Ritsch}},\
  and\ \bibinfo {author} {\bibfnamefont {F.}~\bibnamefont {Piazza}},\
  }\bibfield  {title} {\bibinfo {title} {Emergent quasicrystalline symmetry in
  light-induced quantum phase transitions},\ }\href
  {https://doi.org/10.1103/PhysRevLett.123.210604} {\bibfield  {journal}
  {\bibinfo  {journal} {Phys. Rev. Lett.}\ }\textbf {\bibinfo {volume} {123}},\
  \bibinfo {pages} {210604} (\bibinfo {year} {2019})}\BibitemShut {NoStop}%
\bibitem [{\citenamefont {Socolar}\ \emph {et~al.}(1986)\citenamefont
  {Socolar}, \citenamefont {Lubensky},\ and\ \citenamefont
  {Steinhardt}}]{Socoloar1986phonons}%
  \BibitemOpen
  \bibfield  {author} {\bibinfo {author} {\bibfnamefont {J.~E.~S.}\
  \bibnamefont {Socolar}}, \bibinfo {author} {\bibfnamefont {T.~C.}\
  \bibnamefont {Lubensky}},\ and\ \bibinfo {author} {\bibfnamefont {P.~J.}\
  \bibnamefont {Steinhardt}},\ }\bibfield  {title} {\bibinfo {title} {Phonons,
  phasons, and dislocations in quasicrystals},\ }\href
  {https://doi.org/10.1103/PhysRevB.34.3345} {\bibfield  {journal} {\bibinfo
  {journal} {Phys. Rev. B}\ }\textbf {\bibinfo {volume} {34}},\ \bibinfo
  {pages} {3345} (\bibinfo {year} {1986})}\BibitemShut {NoStop}%
\bibitem [{\citenamefont {Bak}(1986)}]{Bak86}%
  \BibitemOpen
  \bibfield  {author} {\bibinfo {author} {\bibfnamefont {P.}~\bibnamefont
  {Bak}},\ }\bibfield  {title} {\bibinfo {title} {Icosahedral crystals: Where
  are the atoms?},\ }\href {https://doi.org/10.1103/PhysRevLett.56.861}
  {\bibfield  {journal} {\bibinfo  {journal} {Phys. Rev. Lett.}\ }\textbf
  {\bibinfo {volume} {56}},\ \bibinfo {pages} {861} (\bibinfo {year}
  {1986})}\BibitemShut {NoStop}%
\bibitem [{\citenamefont {Guyot}(2001)}]{guyot2001dislocations}%
  \BibitemOpen
  \bibfield  {author} {\bibinfo {author} {\bibfnamefont {P.}~\bibnamefont
  {Guyot}},\ }\bibfield  {title} {\bibinfo {title} {Dislocations in
  quasicrystals},\ }\href {https://doi.org/10.1016/B0-08-043152-6/00407-1}
  {\bibfield  {journal} {\bibinfo  {journal} {Encyclopedia of Materials:
  Science and Technology}\ ,\ \bibinfo {pages} {2291}} (\bibinfo {year}
  {2001})}\BibitemShut {NoStop}%
\bibitem [{\citenamefont {Bendersky}(1985)}]{Bendersky1985quasicystal}%
  \BibitemOpen
  \bibfield  {author} {\bibinfo {author} {\bibfnamefont {L.}~\bibnamefont
  {Bendersky}},\ }\bibfield  {title} {\bibinfo {title} {Quasicrystal with
  one-dimensional translational symmetry and a tenfold rotation axis},\ }\href
  {https://doi.org/10.1103/PhysRevLett.55.1461} {\bibfield  {journal} {\bibinfo
   {journal} {Phys. Rev. Lett.}\ }\textbf {\bibinfo {volume} {55}},\ \bibinfo
  {pages} {1461} (\bibinfo {year} {1985})}\BibitemShut {NoStop}%
\bibitem [{Note1()}]{Note1}%
  \BibitemOpen
  \bibinfo {note} {Discrete diffraction diagrams became the defining property
  of crystals by the International Union of Crystallography since 1992,
  fostered by the discovery of QCs \cite {:es0177}}\BibitemShut {NoStop}%
\bibitem [{\citenamefont {Levine}\ \emph {et~al.}(1985)\citenamefont {Levine},
  \citenamefont {Lubensky}, \citenamefont {Ostlund}, \citenamefont {Ramaswamy},
  \citenamefont {Steinhardt},\ and\ \citenamefont
  {Toner}}]{Levine85elasticity}%
  \BibitemOpen
  \bibfield  {author} {\bibinfo {author} {\bibfnamefont {D.}~\bibnamefont
  {Levine}}, \bibinfo {author} {\bibfnamefont {T.~C.}\ \bibnamefont
  {Lubensky}}, \bibinfo {author} {\bibfnamefont {S.}~\bibnamefont {Ostlund}},
  \bibinfo {author} {\bibfnamefont {S.}~\bibnamefont {Ramaswamy}}, \bibinfo
  {author} {\bibfnamefont {P.~J.}\ \bibnamefont {Steinhardt}},\ and\ \bibinfo
  {author} {\bibfnamefont {J.}~\bibnamefont {Toner}},\ }\bibfield  {title}
  {\bibinfo {title} {Elasticity and dislocations in pentagonal and icosahedral
  quasicrystals},\ }\href {https://doi.org/10.1103/PhysRevLett.54.1520}
  {\bibfield  {journal} {\bibinfo  {journal} {Phys. Rev. Lett.}\ }\textbf
  {\bibinfo {volume} {54}},\ \bibinfo {pages} {1520} (\bibinfo {year}
  {1985})}\BibitemShut {NoStop}%
\bibitem [{Note2()}]{Note2}%
  \BibitemOpen
  \bibinfo {note} {Note that the formulation in \protect \eqref {eq:psir} with
  c.c. accounts for $\pm \protect \mathbf {G}_n$ $\forall n$}\BibitemShut
  {NoStop}%
\bibitem [{\citenamefont {Mermin}\ and\ \citenamefont
  {Troian}(1985)}]{Mermin1985}%
  \BibitemOpen
  \bibfield  {author} {\bibinfo {author} {\bibfnamefont {N.~D.}\ \bibnamefont
  {Mermin}}\ and\ \bibinfo {author} {\bibfnamefont {S.~M.}\ \bibnamefont
  {Troian}},\ }\bibfield  {title} {\bibinfo {title} {Mean-field theory of
  quasicrystalline order},\ }\href
  {https://doi.org/10.1103/PhysRevLett.54.1524} {\bibfield  {journal} {\bibinfo
   {journal} {Phys. Rev. Lett.}\ }\textbf {\bibinfo {volume} {54}},\ \bibinfo
  {pages} {1524} (\bibinfo {year} {1985})}\BibitemShut {NoStop}%
\bibitem [{\citenamefont {Lifshitz}\ and\ \citenamefont
  {Petrich}(1997)}]{Lifshitz1997}%
  \BibitemOpen
  \bibfield  {author} {\bibinfo {author} {\bibfnamefont {R.}~\bibnamefont
  {Lifshitz}}\ and\ \bibinfo {author} {\bibfnamefont {D.~M.}\ \bibnamefont
  {Petrich}},\ }\bibfield  {title} {\bibinfo {title} {Theoretical model for
  faraday waves with multiple-frequency forcing},\ }\href
  {https://doi.org/10.1103/PhysRevLett.79.1261} {\bibfield  {journal} {\bibinfo
   {journal} {Phys. Rev. Lett.}\ }\textbf {\bibinfo {volume} {79}},\ \bibinfo
  {pages} {1261} (\bibinfo {year} {1997})}\BibitemShut {NoStop}%
\bibitem [{\citenamefont {Swift}\ and\ \citenamefont
  {Hohenberg}(1977)}]{Swift1977}%
  \BibitemOpen
  \bibfield  {author} {\bibinfo {author} {\bibfnamefont {J.}~\bibnamefont
  {Swift}}\ and\ \bibinfo {author} {\bibfnamefont {P.~C.}\ \bibnamefont
  {Hohenberg}},\ }\bibfield  {title} {\bibinfo {title} {Hydrodynamic
  fluctuations at the convective instability},\ }\href
  {https://doi.org/10.1103/PhysRevA.15.319} {\bibfield  {journal} {\bibinfo
  {journal} {Phys. Rev. A}\ }\textbf {\bibinfo {volume} {15}},\ \bibinfo
  {pages} {319} (\bibinfo {year} {1977})}\BibitemShut {NoStop}%
\bibitem [{\citenamefont {Cross}\ and\ \citenamefont
  {Hohenberg}(1993)}]{Cross1993}%
  \BibitemOpen
  \bibfield  {author} {\bibinfo {author} {\bibfnamefont {M.~C.}\ \bibnamefont
  {Cross}}\ and\ \bibinfo {author} {\bibfnamefont {P.~C.}\ \bibnamefont
  {Hohenberg}},\ }\bibfield  {title} {\bibinfo {title} {Pattern formation
  outside of equilibrium},\ }\href {https://doi.org/10.1103/RevModPhys.65.851}
  {\bibfield  {journal} {\bibinfo  {journal} {Rev. Mod. Phys.}\ }\textbf
  {\bibinfo {volume} {65}},\ \bibinfo {pages} {851} (\bibinfo {year}
  {1993})}\BibitemShut {NoStop}%
\bibitem [{\citenamefont {Elder}\ \emph {et~al.}(2002)\citenamefont {Elder},
  \citenamefont {Katakowski}, \citenamefont {Haataja},\ and\ \citenamefont
  {Grant}}]{Elder2002}%
  \BibitemOpen
  \bibfield  {author} {\bibinfo {author} {\bibfnamefont {K.~R.}\ \bibnamefont
  {Elder}}, \bibinfo {author} {\bibfnamefont {M.}~\bibnamefont {Katakowski}},
  \bibinfo {author} {\bibfnamefont {M.}~\bibnamefont {Haataja}},\ and\ \bibinfo
  {author} {\bibfnamefont {M.}~\bibnamefont {Grant}},\ }\bibfield  {title}
  {\bibinfo {title} {{Modeling Elasticity in Crystal Growth}},\ }\href
  {https://doi.org/10.1103/PhysRevLett.88.245701} {\bibfield  {journal}
  {\bibinfo  {journal} {Phys. Rev. Lett.}\ }\textbf {\bibinfo {volume} {88}},\
  \bibinfo {pages} {245701} (\bibinfo {year} {2002})}\BibitemShut {NoStop}%
\bibitem [{\citenamefont {Ramakrishnan}\ and\ \citenamefont
  {Yussouff}(1979)}]{Ramakrishnan1979}%
  \BibitemOpen
  \bibfield  {author} {\bibinfo {author} {\bibfnamefont {T.~V.}\ \bibnamefont
  {Ramakrishnan}}\ and\ \bibinfo {author} {\bibfnamefont {M.}~\bibnamefont
  {Yussouff}},\ }\bibfield  {title} {\bibinfo {title} {First-principles
  order-parameter theory of freezing},\ }\href
  {https://doi.org/10.1103/PhysRevB.19.2775} {\bibfield  {journal} {\bibinfo
  {journal} {Phys. Rev. B}\ }\textbf {\bibinfo {volume} {19}},\ \bibinfo
  {pages} {2775} (\bibinfo {year} {1979})}\BibitemShut {NoStop}%
\bibitem [{\citenamefont {Michael~te Vrugt}\ and\ \citenamefont
  {Wittkowski}(2020)}]{Vrugt2020}%
  \BibitemOpen
  \bibfield  {author} {\bibinfo {author} {\bibfnamefont {H.~L.}\ \bibnamefont
  {Michael~te Vrugt}}\ and\ \bibinfo {author} {\bibfnamefont {R.}~\bibnamefont
  {Wittkowski}},\ }\bibfield  {title} {\bibinfo {title} {Classical dynamical
  density functional theory: from fundamentals to applications},\ }\href
  {https://doi.org/10.1080/00018732.2020.1854965} {\bibfield  {journal}
  {\bibinfo  {journal} {Advances in Physics}\ }\textbf {\bibinfo {volume}
  {69}},\ \bibinfo {pages} {121} (\bibinfo {year} {2020})}\BibitemShut
  {NoStop}%
\bibitem [{\citenamefont {Barkan}\ \emph {et~al.}(2011)\citenamefont {Barkan},
  \citenamefont {Diamant},\ and\ \citenamefont {Lifshitz}}]{Barkan2011}%
  \BibitemOpen
  \bibfield  {author} {\bibinfo {author} {\bibfnamefont {K.}~\bibnamefont
  {Barkan}}, \bibinfo {author} {\bibfnamefont {H.}~\bibnamefont {Diamant}},\
  and\ \bibinfo {author} {\bibfnamefont {R.}~\bibnamefont {Lifshitz}},\
  }\bibfield  {title} {\bibinfo {title} {Stability of quasicrystals composed of
  soft isotropic particles},\ }\href
  {https://doi.org/10.1103/PhysRevB.83.172201} {\bibfield  {journal} {\bibinfo
  {journal} {Phys. Rev. B}\ }\textbf {\bibinfo {volume} {83}},\ \bibinfo
  {pages} {172201} (\bibinfo {year} {2011})}\BibitemShut {NoStop}%
\bibitem [{\citenamefont {Rottler}\ \emph {et~al.}(2012)\citenamefont
  {Rottler}, \citenamefont {Greenwood},\ and\ \citenamefont
  {Ziebarth}}]{Rottler_2012}%
  \BibitemOpen
  \bibfield  {author} {\bibinfo {author} {\bibfnamefont {J.}~\bibnamefont
  {Rottler}}, \bibinfo {author} {\bibfnamefont {M.}~\bibnamefont {Greenwood}},\
  and\ \bibinfo {author} {\bibfnamefont {B.}~\bibnamefont {Ziebarth}},\
  }\bibfield  {title} {\bibinfo {title} {Morphology of monolayer films on
  quasicrystalline surfaces from the phase field crystal model},\ }\href
  {https://doi.org/10.1088/0953-8984/24/13/135002} {\bibfield  {journal}
  {\bibinfo  {journal} {J. Phys. Cond. Matter}\ }\textbf {\bibinfo {volume}
  {24}},\ \bibinfo {pages} {135002} (\bibinfo {year} {2012})}\BibitemShut
  {NoStop}%
\bibitem [{\citenamefont {Archer}\ \emph {et~al.}(2013)\citenamefont {Archer},
  \citenamefont {Rucklidge},\ and\ \citenamefont {Knobloch}}]{Archer2013}%
  \BibitemOpen
  \bibfield  {author} {\bibinfo {author} {\bibfnamefont {A.~J.}\ \bibnamefont
  {Archer}}, \bibinfo {author} {\bibfnamefont {A.~M.}\ \bibnamefont
  {Rucklidge}},\ and\ \bibinfo {author} {\bibfnamefont {E.}~\bibnamefont
  {Knobloch}},\ }\bibfield  {title} {\bibinfo {title} {Quasicrystalline order
  and a crystal-liquid state in a soft-core fluid},\ }\href
  {https://doi.org/10.1103/PhysRevLett.111.165501} {\bibfield  {journal}
  {\bibinfo  {journal} {Phys. Rev. Lett.}\ }\textbf {\bibinfo {volume} {111}},\
  \bibinfo {pages} {165501} (\bibinfo {year} {2013})}\BibitemShut {NoStop}%
\bibitem [{\citenamefont {Achim}\ \emph {et~al.}(2014)\citenamefont {Achim},
  \citenamefont {Schmiedeberg},\ and\ \citenamefont {L\"owen}}]{Achim2014}%
  \BibitemOpen
  \bibfield  {author} {\bibinfo {author} {\bibfnamefont {C.~V.}\ \bibnamefont
  {Achim}}, \bibinfo {author} {\bibfnamefont {M.}~\bibnamefont
  {Schmiedeberg}},\ and\ \bibinfo {author} {\bibfnamefont {H.}~\bibnamefont
  {L\"owen}},\ }\bibfield  {title} {\bibinfo {title} {Growth modes of
  quasicrystals},\ }\href {https://doi.org/10.1103/PhysRevLett.112.255501}
  {\bibfield  {journal} {\bibinfo  {journal} {Phys. Rev. Lett.}\ }\textbf
  {\bibinfo {volume} {112}},\ \bibinfo {pages} {255501} (\bibinfo {year}
  {2014})}\BibitemShut {NoStop}%
\bibitem [{\citenamefont {Barkan}\ \emph {et~al.}(2014)\citenamefont {Barkan},
  \citenamefont {Engel},\ and\ \citenamefont {Lifshitz}}]{Barkan2014}%
  \BibitemOpen
  \bibfield  {author} {\bibinfo {author} {\bibfnamefont {K.}~\bibnamefont
  {Barkan}}, \bibinfo {author} {\bibfnamefont {M.}~\bibnamefont {Engel}},\ and\
  \bibinfo {author} {\bibfnamefont {R.}~\bibnamefont {Lifshitz}},\ }\bibfield
  {title} {\bibinfo {title} {Controlled self-assembly of periodic and aperiodic
  cluster crystals},\ }\href {https://doi.org/10.1103/PhysRevLett.113.098304}
  {\bibfield  {journal} {\bibinfo  {journal} {Phys. Rev. Lett.}\ }\textbf
  {\bibinfo {volume} {113}},\ \bibinfo {pages} {098304} (\bibinfo {year}
  {2014})}\BibitemShut {NoStop}%
\bibitem [{\citenamefont {Xue}\ \emph {et~al.}(2022)\citenamefont {Xue},
  \citenamefont {Gong},\ and\ \citenamefont {Tang}}]{xue2022atomic}%
  \BibitemOpen
  \bibfield  {author} {\bibinfo {author} {\bibfnamefont {Q.}~\bibnamefont
  {Xue}}, \bibinfo {author} {\bibfnamefont {S.-a.}\ \bibnamefont {Gong}},\ and\
  \bibinfo {author} {\bibfnamefont {S.}~\bibnamefont {Tang}},\ }\bibfield
  {title} {\bibinfo {title} {Atomic-scale investigation of the pattern
  formation of quasicrystal growth by phase-field crystal simulations},\ }\href
  {https://doi.org/10.1021/acs.cgd.2c00591} {\bibfield  {journal} {\bibinfo
  {journal} {Cryst. Growth Des}\ }\textbf {\bibinfo {volume} {22}},\ \bibinfo
  {pages} {5497} (\bibinfo {year} {2022})}\BibitemShut {NoStop}%
\bibitem [{\citenamefont {Goldenfeld}\ \emph {et~al.}(2005)\citenamefont
  {Goldenfeld}, \citenamefont {Athreya},\ and\ \citenamefont
  {Dantzig}}]{Goldenfeld2005}%
  \BibitemOpen
  \bibfield  {author} {\bibinfo {author} {\bibfnamefont {N.}~\bibnamefont
  {Goldenfeld}}, \bibinfo {author} {\bibfnamefont {B.~P.}\ \bibnamefont
  {Athreya}},\ and\ \bibinfo {author} {\bibfnamefont {J.~A.}\ \bibnamefont
  {Dantzig}},\ }\bibfield  {title} {\bibinfo {title} {{Renormalization group
  approach to multiscale simulation of polycrystalline materials using the
  phase field crystal model}},\ }\href
  {https://doi.org/10.1103/PhysRevE.72.020601} {\bibfield  {journal} {\bibinfo
  {journal} {Phys. Rev. E}\ }\textbf {\bibinfo {volume} {72}},\ \bibinfo
  {pages} {020601(R)} (\bibinfo {year} {2005})}\BibitemShut {NoStop}%
\bibitem [{\citenamefont {Salvalaglio}\ \emph {et~al.}(2019)\citenamefont
  {Salvalaglio}, \citenamefont {Voigt},\ and\ \citenamefont
  {Elder}}]{SalvalaglioNPJ2019}%
  \BibitemOpen
  \bibfield  {author} {\bibinfo {author} {\bibfnamefont {M.}~\bibnamefont
  {Salvalaglio}}, \bibinfo {author} {\bibfnamefont {A.}~\bibnamefont {Voigt}},\
  and\ \bibinfo {author} {\bibfnamefont {K.~R.}\ \bibnamefont {Elder}},\
  }\bibfield  {title} {\bibinfo {title} {{Closing the gap between atomic-scale
  lattice deformations and continuum elasticity}},\ }\href
  {https://doi.org/10.1038/s41524-019-0185-0} {\bibfield  {journal} {\bibinfo
  {journal} {npj Comput. Mater.}\ }\textbf {\bibinfo {volume} {5}},\ \bibinfo
  {pages} {48} (\bibinfo {year} {2019})}\BibitemShut {NoStop}%
\bibitem [{\citenamefont {Salvalaglio}\ and\ \citenamefont
  {Elder}(2022)}]{salvalaglio2022coarse}%
  \BibitemOpen
  \bibfield  {author} {\bibinfo {author} {\bibfnamefont {M.}~\bibnamefont
  {Salvalaglio}}\ and\ \bibinfo {author} {\bibfnamefont {K.~R.}\ \bibnamefont
  {Elder}},\ }\bibfield  {title} {\bibinfo {title} {Coarse-grained modeling of
  crystals by the amplitude expansion of the phase-field crystal model: an
  overview},\ }\href {https://doi.org/10.1088/1361-651x/ac681e} {\bibfield
  {journal} {\bibinfo  {journal} {Model. Simul. Mater. Sci. Eng.}\ }\textbf
  {\bibinfo {volume} {30}},\ \bibinfo {pages} {053001} (\bibinfo {year}
  {2022})}\BibitemShut {NoStop}%
\bibitem [{\citenamefont {Socolar}\ and\ \citenamefont
  {Steinhardt}(1986)}]{Socolar86unitcell}%
  \BibitemOpen
  \bibfield  {author} {\bibinfo {author} {\bibfnamefont {J.~E.~S.}\
  \bibnamefont {Socolar}}\ and\ \bibinfo {author} {\bibfnamefont {P.~J.}\
  \bibnamefont {Steinhardt}},\ }\bibfield  {title} {\bibinfo {title}
  {Quasicrystals. ii. unit-cell configurations},\ }\href
  {https://doi.org/10.1103/PhysRevB.34.617} {\bibfield  {journal} {\bibinfo
  {journal} {Phys. Rev. B}\ }\textbf {\bibinfo {volume} {34}},\ \bibinfo
  {pages} {617} (\bibinfo {year} {1986})}\BibitemShut {NoStop}%
\bibitem [{\citenamefont {Steinhardt}\ \emph {et~al.}(1998)\citenamefont
  {Steinhardt}, \citenamefont {Jeong}, \citenamefont {Saitoh}, \citenamefont
  {Tanaka}, \citenamefont {Abe},\ and\ \citenamefont
  {Tsai}}]{steinhardt1998experimental}%
  \BibitemOpen
  \bibfield  {author} {\bibinfo {author} {\bibfnamefont {P.~J.}\ \bibnamefont
  {Steinhardt}}, \bibinfo {author} {\bibfnamefont {H.-C.}\ \bibnamefont
  {Jeong}}, \bibinfo {author} {\bibfnamefont {K.}~\bibnamefont {Saitoh}},
  \bibinfo {author} {\bibfnamefont {M.}~\bibnamefont {Tanaka}}, \bibinfo
  {author} {\bibfnamefont {E.}~\bibnamefont {Abe}},\ and\ \bibinfo {author}
  {\bibfnamefont {A.}~\bibnamefont {Tsai}},\ }\bibfield  {title} {\bibinfo
  {title} {Experimental verification of the quasi-unit-cell model of
  quasicrystal structure},\ }\href {https://doi.org/10.1038/23902} {\bibfield
  {journal} {\bibinfo  {journal} {Nature}\ }\textbf {\bibinfo {volume} {396}},\
  \bibinfo {pages} {55} (\bibinfo {year} {1998})}\BibitemShut {NoStop}%
\bibitem [{Sup()}]{SupplementalMaterial}%
  \BibitemOpen
  \href@noop {} {\bibinfo {title} {{See supplemental material (SM) for: a
  review of the definitions of $\mathcal{L}$ and
  $\mathbf{G}_n^{||},\mathbf{G}_n^{\perp}$ for decagonal QCs; details for the
  determination of $\Lambda$ and the tiling; full expressions for the free
  energy for decagonal and icosahedral QCs; all derivation steps for strain and
  stress fields as well as dislocation velocity equations; numerical methods}.
  {SM includes Refs}.~\cite{hirvonen_PhysRevB.100.165412,skogvoll2021_10.1103,Benoit-Marechal_2024}}}\BibitemShut
  {NoStop}%
\bibitem [{\citenamefont {Yeon}\ \emph {et~al.}(2010)\citenamefont {Yeon},
  \citenamefont {Huang}, \citenamefont {Elder},\ and\ \citenamefont
  {Thornton}}]{Yeon2010}%
  \BibitemOpen
  \bibfield  {author} {\bibinfo {author} {\bibfnamefont {D.-H.}\ \bibnamefont
  {Yeon}}, \bibinfo {author} {\bibfnamefont {Z.-F.}\ \bibnamefont {Huang}},
  \bibinfo {author} {\bibfnamefont {K.~R.}\ \bibnamefont {Elder}},\ and\
  \bibinfo {author} {\bibfnamefont {K.}~\bibnamefont {Thornton}},\ }\bibfield
  {title} {\bibinfo {title} {Density-amplitude formulation of the phase-field
  crystal model for two-phase coexistence in two and three dimensions},\ }\href
  {https://doi.org/10.1080/14786430903164572} {\bibfield  {journal} {\bibinfo
  {journal} {Phil. Mag.}\ }\textbf {\bibinfo {volume} {90}},\ \bibinfo {pages}
  {237} (\bibinfo {year} {2010})}\BibitemShut {NoStop}%
\bibitem [{\citenamefont {Ding}\ \emph {et~al.}(1993)\citenamefont {Ding},
  \citenamefont {Yang}, \citenamefont {Hu},\ and\ \citenamefont
  {Wang}}]{DingPRB93}%
  \BibitemOpen
  \bibfield  {author} {\bibinfo {author} {\bibfnamefont {D.~H.}\ \bibnamefont
  {Ding}}, \bibinfo {author} {\bibfnamefont {W.}~\bibnamefont {Yang}}, \bibinfo
  {author} {\bibfnamefont {C.}~\bibnamefont {Hu}},\ and\ \bibinfo {author}
  {\bibfnamefont {R.}~\bibnamefont {Wang}},\ }\bibfield  {title} {\bibinfo
  {title} {Generalized elasticity theory of quasicrystals},\ }\href
  {https://doi.org/10.1103/PhysRevB.48.7003} {\bibfield  {journal} {\bibinfo
  {journal} {Phys. Rev. B}\ }\textbf {\bibinfo {volume} {48}},\ \bibinfo
  {pages} {7003} (\bibinfo {year} {1993})}\BibitemShut {NoStop}%
\bibitem [{\citenamefont {Chernikov}\ \emph {et~al.}(1998)\citenamefont
  {Chernikov}, \citenamefont {Ott}, \citenamefont {Bianchi}, \citenamefont
  {Migliori},\ and\ \citenamefont {Darling}}]{Chernikov1998}%
  \BibitemOpen
  \bibfield  {author} {\bibinfo {author} {\bibfnamefont {M.~A.}\ \bibnamefont
  {Chernikov}}, \bibinfo {author} {\bibfnamefont {H.~R.}\ \bibnamefont {Ott}},
  \bibinfo {author} {\bibfnamefont {A.}~\bibnamefont {Bianchi}}, \bibinfo
  {author} {\bibfnamefont {A.}~\bibnamefont {Migliori}},\ and\ \bibinfo
  {author} {\bibfnamefont {T.~W.}\ \bibnamefont {Darling}},\ }\bibfield
  {title} {\bibinfo {title} {Elastic moduli of a single quasicrystal of
  decagonal al-ni-co: Evidence for transverse elastic isotropy},\ }\href
  {https://doi.org/10.1103/PhysRevLett.80.321} {\bibfield  {journal} {\bibinfo
  {journal} {Phys. Rev. Lett.}\ }\textbf {\bibinfo {volume} {80}},\ \bibinfo
  {pages} {321} (\bibinfo {year} {1998})}\BibitemShut {NoStop}%
\bibitem [{\citenamefont {Han}\ \emph {et~al.}(2021)\citenamefont {Han},
  \citenamefont {Wang}, \citenamefont {Cadotte}, \citenamefont {Xi},
  \citenamefont {Parsamehr}, \citenamefont {Xiao}, \citenamefont {Glotzer},\
  and\ \citenamefont {Shahani}}]{han2021formation}%
  \BibitemOpen
  \bibfield  {author} {\bibinfo {author} {\bibfnamefont {I.}~\bibnamefont
  {Han}}, \bibinfo {author} {\bibfnamefont {K.~L.}\ \bibnamefont {Wang}},
  \bibinfo {author} {\bibfnamefont {A.~T.}\ \bibnamefont {Cadotte}}, \bibinfo
  {author} {\bibfnamefont {Z.}~\bibnamefont {Xi}}, \bibinfo {author}
  {\bibfnamefont {H.}~\bibnamefont {Parsamehr}}, \bibinfo {author}
  {\bibfnamefont {X.}~\bibnamefont {Xiao}}, \bibinfo {author} {\bibfnamefont
  {S.~C.}\ \bibnamefont {Glotzer}},\ and\ \bibinfo {author} {\bibfnamefont
  {A.~J.}\ \bibnamefont {Shahani}},\ }\bibfield  {title} {\bibinfo {title}
  {Formation of a single quasicrystal upon collision of multiple grains},\
  }\href {https://doi.org/10.1038/s41467-021-26070-9} {\bibfield  {journal}
  {\bibinfo  {journal} {Nature Communications}\ }\textbf {\bibinfo {volume}
  {12}},\ \bibinfo {pages} {5790} (\bibinfo {year} {2021})}\BibitemShut
  {NoStop}%
\bibitem [{\citenamefont {De}\ and\ \citenamefont
  {Pelcovits}(1987)}]{PialiPRB1987}%
  \BibitemOpen
  \bibfield  {author} {\bibinfo {author} {\bibfnamefont {P.}~\bibnamefont
  {De}}\ and\ \bibinfo {author} {\bibfnamefont {R.~A.}\ \bibnamefont
  {Pelcovits}},\ }\bibfield  {title} {\bibinfo {title} {Linear elasticity
  theory of pentagonal quasicrystals},\ }\href
  {https://doi.org/10.1103/PhysRevB.35.8609} {\bibfield  {journal} {\bibinfo
  {journal} {Phys. Rev. B}\ }\textbf {\bibinfo {volume} {35}},\ \bibinfo
  {pages} {8609} (\bibinfo {year} {1987})}\BibitemShut {NoStop}%
\bibitem [{\citenamefont {Praetorius}\ \emph {et~al.}(2019)\citenamefont
  {Praetorius}, \citenamefont {Salvalaglio},\ and\ \citenamefont
  {Voigt}}]{Praetorius2019}%
  \BibitemOpen
  \bibfield  {author} {\bibinfo {author} {\bibfnamefont {S.}~\bibnamefont
  {Praetorius}}, \bibinfo {author} {\bibfnamefont {M.}~\bibnamefont
  {Salvalaglio}},\ and\ \bibinfo {author} {\bibfnamefont {A.}~\bibnamefont
  {Voigt}},\ }\bibfield  {title} {\bibinfo {title} {{An efficient numerical
  framework for the amplitude expansion of the phase-field crystal model}},\
  }\href {https://doi.org/10.1088/1361-651X/ab1508} {\bibfield  {journal}
  {\bibinfo  {journal} {Model. Simul. Mater. Sci. Eng.}\ }\textbf {\bibinfo
  {volume} {27}},\ \bibinfo {pages} {044004} (\bibinfo {year}
  {2019})}\BibitemShut {NoStop}%
\bibitem [{\citenamefont {Hatwalne}\ \emph {et~al.}(1989)\citenamefont
  {Hatwalne}, \citenamefont {Krishnamurthy}, \citenamefont {Pandit},\ and\
  \citenamefont {Ramaswamy}}]{QCsmallGB_PhysRevLett.62.2699}%
  \BibitemOpen
  \bibfield  {author} {\bibinfo {author} {\bibfnamefont {Y.}~\bibnamefont
  {Hatwalne}}, \bibinfo {author} {\bibfnamefont {H.~R.}\ \bibnamefont
  {Krishnamurthy}}, \bibinfo {author} {\bibfnamefont {R.}~\bibnamefont
  {Pandit}},\ and\ \bibinfo {author} {\bibfnamefont {S.}~\bibnamefont
  {Ramaswamy}},\ }\bibfield  {title} {\bibinfo {title} {Small-angle grain
  boundaries in quasicrystals},\ }\href
  {https://doi.org/10.1103/PhysRevLett.62.2699} {\bibfield  {journal} {\bibinfo
   {journal} {Phys. Rev. Lett.}\ }\textbf {\bibinfo {volume} {62}},\ \bibinfo
  {pages} {2699} (\bibinfo {year} {1989})}\BibitemShut {NoStop}%
\bibitem [{\citenamefont {Nagao}\ \emph {et~al.}(2015)\citenamefont {Nagao},
  \citenamefont {Inuzuka}, \citenamefont {Nishimoto},\ and\ \citenamefont
  {Edagawa}}]{NagaoPRL2015}%
  \BibitemOpen
  \bibfield  {author} {\bibinfo {author} {\bibfnamefont {K.}~\bibnamefont
  {Nagao}}, \bibinfo {author} {\bibfnamefont {T.}~\bibnamefont {Inuzuka}},
  \bibinfo {author} {\bibfnamefont {K.}~\bibnamefont {Nishimoto}},\ and\
  \bibinfo {author} {\bibfnamefont {K.}~\bibnamefont {Edagawa}},\ }\bibfield
  {title} {\bibinfo {title} {Experimental observation of quasicrystal growth},\
  }\href {https://doi.org/10.1103/PhysRevLett.115.075501} {\bibfield  {journal}
  {\bibinfo  {journal} {Phys. Rev. Lett.}\ }\textbf {\bibinfo {volume} {115}},\
  \bibinfo {pages} {075501} (\bibinfo {year} {2015})}\BibitemShut {NoStop}%
\bibitem [{\citenamefont {Skaugen}\ \emph {et~al.}(2018)\citenamefont
  {Skaugen}, \citenamefont {Angheluta},\ and\ \citenamefont
  {Vi\~nals}}]{Skaugen2018b}%
  \BibitemOpen
  \bibfield  {author} {\bibinfo {author} {\bibfnamefont {A.}~\bibnamefont
  {Skaugen}}, \bibinfo {author} {\bibfnamefont {L.}~\bibnamefont {Angheluta}},\
  and\ \bibinfo {author} {\bibfnamefont {J.}~\bibnamefont {Vi\~nals}},\
  }\bibfield  {title} {\bibinfo {title} {Dislocation dynamics and crystal
  plasticity in the phase-field crystal model},\ }\href
  {https://doi.org/10.1103/PhysRevB.97.054113} {\bibfield  {journal} {\bibinfo
  {journal} {Phys. Rev. B}\ }\textbf {\bibinfo {volume} {97}},\ \bibinfo
  {pages} {054113} (\bibinfo {year} {2018})}\BibitemShut {NoStop}%
\bibitem [{\citenamefont {Salvalaglio}\ \emph {et~al.}(2021)\citenamefont
  {Salvalaglio}, \citenamefont {Voigt}, \citenamefont {Huang},\ and\
  \citenamefont {Elder}}]{salvalaglioPRL2021}%
  \BibitemOpen
  \bibfield  {author} {\bibinfo {author} {\bibfnamefont {M.}~\bibnamefont
  {Salvalaglio}}, \bibinfo {author} {\bibfnamefont {A.}~\bibnamefont {Voigt}},
  \bibinfo {author} {\bibfnamefont {Z.-F.}\ \bibnamefont {Huang}},\ and\
  \bibinfo {author} {\bibfnamefont {K.~R.}\ \bibnamefont {Elder}},\ }\bibfield
  {title} {\bibinfo {title} {Mesoscale defect motion in binary systems: Effects
  of compositional strain and cottrell atmospheres},\ }\href
  {https://doi.org/10.1103/PhysRevLett.126.185502} {\bibfield  {journal}
  {\bibinfo  {journal} {Phys. Rev. Lett.}\ }\textbf {\bibinfo {volume} {126}},\
  \bibinfo {pages} {185502} (\bibinfo {year} {2021})}\BibitemShut {NoStop}%
\bibitem [{\citenamefont {Skogvoll}\ \emph {et~al.}(2022)\citenamefont
  {Skogvoll}, \citenamefont {Angheluta}, \citenamefont {Skaugen}, \citenamefont
  {Salvalaglio},\ and\ \citenamefont {Viñals}}]{SKOGVOLL2022104932}%
  \BibitemOpen
  \bibfield  {author} {\bibinfo {author} {\bibfnamefont {V.}~\bibnamefont
  {Skogvoll}}, \bibinfo {author} {\bibfnamefont {L.}~\bibnamefont {Angheluta}},
  \bibinfo {author} {\bibfnamefont {A.}~\bibnamefont {Skaugen}}, \bibinfo
  {author} {\bibfnamefont {M.}~\bibnamefont {Salvalaglio}},\ and\ \bibinfo
  {author} {\bibfnamefont {J.}~\bibnamefont {Viñals}},\ }\bibfield  {title}
  {\bibinfo {title} {A phase field crystal theory of the kinematics of
  dislocation lines},\ }\href
  {https://doi.org/https://doi.org/10.1016/j.jmps.2022.104932} {\bibfield
  {journal} {\bibinfo  {journal} {J. Mech. Phys. of Solids}\ }\textbf {\bibinfo
  {volume} {166}},\ \bibinfo {pages} {104932} (\bibinfo {year}
  {2022})}\BibitemShut {NoStop}%
\bibitem [{Note3()}]{Note3}%
  \BibitemOpen
  \bibinfo {note} {This holds true in 2D, see Ref~\cite {SKOGVOLL2022104932}
  for extensions to 3D}\BibitemShut {NoStop}%
\bibitem [{\citenamefont {Skogvoll}\ \emph {et~al.}(2023)\citenamefont
  {Skogvoll}, \citenamefont {R{\o}nning}, \citenamefont {Salvalaglio},\ and\
  \citenamefont {Angheluta}}]{SkogvollNPJ2023}%
  \BibitemOpen
  \bibfield  {author} {\bibinfo {author} {\bibfnamefont {V.}~\bibnamefont
  {Skogvoll}}, \bibinfo {author} {\bibfnamefont {J.}~\bibnamefont
  {R{\o}nning}}, \bibinfo {author} {\bibfnamefont {M.}~\bibnamefont
  {Salvalaglio}},\ and\ \bibinfo {author} {\bibfnamefont {L.}~\bibnamefont
  {Angheluta}},\ }\bibfield  {title} {\bibinfo {title} {A unified field theory
  of topological defects and non-linear local excitations},\ }\href
  {https://doi.org/10.1038/s41524-023-01077-6} {\bibfield  {journal} {\bibinfo
  {journal} {npj Comput. Mater.}\ }\textbf {\bibinfo {volume} {9}},\ \bibinfo
  {pages} {122} (\bibinfo {year} {2023})}\BibitemShut {NoStop}%
\bibitem [{\citenamefont {Mazenko}(1997)}]{Mazenko97}%
  \BibitemOpen
  \bibfield  {author} {\bibinfo {author} {\bibfnamefont {G.~F.}\ \bibnamefont
  {Mazenko}},\ }\bibfield  {title} {\bibinfo {title} {Vortex velocities in the
  $\mathit{O}(\mathit{n})$ symmetric time-dependent ginzburg-landau model},\
  }\href {https://doi.org/10.1103/PhysRevLett.78.401} {\bibfield  {journal}
  {\bibinfo  {journal} {Phys. Rev. Lett.}\ }\textbf {\bibinfo {volume} {78}},\
  \bibinfo {pages} {401} (\bibinfo {year} {1997})}\BibitemShut {NoStop}%
\bibitem [{\citenamefont {Lubensky}\ \emph {et~al.}(1986)\citenamefont
  {Lubensky}, \citenamefont {Ramaswamy},\ and\ \citenamefont
  {Toner}}]{Lubensky86dislomotion}%
  \BibitemOpen
  \bibfield  {author} {\bibinfo {author} {\bibfnamefont {T.~C.}\ \bibnamefont
  {Lubensky}}, \bibinfo {author} {\bibfnamefont {S.}~\bibnamefont
  {Ramaswamy}},\ and\ \bibinfo {author} {\bibfnamefont {J.}~\bibnamefont
  {Toner}},\ }\bibfield  {title} {\bibinfo {title} {Dislocation motion in
  quasicrystals and implications for macroscopic properties},\ }\href
  {https://doi.org/10.1103/PhysRevB.33.7715} {\bibfield  {journal} {\bibinfo
  {journal} {Phys. Rev. B}\ }\textbf {\bibinfo {volume} {33}},\ \bibinfo
  {pages} {7715} (\bibinfo {year} {1986})}\BibitemShut {NoStop}%
\bibitem [{\citenamefont {Agiasofitou}\ \emph {et~al.}(2010)\citenamefont
  {Agiasofitou}, \citenamefont {Lazar},\ and\ \citenamefont
  {Kirchner}}]{Agiasofitou_2010}%
  \BibitemOpen
  \bibfield  {author} {\bibinfo {author} {\bibfnamefont {E.}~\bibnamefont
  {Agiasofitou}}, \bibinfo {author} {\bibfnamefont {M.}~\bibnamefont {Lazar}},\
  and\ \bibinfo {author} {\bibfnamefont {H.}~\bibnamefont {Kirchner}},\
  }\bibfield  {title} {\bibinfo {title} {Generalized dynamics of moving
  dislocations in quasicrystals},\ }\href
  {https://doi.org/10.1088/0953-8984/22/49/495401} {\bibfield  {journal}
  {\bibinfo  {journal} {J. Phys. Cond. Matter}\ }\textbf {\bibinfo {volume}
  {22}},\ \bibinfo {pages} {495401} (\bibinfo {year} {2010})}\BibitemShut
  {NoStop}%
\bibitem [{:es(1992)}]{:es0177}%
  \BibitemOpen
  \bibfield  {title} {\bibinfo {title} {{Report of the Executive Committee for
  1991}},\ }\href {https://doi.org/10.1107/S0108767392008328} {\bibfield
  {journal} {\bibinfo  {journal} {Acta Crystallographica Section A}\ }\textbf
  {\bibinfo {volume} {48}},\ \bibinfo {pages} {922} (\bibinfo {year}
  {1992})}\BibitemShut {NoStop}%
\bibitem [{\citenamefont {Hirvonen}\ \emph {et~al.}(2019)\citenamefont
  {Hirvonen}, \citenamefont {Heinonen}, \citenamefont {Dong}, \citenamefont
  {Fan}, \citenamefont {Elder},\ and\ \citenamefont
  {Ala-Nissila}}]{hirvonen_PhysRevB.100.165412}%
  \BibitemOpen
  \bibfield  {author} {\bibinfo {author} {\bibfnamefont {P.}~\bibnamefont
  {Hirvonen}}, \bibinfo {author} {\bibfnamefont {V.}~\bibnamefont {Heinonen}},
  \bibinfo {author} {\bibfnamefont {H.}~\bibnamefont {Dong}}, \bibinfo {author}
  {\bibfnamefont {Z.}~\bibnamefont {Fan}}, \bibinfo {author} {\bibfnamefont
  {K.~R.}\ \bibnamefont {Elder}},\ and\ \bibinfo {author} {\bibfnamefont
  {T.}~\bibnamefont {Ala-Nissila}},\ }\bibfield  {title} {\bibinfo {title}
  {Phase-field crystal model for heterostructures},\ }\href
  {https://doi.org/10.1103/PhysRevB.100.165412} {\bibfield  {journal} {\bibinfo
   {journal} {Phys. Rev. B}\ }\textbf {\bibinfo {volume} {100}},\ \bibinfo
  {pages} {165412} (\bibinfo {year} {2019})}\BibitemShut {NoStop}%
\bibitem [{\citenamefont {Skogvoll}\ \emph {et~al.}(2021)\citenamefont
  {Skogvoll}, \citenamefont {Skaugen},\ and\ \citenamefont
  {Angheluta}}]{skogvoll2021_10.1103}%
  \BibitemOpen
  \bibfield  {author} {\bibinfo {author} {\bibfnamefont {V.}~\bibnamefont
  {Skogvoll}}, \bibinfo {author} {\bibfnamefont {A.}~\bibnamefont {Skaugen}},\
  and\ \bibinfo {author} {\bibfnamefont {L.}~\bibnamefont {Angheluta}},\
  }\bibfield  {title} {\bibinfo {title} {Stress in ordered systems:
  Ginzburg-landau-type density field theory},\ }\href
  {https://doi.org/10.1103/PhysRevB.103.224107} {\bibfield  {journal} {\bibinfo
   {journal} {Phys. Rev. B}\ }\textbf {\bibinfo {volume} {103}},\ \bibinfo
  {pages} {224107} (\bibinfo {year} {2021})}\BibitemShut {NoStop}%
\bibitem [{\citenamefont {Benoit-Maréchal}\ and\ \citenamefont
  {Salvalaglio}(2024)}]{Benoit-Marechal_2024}%
  \BibitemOpen
  \bibfield  {author} {\bibinfo {author} {\bibfnamefont {L.}~\bibnamefont
  {Benoit-Maréchal}}\ and\ \bibinfo {author} {\bibfnamefont {M.}~\bibnamefont
  {Salvalaglio}},\ }\bibfield  {title} {\bibinfo {title} {{Gradient elasticity
  in Swift–Hohenberg and phase-field crystal models}},\ }\href
  {https://doi.org/10.1088/1361-651X/ad42bb} {\bibfield  {journal} {\bibinfo
  {journal} {Model. Simul. Mater. Sci. Eng.}\ }\textbf {\bibinfo {volume}
  {32}},\ \bibinfo {pages} {055005} (\bibinfo {year} {2024})}\BibitemShut
  {NoStop}%
\end{thebibliography}

\begin{thebibliography}{19}%
\makeatletter
\providecommand \@ifxundefined [1]{%
 \@ifx{#1\undefined}
}%
\providecommand \@ifnum [1]{%
 \ifnum #1\expandafter \@firstoftwo
 \else \expandafter \@secondoftwo
 \fi
}%
\providecommand \@ifx [1]{%
 \ifx #1\expandafter \@firstoftwo
 \else \expandafter \@secondoftwo
 \fi
}%
\providecommand \natexlab [1]{#1}%
\providecommand \enquote  [1]{``#1''}%
\providecommand \bibnamefont  [1]{#1}%
\providecommand \bibfnamefont [1]{#1}%
\providecommand \citenamefont [1]{#1}%
\providecommand \href@noop [0]{\@secondoftwo}%
\providecommand \href [0]{\begingroup \@sanitize@url \@href}%
\providecommand \@href[1]{\@@startlink{#1}\@@href}%
\providecommand \@@href[1]{\endgroup#1\@@endlink}%
\providecommand \@sanitize@url [0]{\catcode `\\12\catcode `\$12\catcode
  `\&12\catcode `\#12\catcode `\^12\catcode `\_12\catcode `\%12\relax}%
\providecommand \@@startlink[1]{}%
\providecommand \@@endlink[0]{}%
\providecommand \url  [0]{\begingroup\@sanitize@url \@url }%
\providecommand \@url [1]{\endgroup\@href {#1}{\urlprefix }}%
\providecommand \urlprefix  [0]{URL }%
\providecommand \Eprint [0]{\href }%
\providecommand \doibase [0]{http://dx.doi.org/}%
\providecommand \selectlanguage [0]{\@gobble}%
\providecommand \bibinfo  [0]{\@secondoftwo}%
\providecommand \bibfield  [0]{\@secondoftwo}%
\providecommand \translation [1]{[#1]}%
\providecommand \BibitemOpen [0]{}%
\providecommand \bibitemStop [0]{}%
\providecommand \bibitemNoStop [0]{.\EOS\space}%
\providecommand \EOS [0]{\spacefactor3000\relax}%
\providecommand \BibitemShut  [1]{\csname bibitem#1\endcsname}%
\let\auto@bib@innerbib\@empty
%</preamble>
\bibitem [{\citenamefont {Levine}\ \emph {et~al.}(1985)\citenamefont {Levine},
  \citenamefont {Lubensky}, \citenamefont {Ostlund}, \citenamefont {Ramaswamy},
  \citenamefont {Steinhardt},\ and\ \citenamefont
  {Toner}}]{SMLevine85elasticity}%
  \BibitemOpen
  \bibfield  {author} {\bibinfo {author} {\bibfnamefont {D.}~\bibnamefont
  {Levine}}, \bibinfo {author} {\bibfnamefont {T.~C.}\ \bibnamefont
  {Lubensky}}, \bibinfo {author} {\bibfnamefont {S.}~\bibnamefont {Ostlund}},
  \bibinfo {author} {\bibfnamefont {S.}~\bibnamefont {Ramaswamy}}, \bibinfo
  {author} {\bibfnamefont {P.~J.}\ \bibnamefont {Steinhardt}}, \ and\ \bibinfo
  {author} {\bibfnamefont {J.}~\bibnamefont {Toner}},\ }\href {\doibase
  10.1103/PhysRevLett.54.1520} {\bibfield  {journal} {\bibinfo  {journal}
  {Phys. Rev. Lett.}\ }\textbf {\bibinfo {volume} {54}},\ \bibinfo {pages}
  {1520} (\bibinfo {year} {1985})}\BibitemShut {NoStop}%
\bibitem [{\citenamefont {Salvalaglio}\ and\ \citenamefont
  {Elder}(2022)}]{SMsalvalaglio2022coarse}%
  \BibitemOpen
  \bibfield  {author} {\bibinfo {author} {\bibfnamefont {M.}~\bibnamefont
  {Salvalaglio}}\ and\ \bibinfo {author} {\bibfnamefont {K.~R.}\ \bibnamefont
  {Elder}},\ }\href {\doibase 10.1088/1361-651x/ac681e} {\bibfield  {journal}
  {\bibinfo  {journal} {Model. Simul. Mater. Sci. Eng.}\ }\textbf {\bibinfo
  {volume} {30}},\ \bibinfo {pages} {053001} (\bibinfo {year}
  {2022})}\BibitemShut {NoStop}%
\bibitem [{\citenamefont {Socolar}\ and\ \citenamefont
  {Steinhardt}(1986)}]{SMSocolar86unitcell}%
  \BibitemOpen
  \bibfield  {author} {\bibinfo {author} {\bibfnamefont {J.~E.~S.}\
  \bibnamefont {Socolar}}\ and\ \bibinfo {author} {\bibfnamefont {P.~J.}\
  \bibnamefont {Steinhardt}},\ }\href {\doibase 10.1103/PhysRevB.34.617}
  {\bibfield  {journal} {\bibinfo  {journal} {Phys. Rev. B}\ }\textbf {\bibinfo
  {volume} {34}},\ \bibinfo {pages} {617} (\bibinfo {year} {1986})}\BibitemShut
  {NoStop}%
\bibitem [{\citenamefont {Steinhardt}\ \emph {et~al.}(1998)\citenamefont
  {Steinhardt}, \citenamefont {Jeong}, \citenamefont {Saitoh}, \citenamefont
  {Tanaka}, \citenamefont {Abe},\ and\ \citenamefont
  {Tsai}}]{SMsteinhardt1998experimental}%
  \BibitemOpen
  \bibfield  {author} {\bibinfo {author} {\bibfnamefont {P.~J.}\ \bibnamefont
  {Steinhardt}}, \bibinfo {author} {\bibfnamefont {H.-C.}\ \bibnamefont
  {Jeong}}, \bibinfo {author} {\bibfnamefont {K.}~\bibnamefont {Saitoh}},
  \bibinfo {author} {\bibfnamefont {M.}~\bibnamefont {Tanaka}}, \bibinfo
  {author} {\bibfnamefont {E.}~\bibnamefont {Abe}}, \ and\ \bibinfo {author}
  {\bibfnamefont {A.}~\bibnamefont {Tsai}},\ }\href {\doibase 10.1038/23902}
  {\bibfield  {journal} {\bibinfo  {journal} {Nature}\ }\textbf {\bibinfo
  {volume} {396}},\ \bibinfo {pages} {55} (\bibinfo {year} {1998})}\BibitemShut
  {NoStop}%
\bibitem [{\citenamefont {Hirvonen}\ \emph {et~al.}(2019)\citenamefont
  {Hirvonen}, \citenamefont {Heinonen}, \citenamefont {Dong}, \citenamefont
  {Fan}, \citenamefont {Elder},\ and\ \citenamefont
  {Ala-Nissila}}]{SMhirvonen_PhysRevB.100.165412}%
  \BibitemOpen
  \bibfield  {author} {\bibinfo {author} {\bibfnamefont {P.}~\bibnamefont
  {Hirvonen}}, \bibinfo {author} {\bibfnamefont {V.}~\bibnamefont {Heinonen}},
  \bibinfo {author} {\bibfnamefont {H.}~\bibnamefont {Dong}}, \bibinfo {author}
  {\bibfnamefont {Z.}~\bibnamefont {Fan}}, \bibinfo {author} {\bibfnamefont
  {K.~R.}\ \bibnamefont {Elder}}, \ and\ \bibinfo {author} {\bibfnamefont
  {T.}~\bibnamefont {Ala-Nissila}},\ }\href {\doibase
  10.1103/PhysRevB.100.165412} {\bibfield  {journal} {\bibinfo  {journal}
  {Phys. Rev. B}\ }\textbf {\bibinfo {volume} {100}},\ \bibinfo {pages}
  {165412} (\bibinfo {year} {2019})}\BibitemShut {NoStop}%
\bibitem [{\citenamefont {Skogvoll}\ \emph {et~al.}(2021)\citenamefont
  {Skogvoll}, \citenamefont {Skaugen},\ and\ \citenamefont
  {Angheluta}}]{SMskogvoll2021_10.1103}%
  \BibitemOpen
  \bibfield  {author} {\bibinfo {author} {\bibfnamefont {V.}~\bibnamefont
  {Skogvoll}}, \bibinfo {author} {\bibfnamefont {A.}~\bibnamefont {Skaugen}}, \
  and\ \bibinfo {author} {\bibfnamefont {L.}~\bibnamefont {Angheluta}},\ }\href
  {\doibase 10.1103/PhysRevB.103.224107} {\bibfield  {journal} {\bibinfo
  {journal} {Phys. Rev. B}\ }\textbf {\bibinfo {volume} {103}},\ \bibinfo
  {pages} {224107} (\bibinfo {year} {2021})}\BibitemShut {NoStop}%
\bibitem [{\citenamefont {Achim}\ \emph {et~al.}(2014)\citenamefont {Achim},
  \citenamefont {Schmiedeberg},\ and\ \citenamefont {L\"owen}}]{SMAchim2014}%
  \BibitemOpen
  \bibfield  {author} {\bibinfo {author} {\bibfnamefont {C.~V.}\ \bibnamefont
  {Achim}}, \bibinfo {author} {\bibfnamefont {M.}~\bibnamefont {Schmiedeberg}},
  \ and\ \bibinfo {author} {\bibfnamefont {H.}~\bibnamefont {L\"owen}},\ }\href
  {\doibase 10.1103/PhysRevLett.112.255501} {\bibfield  {journal} {\bibinfo
  {journal} {Phys. Rev. Lett.}\ }\textbf {\bibinfo {volume} {112}},\ \bibinfo
  {pages} {255501} (\bibinfo {year} {2014})}\BibitemShut {NoStop}%
\bibitem [{\citenamefont {Goldenfeld}\ \emph {et~al.}(2005)\citenamefont
  {Goldenfeld}, \citenamefont {Athreya},\ and\ \citenamefont
  {Dantzig}}]{SMGoldenfeld2005}%
  \BibitemOpen
  \bibfield  {author} {\bibinfo {author} {\bibfnamefont {N.}~\bibnamefont
  {Goldenfeld}}, \bibinfo {author} {\bibfnamefont {B.~P.}\ \bibnamefont
  {Athreya}}, \ and\ \bibinfo {author} {\bibfnamefont {J.~A.}\ \bibnamefont
  {Dantzig}},\ }\href {\doibase 10.1103/PhysRevE.72.020601} {\bibfield
  {journal} {\bibinfo  {journal} {Phys. Rev. E}\ }\textbf {\bibinfo {volume}
  {72}},\ \bibinfo {pages} {020601(R)} (\bibinfo {year} {2005})}\BibitemShut
  {NoStop}%
\bibitem [{\citenamefont {Mermin}\ and\ \citenamefont
  {Troian}(1985)}]{SMMermin1985}%
  \BibitemOpen
  \bibfield  {author} {\bibinfo {author} {\bibfnamefont {N.~D.}\ \bibnamefont
  {Mermin}}\ and\ \bibinfo {author} {\bibfnamefont {S.~M.}\ \bibnamefont
  {Troian}},\ }\href {\doibase 10.1103/PhysRevLett.54.1524} {\bibfield
  {journal} {\bibinfo  {journal} {Phys. Rev. Lett.}\ }\textbf {\bibinfo
  {volume} {54}},\ \bibinfo {pages} {1524} (\bibinfo {year}
  {1985})}\BibitemShut {NoStop}%
\bibitem [{\citenamefont {Lifshitz}\ and\ \citenamefont
  {Petrich}(1997)}]{SMLifshitz1997}%
  \BibitemOpen
  \bibfield  {author} {\bibinfo {author} {\bibfnamefont {R.}~\bibnamefont
  {Lifshitz}}\ and\ \bibinfo {author} {\bibfnamefont {D.~M.}\ \bibnamefont
  {Petrich}},\ }\href {\doibase 10.1103/PhysRevLett.79.1261} {\bibfield
  {journal} {\bibinfo  {journal} {Phys. Rev. Lett.}\ }\textbf {\bibinfo
  {volume} {79}},\ \bibinfo {pages} {1261} (\bibinfo {year}
  {1997})}\BibitemShut {NoStop}%
\bibitem [{\citenamefont {Levine}\ and\ \citenamefont
  {Steinhardt}(1984)}]{SMLevine1984quasicrystals}%
  \BibitemOpen
  \bibfield  {author} {\bibinfo {author} {\bibfnamefont {D.}~\bibnamefont
  {Levine}}\ and\ \bibinfo {author} {\bibfnamefont {P.~J.}\ \bibnamefont
  {Steinhardt}},\ }\href {\doibase 10.1103/PhysRevLett.53.2477} {\bibfield
  {journal} {\bibinfo  {journal} {Phys. Rev. Lett.}\ }\textbf {\bibinfo
  {volume} {53}},\ \bibinfo {pages} {2477} (\bibinfo {year}
  {1984})}\BibitemShut {NoStop}%
\bibitem [{\citenamefont {Benoit-Maréchal}\ and\ \citenamefont
  {Salvalaglio}(2024)}]{SMBenoit-Marechal_2024}%
  \BibitemOpen
  \bibfield  {author} {\bibinfo {author} {\bibfnamefont {L.}~\bibnamefont
  {Benoit-Maréchal}}\ and\ \bibinfo {author} {\bibfnamefont {M.}~\bibnamefont
  {Salvalaglio}},\ }\href {\doibase 10.1088/1361-651X/ad42bb} {\bibfield
  {journal} {\bibinfo  {journal} {Model. Simul. Mater. Sci. Eng.}\ }\textbf
  {\bibinfo {volume} {32}},\ \bibinfo {pages} {055005} (\bibinfo {year}
  {2024})}\BibitemShut {NoStop}%
\bibitem [{\citenamefont {De}\ and\ \citenamefont
  {Pelcovits}(1987)}]{SMPialiPRB1987}%
  \BibitemOpen
  \bibfield  {author} {\bibinfo {author} {\bibfnamefont {P.}~\bibnamefont
  {De}}\ and\ \bibinfo {author} {\bibfnamefont {R.~A.}\ \bibnamefont
  {Pelcovits}},\ }\href {\doibase 10.1103/PhysRevB.35.8609} {\bibfield
  {journal} {\bibinfo  {journal} {Phys. Rev. B}\ }\textbf {\bibinfo {volume}
  {35}},\ \bibinfo {pages} {8609} (\bibinfo {year} {1987})}\BibitemShut
  {NoStop}%
\bibitem [{\citenamefont {Ding}\ \emph {et~al.}(1993)\citenamefont {Ding},
  \citenamefont {Yang}, \citenamefont {Hu},\ and\ \citenamefont
  {Wang}}]{SMDingPRB93}%
  \BibitemOpen
  \bibfield  {author} {\bibinfo {author} {\bibfnamefont {D.-h.}\ \bibnamefont
  {Ding}}, \bibinfo {author} {\bibfnamefont {W.}~\bibnamefont {Yang}}, \bibinfo
  {author} {\bibfnamefont {C.}~\bibnamefont {Hu}}, \ and\ \bibinfo {author}
  {\bibfnamefont {R.}~\bibnamefont {Wang}},\ }\href {\doibase
  10.1103/PhysRevB.48.7003} {\bibfield  {journal} {\bibinfo  {journal} {Phys.
  Rev. B}\ }\textbf {\bibinfo {volume} {48}},\ \bibinfo {pages} {7003}
  (\bibinfo {year} {1993})}\BibitemShut {NoStop}%
\bibitem [{\citenamefont {Salvalaglio}\ \emph {et~al.}(2019)\citenamefont
  {Salvalaglio}, \citenamefont {Voigt},\ and\ \citenamefont
  {Elder}}]{SMSalvalaglioNPJ2019}%
  \BibitemOpen
  \bibfield  {author} {\bibinfo {author} {\bibfnamefont {M.}~\bibnamefont
  {Salvalaglio}}, \bibinfo {author} {\bibfnamefont {A.}~\bibnamefont {Voigt}},
  \ and\ \bibinfo {author} {\bibfnamefont {K.~R.}\ \bibnamefont {Elder}},\
  }\href {\doibase 10.1038/s41524-019-0185-0} {\bibfield  {journal} {\bibinfo
  {journal} {npj Comput. Mater.}\ }\textbf {\bibinfo {volume} {5}},\ \bibinfo
  {pages} {48} (\bibinfo {year} {2019})}\BibitemShut {NoStop}%
\bibitem [{\citenamefont {Mazenko}(1997)}]{SMMazenko97}%
  \BibitemOpen
  \bibfield  {author} {\bibinfo {author} {\bibfnamefont {G.~F.}\ \bibnamefont
  {Mazenko}},\ }\href {\doibase 10.1103/PhysRevLett.78.401} {\bibfield
  {journal} {\bibinfo  {journal} {Phys. Rev. Lett.}\ }\textbf {\bibinfo
  {volume} {78}},\ \bibinfo {pages} {401} (\bibinfo {year} {1997})}\BibitemShut
  {NoStop}%
\bibitem [{\citenamefont {Skaugen}\ \emph {et~al.}(2018)\citenamefont
  {Skaugen}, \citenamefont {Angheluta},\ and\ \citenamefont
  {Vi\~nals}}]{SMSkaugen2018b}%
  \BibitemOpen
  \bibfield  {author} {\bibinfo {author} {\bibfnamefont {A.}~\bibnamefont
  {Skaugen}}, \bibinfo {author} {\bibfnamefont {L.}~\bibnamefont {Angheluta}},
  \ and\ \bibinfo {author} {\bibfnamefont {J.}~\bibnamefont {Vi\~nals}},\
  }\href {\doibase 10.1103/PhysRevB.97.054113} {\bibfield  {journal} {\bibinfo
  {journal} {Phys. Rev. B}\ }\textbf {\bibinfo {volume} {97}},\ \bibinfo
  {pages} {054113} (\bibinfo {year} {2018})}\BibitemShut {NoStop}%
\bibitem [{\citenamefont {Skogvoll}\ \emph {et~al.}(2022)\citenamefont
  {Skogvoll}, \citenamefont {Angheluta}, \citenamefont {Skaugen}, \citenamefont
  {Salvalaglio},\ and\ \citenamefont {Viñals}}]{SMSKOGVOLL2022104932}%
  \BibitemOpen
  \bibfield  {author} {\bibinfo {author} {\bibfnamefont {V.}~\bibnamefont
  {Skogvoll}}, \bibinfo {author} {\bibfnamefont {L.}~\bibnamefont {Angheluta}},
  \bibinfo {author} {\bibfnamefont {A.}~\bibnamefont {Skaugen}}, \bibinfo
  {author} {\bibfnamefont {M.}~\bibnamefont {Salvalaglio}}, \ and\ \bibinfo
  {author} {\bibfnamefont {J.}~\bibnamefont {Viñals}},\ }\href {\doibase
  https://doi.org/10.1016/j.jmps.2022.104932} {\bibfield  {journal} {\bibinfo
  {journal} {J. Mech. Phys. of Solids}\ }\textbf {\bibinfo {volume} {166}},\
  \bibinfo {pages} {104932} (\bibinfo {year} {2022})}\BibitemShut {NoStop}%
\bibitem [{\citenamefont {Skogvoll}\ \emph {et~al.}(2023)\citenamefont
  {Skogvoll}, \citenamefont {R{\o}nning}, \citenamefont {Salvalaglio},\ and\
  \citenamefont {Angheluta}}]{SMSkogvollNPJ2023}%
  \BibitemOpen
  \bibfield  {author} {\bibinfo {author} {\bibfnamefont {V.}~\bibnamefont
  {Skogvoll}}, \bibinfo {author} {\bibfnamefont {J.}~\bibnamefont
  {R{\o}nning}}, \bibinfo {author} {\bibfnamefont {M.}~\bibnamefont
  {Salvalaglio}}, \ and\ \bibinfo {author} {\bibfnamefont {L.}~\bibnamefont
  {Angheluta}},\ }\href {\doibase 10.1038/s41524-023-01077-6} {\bibfield
  {journal} {\bibinfo  {journal} {npj Comput. Mater.}\ }\textbf {\bibinfo
  {volume} {9}},\ \bibinfo {pages} {122} (\bibinfo {year} {2023})}\BibitemShut
  {NoStop}%
\end{thebibliography}

\iffalse
%apsrev4-2.bst 2019-01-14 (MD) hand-edited version of apsrev4-1.bst
%Control: key (0)
%Control: author (8) initials jnrlst
%Control: editor formatted (1) identically to author
%Control: production of article title (0) allowed
%Control: page (0) single
%Control: year (1) truncated
%Control: production of eprint (0) enabled
%
\fi

\clearpage
\newpage

\onecolumngrid

\begin{center}
  \textbf{\large \hspace{5pt} Supporting Information (SI) Text for: \\ \vspace{0.2cm} Dynamics of microvesicle formation driven by glycocalyx \\}
\vspace{0.4cm}   Ke Xiao$^{1,2}$ and Chen-Xu Wu$^{2}$ \\[.1cm]
  {\itshape \small
  ${}^1$Department of Mechanical and Aerospace Engineering, University of California San Diego, La Jolla CA 92093, USA
  \\
  ${}^2$Fujian Provincial Key Lab for Soft Functional Materials Research, Research Institute for Biomimetics and Soft Matter, Department of Physics, College of Physical Science and Technology, Xiamen University, Xiamen 361005, People's Republic of China
 }
  %\vspace{0.1cm}{\small
  % ${}^*$kexiao@ucsd.edu\\
  % ${}^\dagger$prangamani@health.ucsd.edu\\
  % ${}^\ddagger$prangamani@health.ucsd.edu}
%(Dated: \today)\\
\vspace{0.5cm}
\end{center}

\setcounter{equation}{0}
\setcounter{figure}{0}
\setcounter{table}{0}
\setcounter{page}{1}

\renewcommand{\thesection}{S-\Roman{section}}
\renewcommand{\theequation}{S-\arabic{equation}}
\renewcommand{\thefigure}{S-\arabic{figure}}
\renewcommand{\bibnumfmt}[1]{[S#1]}
\renewcommand{\citenumfont}[1]{S#1}

In this Supporting Information (SI), we provide a detailed description of both the deterministic and stochastic models used to study the process of MV formation. 

\section{1. A Deterministic Description}
\label{Deterministic model}
In order to obtain the equation governing the formation dynamics of the MV, we employ the Onsager variational principle~\cite{Onsager1931_1,Onsager1931_2} to derive the deterministic dynamical equations. 
According to the Onsager variational principle, it requires the construction of a Rayleighian $\mathcal{R}=\dot{\mathcal{F}}_{\rm tot}+\mathcal{D}_{\rm tot}$ with $\dot{\mathcal{F}}_{\rm tot}$ the time derivative of the total free
energy and $\mathcal{D}_{\rm tot}$ the energy dissipation function. 
The kinetic equations are then obtained by minimizing the Rayleighian $\mathcal{R}$.  
In the following, we provide a detailed derivation of the total free energy, the total energy dissipation function, and the resulting dynamic equation presented in the main text. 
For completeness, we recapitulate the derivation of the total free energy; further details can be found in Refs.~\cite{Zhulina2006,K.Xiao2025PNAS,K.Xiao2026}. 
A key distinction from the previous model~\cite{K.Xiao2025PNAS,K.Xiao2026} is that our derivation here accounts for the ionic energy contribution, which was not considered previously. 

\subsection{1.1 Total free energy} 
\label{Total_free_energy}
To estimate the total fee energy described in Eq.~(\ref{F_tot}) in the main text, we consider a glycocalyx-membrane-cortex system, as shown in Fig.~\ref{fig:figure_1}. 
The initial configuration consists of a flat membrane grafted with glycocalyx biopolymers on its outer surface and a thin layer of actin cortex attached beneath the membrane. 
During MV formation, we assume that the membrane domain covered by glycocalyx biopolymers with a fixed surface area $A_0=\pi R_0^2$ is deformed into a family of spherical caps of radius $R_s$ and opening angle $2\alpha$, where $R_0$ denotes the in-plane radius of the flat membrane. 
Since the area $A_0$ covered by the glycocalyx is directly related to $R_0$, therefore, $R_0$ serves as a measure of the coverage of glycocalyx. The radius of the base and the height of the spherical cap are denoted by $R_{\rm b}$ and $h$, respectively. The membrane configuration is conveniently described by a single shape parameter $\eta$, defined as the area ratio between the spherical cap and a complete sphere. 
When $\eta=1$ the membrane morphology corresponds to a sphere that
implies a fully formed MV. 
The special geometries also allow us to express the radius, opening angle $2\alpha$, base radius, and height of the spherical cap in terms of the shape parameter $\eta$, which are given by $R_{\rm s}=R_0/(2\sqrt{\eta})$, $\cos\alpha=1-2\eta$, $R_{\rm b}=R_0\sqrt{1-\eta}$, and $h=R_0\sqrt{\eta}$, respectively.  

\subsubsection{1.1.1 Elastic energy of cell membrane}
By treating the membrane as a thin elastic layer, the bending energy density is given by $f_{\rm bending}=\kappa(2H-c_0)/2$, which involves the local mean curvature $H=(c_1+c_2)/2$, spontaneous curvature $c_0$ and the bending modulus $\kappa$. 
For the spherical cap geometry considered here, the two principal curvatures are equal $c_1=c_2=1/R_{\rm s}$. 
Integrating $f_{\text{bending}}$ across the entire membrane surface yields the energy contribution due to membrane bending: 
\begin{align}
\mathcal{F}_{\rm bending}=8\pi\kappa\biggl(\sqrt{\eta}-\dfrac{R_0c_0}{4}\biggr)^2.
\label{F_bending_s}
\end{align}
The deformation of the membrane from its initially flat configuration also involves a contribution of surface energy due to the tension $\mathcal{F}_{\rm tension} = \sigma \Delta A$, where $\sigma$ is the membrane tension and $\Delta A$ is the excess area relative to the flat state. For the spherical cap geometry, the excess area is given by $\Delta A=A_0-\pi R_{\rm b}^2$, and hence the contribution of the tension to the energy is given by
\begin{align}
\mathcal{F}_{\rm tension}=\pi\sigma\eta R_0^2.
\label{F_tension_s}
\end{align}

\subsubsection{1.1.2 Energy contribution of glycocalyx polymer} 
\begin{figure*}[h!]
\centering
\includegraphics[width=\linewidth]{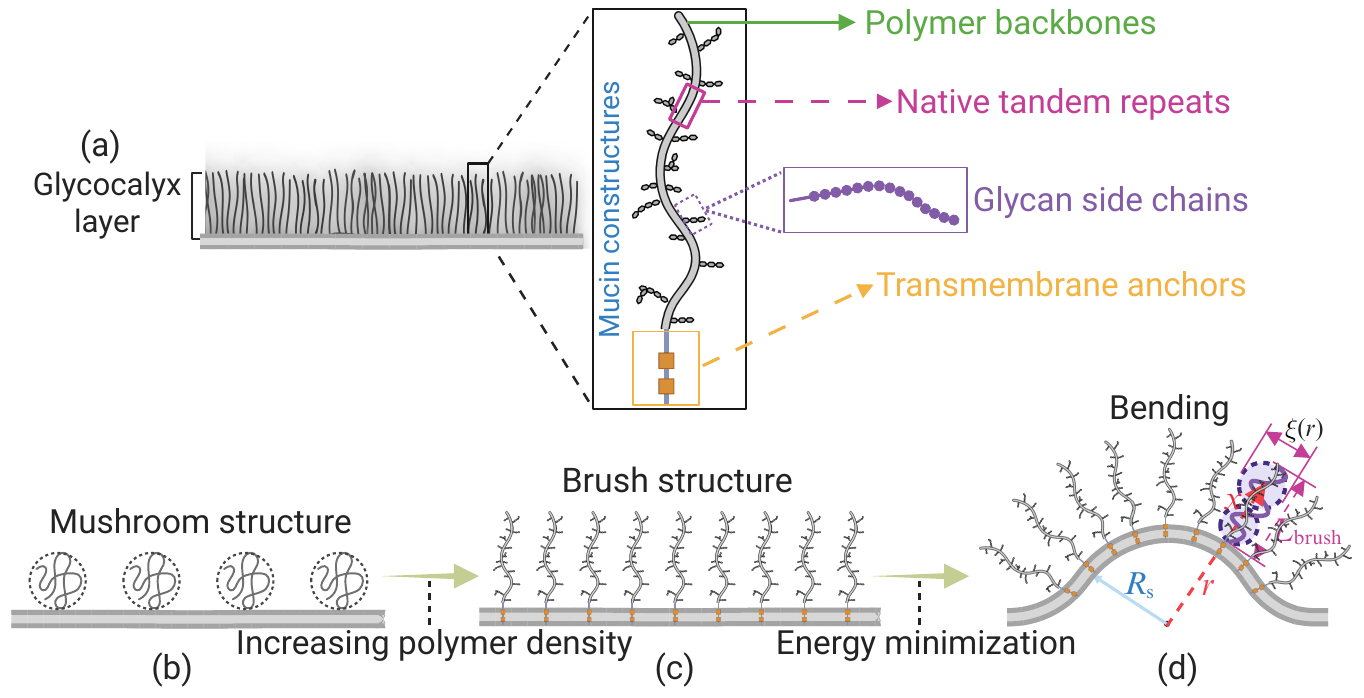}
\caption{Schematic diagram of (a) glycocalyx, a thick layer of heavily glycocylated transmembrane proteins concentrated on the cell surface in a polymer brush structure. 
The enlarged box is the schematic depiction of a bottlebrush structured mucin polymer which is a heavy protein backbone attached to the membrane via a tansmembrane anchor. In mucin polymers, it is usually decorated with a higher density of short glycan side chains on the polymer backbones. 
(b) Glycopolymers densely grafted on the cell surface form a compact structure phenotype: mushroom regime at low densities. 
(c) Upon increasing the glycopolymer density exceed a threshold, polymers extend out into a brush like structure: brush regime to avoid each other and to reduce crowding. 
(d) At high polymer density, the stretched polymer chains lead to a decrease of the chain entropy and a related increase of the chain free energy. To reduce the entropy cost of the system, bending of the membrane provides more space and gives more orientational degrees of freedom for the polymer chains. Thus, the increased order imposes an entropic penalty on the system, which is reduced via bending of the membrane (producing a positive membrane curvature along the whole area covered by the glycocalyx).} \label{fig:schematic polymer}
\end{figure*}

%\subsubsection*{Description of glycocalyx} 
The cell membrane of eukaryotic cells often displays a brush-like complex layer on its outer surface called the glycocalyx~\cite{GODULA20191672}. 
The glycocalyx architecture varies considerably between different cell types due to its richness in components such as proteoglycans, glycolipids, glycoproteins, and proteins~\cite{Leonhard2020.00253}. 
Structurally, the core components of the glycocalyx, a transmembrane fusion proteins, are arranged into two functionally distinct parts: a lower transmembrane domain (anchor) without cytoplasmic tail ($\Delta$CT) and a native tandem repeat (TR) ectodomain of transmembrane mucin, MUC1 (Mucin 1) which is usually regarded as a polymer-based backbone that extends through and above the lower transmembrane domain [see Fig.~\ref{fig:schematic polymer}(a)].  
Hence, the length of transmembrane mucins can be tuned by varying the number of tandem repeats. 
These mucin polymer backbones are heavily glycosylated, forming bottlebrush-like molecular structures due to the attachment of sugar glycan side chains. 
Two major classes of glycoproteins can be divided depending on the attachment of the sugar glycan to the protein: If the mucin polymer backbone is glycosylated with the $O$-linked serine or threonine sugar side chain, one speaks of $O$-glycosylation and $O$-glycans. 
If the glycan is attached via a nitrogen atom of an asparagine side chain, one speaks of $N$-glycosylation and $N$-glycans. 
The resulting large $N$- or $O$-terminal extracellular domain with heavily glycosylated branches renders a bottle-brush structure that can extend hundreds of nanometers from the cell surface. 
Thus, the dense array of the extended and heavily glycosylated transmembrane mucins forms a thick, polymer-based structural network on the membrane surface—the glycocalyx.

In particular, the crowding of large glycosylated proteins has been shown to regulate the shape of the underlying bilayer plasma membrane~\cite{SHURER20191757}. 
According to polymer physics, two distinct regimes can be defined depending on the grafting density of glycocalyx polymers on the cell membrane: the mushroom regime [see Fig.~\ref{fig:schematic polymer}(b)] and the brush regime [see Fig.~\ref{fig:schematic polymer}(c)]. 
Shurer \textit{et al}.~\cite{SHURER20191757} have reported that mucins are in the coiled mushroom state at low grafting densities. 
In this work, we focus on densely grafted regions of the membrane and model the glycocalyx in the brush regime.

Here, we consider a glycocalyx layer grafted on a membrane of radius $R_{\rm s}$, forming a brush-like structure of thickness $L_{\rm brush}$, as shown in Fig.~\ref{fig:schematic polymer}(d). From a coarse-grained perspective, the polymer brush can be treated as an array of blobs. 
The size of each blob, $\xi$, at a given position $r$ equals the square root of the local area per chain $s(r)$, where $r=x+R_{\rm s}$ is the radial distance from the center of the spherical membrane and $x$ is the distance measured from the surface of the membrane. 
Thus, the size of the blob $\xi=\sqrt{s(r)}$ grows as a function of $r$, and the grafting density of glycocalyx polymer brush on the membrane surface can be obtained as $\rho=1/\xi^2$. 
Assuming that the glycopolymers layer is extended non-uniformly but equally in height, $L_{\rm brush}$, then the area per chain at distance $x$ from the membrane surface is given by~\cite{Zhulina2006}
\begin{align}
s(r)=s(x+R_{\rm s})=s(R_{\rm s})\biggl(\frac{r}{R_{\rm s}}\biggr)^{i-1},~~~(i=1,2,3). \label{Sr}
\end{align}
The index $i=1,~2,~3$ indicates planar, cylindrical, and spherical membranes, respectively. When the membrane is bent, the changes in polymer configuration give rise to the local extension of the polymer chain. Here, the local chain extension at a height $r$ is characterized by ${\rm d}r/{\rm d}n$, where the variable $n$ denotes the current monomer.
This local extension is related to the local density profile of monomers $c_{\rm p}(r)$ by~\cite{Zhulina2006}
\begin{align}
c_{\rm p}(r)=\frac{{\rm d}n}{s(r){\rm d}r}.\label{cp_r}
\end{align}
Then, the thickness of the brush, $L_{\rm brush}$, is found from the conservation condition
\begin{align}
\int_{R_{\rm s}}^{R_{\rm s}+L_{\rm brush}}c_{\rm p}(r)s(r){\rm d}r=N\label{conservation condition}
\end{align}
for the total number of monomeric segments $N$~\cite{Zhulina2006}. As a result, in the brush regime, the energy contribution originating from glycocalyx polymers includes three terms~\cite{Zhulina2006,SHURER20191757}: the elastic energy of the polymer chain ($\mathcal{F}_{\rm glycocalyx}^{\rm elastic}$), the free energy caused by the excluded volume interactions of polymer monomers ($\mathcal{F}_{\rm glycocalyx}^{\rm EV}$) and the electrostatic contribution for the ionic brush ($\mathcal{F}_{\rm glycocalyx}^{\rm ion}$). Based on the hypotheses in the main text, according to Ref.~\cite{Zhulina2006}, the elastic energy per chain in the brush can be written as
\begin{align}
\mathcal{F}_{\rm glycocalyx}^{\rm elastic}&=\int_{R_{\rm s}}^{R_{\rm s}+L_{\rm brush}}f_{\rm glycocalyx}^{\rm elastic}s(r){\rm d} r \notag\\
&=\int_{R_{\rm s}}^{R_{\rm s}+L_{\rm brush}}\frac{3k_BT}{2a^2c_{\rm p}(r)s^2(r)}s(r){\rm d} r,\label{F_els}
\end{align}
where $a$ is the average effective monomeric segment size and $f_{\rm glycocalyx}^{\rm elastic}$ is the elastic energy density of the polymer chain.
Within the mean-field approximation, the energy density of the excluded volume interactions (van der Waals interactions) between monomers can be modeled in terms of the virial expansion
\begin{align}
f_{\rm glycocalyx}^{\rm EV}=k_BT[vc_{\rm p}^2(r)+wc_{\rm p}^3(r)+\cdots],\label{f_EV}
\end{align}
where $v$ and $w$ are the second and third virial coefficients, respectively. 
As a result, the excluded volume interaction between monomers per chain in the brush is given by
\begin{align}
\mathcal{F}_{\rm glycocalyx}^{\rm EV}=k_BT\int_{R_{\rm s}}^{R_{\rm s}+L_{\rm brush}}[vc_{\rm p}^2(r)+wc_{\rm p}^3(r)+\cdots]s(r){\rm d} r.\label{F_EV}
\end{align}
By combining the mean field and the local electroneutrality approximations~\cite{Zhulina2002}, the ionic contribution to the free energy is given by~\cite{Zhulina2006,SHURER20191757,Zhulina2002,FengYuan2023}
\begin{align}
\mathcal{F}_{\rm glycocalyx}^{\rm ion}&=\int_{R_{\rm s}}^{R_{\rm s}+L_{\rm brush}}f_{\rm glycocalyx}^{\rm ion}s(r){\rm d} r \notag\\
&=k_BT\displaystyle{\int_{R_{\rm s}}^{R_{\rm s}+L_{\rm brush}}}\biggl[\frac{\alpha_{\rm ion}^2}{2I_{\rm ion}}c_{\rm p}^2(r)\biggr]s(r){\rm d}r, \label{F_ion}
\end{align}
where $\alpha_{\rm ion}=\phi_{\rm b}\cdot N_{\rm ion}$ is the degree of ionization of each segment, and $I_{\rm ion}=\sum_jc_j^{\rm bulk}$ is the ion concentration (ionic strength) in the bulk solution, which is multiplied by the Avogadro's constant $N_{\rm A}$ in the simulations.
Here, $\phi_{\rm b}$ is the degree of ionization of each amino acid in the glycocalyx, and $N_{\rm ion}=a/a_{\rm m}$ denotes the number of
ionizable sites per segment with $a_{\rm m}$ being the size of a monomer. Note that the degree of ionization of an acid group in the bulk solution (i.e., outside the polymer brush layer), $\phi_{\rm b}$, is determined by the bulk concentration of protons in the form~\cite{Zhulina1996,Zhulina1995} 
\begin{align}
\frac{\phi_{\rm b}}{1-\phi_{\rm b}}=\frac{K_{\rm d}}{c_{\rm H}^{\rm bulk}}, \label{degree_of_ionization}
\end{align}
where $K_{\rm d}$ is the dissociation constant and $c_{\rm H}^{\rm bulk}$ is the bulk concentration of protons, which is related to the ${\rm pH}$ value of the buffer solution of the form ${\rm pH}=-\log_{10}[c_{\rm H}^{\rm bulk}]$. 
For simplicity, we ignore the possible dependence of $K_{\rm d}$ on other parameters and assume that $K_{\rm d}$ is a constant. 
The dimensionality of $K_{\rm d}$ is equal to that of $c_{\rm H}^{\rm bulk}$.

Moving forward, the cubic and higher terms in Eq.~(\ref{F_EV}) are neglected. Therefore, the sum of Eq.~(\ref{F_els}), Eq.~(\ref{F_EV}) and Eq.~(\ref{F_ion}) yields the energy contribution of glycocalyx polymers
\begin{align}
\mathcal{F}_{\rm glycocalyx}=&N_{\rm p}\Bigl(\mathcal{F}_{\rm glycocalyx}^{\rm elastic}+\mathcal{F}_{\rm glycocalyx}^{\rm EV}+\mathcal{F}_{\rm glycocalyx}^{\rm ion}\Bigr) \notag\\
=&N_{\rm p}k_BT\int_{R_{\rm s}}^{R_{\rm s}+L_{\rm brush}}\biggl[\frac{3}{2a^2c_{\rm p}(r)s^2(r)}+vc_{\rm p}^2(r)+\frac{\alpha_{\rm ion}^2}{2I_{\rm ion}}c_{\rm p}^2(r)\biggr]s(r){\rm d} r \notag\\
=&N_{\rm p}k_BT\int_{R_{\rm s}}^{R_{\rm s}+L_{\rm brush}}\biggl[\frac{3}{2a^2c_{\rm p}(r)s^2(r)}+Wc_{\rm p}^2(r)\biggr]s(r){\rm d} r, \label{F_els+F_EV}
\end{align}
where $N_{\rm p}=\rho A_0$ is the number of polymer chains grafted to the membrane and $W=v+\dfrac{\alpha_{\rm ion}^2}{2I_{\rm ion}}$. Here, $A_0$ stands for the area of the membrane patch covered by glycocalyx polymers and $\rho$ is the density.

To calculate the free energy, we need to further determine the local concentration of monomers $c_{\rm p}(r)$ and the brush thickness $L_{\rm brush}$. On a planar membrane surface, the planar brush area per chain, $s(r)$, is constant, i.e., $s(r)=s$.
Minimizing the free energy of a planar brush $F_{\rm glycocalyx}$ with respect to $c_{\rm p}$ by using $c_{\rm p}(r)\delta f[c_{\rm p}(r),r]/\delta c_{\rm p}(r)-f[c_{\rm p}(r),r]=0$ leads to the equilibrium polymer concentration $c_{\rm p}^{\rm flat}$ (see Ref.~\cite{Zhulina2006} for detailed steps)
\begin{align}
c_{\rm p}^{\rm flat}=\biggl(\frac{3}{Ws^2a^2}\biggr)^{1/3}. \label{c_p^flat}
\end{align}
Here, $f[c_{\rm p}(r),r]=f_{\rm glycocalyx}^{\rm elastic}+f_{\rm glycocalyx}^{\rm EV}+f_{\rm glycocalyx}^{\rm ion}$.  Using the conservation condition defined by Eq.~(\ref{conservation condition}) yields the brush thickness $L_{\rm brush}^{\rm flat}$ (see Ref.~\cite{Zhulina2006} for detailed steps)
\begin{align}
L_{\rm brush}^{\rm flat}=N\cdot\biggl(\frac{Wa^2}{3s}\biggr)^{1/3}, \label{H^flat}
\end{align}
where we can see that the thickness of the polymer brush is proportional to the total number of monomeric segments $N$.
In the following, we use the number of monomeric segments, $N$, to capture the length of the polymer.
Substituting Eq.~(\ref{c_p^flat}) and Eq.~(\ref{H^flat}) into Eq.~(\ref{F_els+F_EV}) yields the energy contribution of glycocalyx polymers on a planar membrane surface
\begin{align}
\mathcal{F}_{\rm glycocalyx}^{\rm flat}=N_{\rm p}\frac{9N}{2\beta}\biggl(\frac{W}{3sa}\biggr)^{2/3}.\label{F_polymer_flat}
\end{align}
Thus, even for a flat membrane, the energy contribution of the glycocalyx is directly proportional to the extent of grafting $N_{\rm p}$ and the length of the polymer brush $N$. 

On a spherical membrane surface, the corresponding monomer density profile $c_{\rm p}(r)$ and brush thickness $L_{\rm brush}$ are, respectively, expressed as (see Ref.~\cite{Zhulina2006} for a detailed calculation)
\begin{align}
c_{\rm p}(r)=\biggl[\dfrac{3}{Ws^2(r)a^2}\biggr]^{1/3}=c_{\rm p}^{\rm flat}\biggl(\dfrac{R_{\rm s}}{r}\biggr)^{4/3},\label{c_p_S_C}
\end{align}
and
\begin{align}
L_{\rm brush}=R_{\rm s}\biggl(1+\dfrac{5}{3}\cdot\dfrac{L_{\rm brush}^{\rm flat}}{R_{\rm s}}\biggr)^{3/5}-R_{\rm s}.\label{H_S_C}
\end{align}
Similarly, substituting Eq.~(\ref{c_p_S_C}) and Eq.~(\ref{H_S_C}) into Eq.~(\ref{F_els+F_EV}) leads to the energy contribution of glycocalyx polymers on a spherical membrane surface as
\begin{align}
\mathcal{F}_{\rm glycocalyx}=&N_{\rm p}(F_{\rm glycocalyx}^{\rm elastic}+F_{\rm glycocalyx}^{\rm EV}+F_{\rm glycocalyx}^{\rm ion}) \notag\\
=&N_{\rm p}\frac{9R_{\rm s}}{2\beta}\biggl(\frac{3\sqrt{W}}{\sqrt{s} a^2}\biggr)^{2/3}\biggl[\biggl(1+\frac{5N}{3R_{\rm s}}\cdot\Bigl(\frac{Wa^2}{3s}\Bigr)^{1/3}\biggr)^{1/5}-1\biggr].
\label{F_glycocalyx_s}
\end{align}

\subsubsection{1.1.3 Energy associated with membrane-cortex adhesion} 
To estimate the energy associated with membrane-cortex adhesion, we assume that the membrane-cortex adhesion energy is quadratic with respect to the membrane-cortex linker deformation and is thus given by 
\begin{align}
\mathcal{F}_{\rm ad}=\int\frac{k}{2}\rho_{\rm b}z^2~{\rm d} A, \label{F_binding}
\end{align}
where $k$ is the elastic stiffness (spring constant) of each membrane-cortex linker, $\rho_{\rm b}$ is the local density of the bound spring-like linkers and $z$ is the stretching length of the linkers. 
For simplicity, we further assume that (i) the density of the bound spring-like linkers is equal to that of the available linkers $\rho_0$, i.e., $\rho_{\rm b}=\rho_0$; 
(ii) the adhesion bond breaks when the membrane-cortex adhesion bond reaches its maximum length $l_{\rm c}$. 
\begin{figure*}[h!]
\centering
\includegraphics[width=\linewidth]{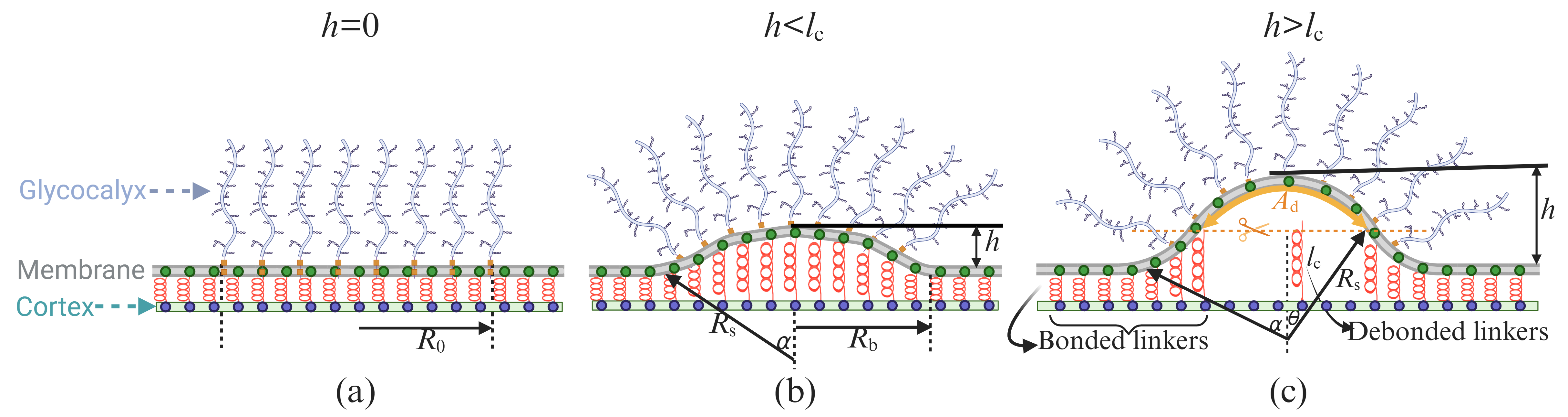}
\caption{Schematic of the glycocalyx-membrane-cortex system for MV formation. (a) A patch of glycocalyx with in-plane radius $R_0$, represented as a layer of brushlike polymer, is anchored on an initially flat membrane. The cortex beneath the membrane is attached to the membrane through linker proteins, which are represented by springs. 
(b) The small deformation state where the membrane takes the shape of a spherical cap with its radius of curvature, base radius, opening angle, and height denoted by $R_{\rm s}$, $R_{\rm b}$, $2\alpha$, and $h$. The linkers are attached to the membrane when the depth of the spherical cap is smaller than the maximum extension length of the membrane-cortex spring, i.e., $h<l_{\rm c}$. 
(c) In the case of $h>l_{\rm c}$, the spherical cap-shaped membrane is divided into two parts: the membrane-cortex adhesion region below the dashed
line with bonded linkers and the membrane-cortex detached region above the dashed line with debonded linkers. 
We denote the opening angle and area of the membrane-cortex detached region by $2\theta$ and $A_{\rm d}$.}
\label{fig:schematic cortex}
\end{figure*}

When the membrane deformation is small, the maximum height of the deformed membrane is smaller than the maximum extension length of the linkers ($R_0\sqrt{\eta}=h\leq l_{\rm c}$ which corresponds to $\eta\leq(l_{\rm c}/R_0)^2$, see Fig.~\ref{fig:schematic cortex}), the membrane-cortex adhesion energy can be calculated as 
\begin{align}
\mathcal{F}_{\rm ad}=&\int\frac{k}{2}\rho_{\rm 0}z^2~{\rm d}A \notag\\
=&\int_0^\alpha\frac{k}{2}\rho_{\rm 0}R_{\rm s}^2(\cos\Theta-\cos\alpha)^2 2\pi R_{\rm s}^2\sin\Theta ~{\rm d}\Theta \notag\\
=&\frac{\eta}{6}\pi k\rho_{\rm 0}R_0^4,\label{h<l_c}
\end{align}
where $\cos\alpha=1-2\eta$ with $\alpha$ being the opening angle of the spherical cap and $\Theta$ the integral variable. 

If the membrane is deformed into a spherical cap with its maximum height larger than the maximum extension length of the linkers, that is, $h>l_{\rm c}$ (see Fig.~\ref{fig:schematic cortex}). 
Then it means that part of the linkers are broken (see the regime above the dashed line in Fig.~\ref{fig:schematic cortex}) and some of the linkers are still attached to the membrane but with some degree of extension (see the regime below the dashed line in Fig.~\ref{fig:schematic cortex}) for the deformed membrane. 
Thereby, we can treat the membrane as two parts, one with the bond linkers broken, and the other one with the linkers binding on it. 
As a result, in the case of $h>l_{\rm c}$, the membrane-cortex adhesion energy consists of two parts, the first part is the one with the linker broken and the second part is the detached one with the linkers still attached but extended. 

When the membrane is detached from the cortex, the total number of broken adhesion bonds can be approximated by $N_{\rm b}=\rho_{\rm b} A_{\rm d}(\eta)$, where $A_{\rm d}(\eta)$ is the area of the detached membrane, which represents the area where the adhesion between the membrane and the cortex has been disrupted, as shown in Fig.~\ref{fig:schematic cortex}. 
Specifically, if we assume that the detachment between the membrane and the cortex occurs once their separation reaches $l_{\rm c}$, then $A_{\rm d}(\eta)$ can be shown to take the form of 
\begin{align}
A_{\rm d}(\eta)=2\pi R_{\rm s}^2\biggl(2\eta-\dfrac{l_{\rm c}}{R_{\rm s}}\biggr)=\pi R_0^2\biggl(1-\dfrac{l_{\rm c}}{R_0\sqrt{\eta}}\biggr).\label{A_d}
\end{align}
Therefore, the energy contribution from the first part can be obtained as 
\begin{align}
\mathcal{F}_{\rm ad}^{\rm broken}=&N_{\rm b}\frac{1}{2}kl_{\rm c}^2=\frac{k}{2}\rho_{\rm 0}l_{\rm c}^2A_{\rm d}(\eta) \notag\\
=&\frac{1}{2}\pi k\rho_{\rm 0}l_{\rm c}^2R_0^2\biggl(1-\frac{l_{\rm c}}{R_0\sqrt{\eta}}\biggr).\label{F_binding_broken}
\end{align}
The energy contribution from the second part can be obtained as
\begin{align}
\mathcal{F}_{\rm ad}^{\rm extended}=&\int\frac{k}{2}\rho_{\rm 0}z^2{\rm d}A \notag\\
=&\int_\theta^\alpha\frac{k}{2}\rho_{\rm 0}R_{\rm s}^2(\cos\Theta-\cos\alpha)^2 2\pi R_{\rm s}^2\sin\Theta {\rm d}\Theta \notag\\
=&\frac{1}{6\sqrt{\eta}}\pi k\rho_{\rm 0}R_0l_{\rm c}^3.\label{F_binding_extended}
\end{align}
Consequently, the membrane-cortex adhesion energy can be expressed as 
\begin{align}
\mathcal{F}_{\rm ad}=\begin{cases}
\dfrac{\eta}{6}\pi k\rho_{\rm 0}R_0^4, \quad &\eta\leq (l_{\rm c}/R_0)^2 \\
\dfrac{1}{2}\pi k\rho_{\rm 0}R_0l_{\rm c}^2\biggl(R_0-\dfrac{2l_{\rm c}}{3\sqrt{\eta}}\biggr). \quad &\eta> (l_{\rm c}/R_0)^2
\end{cases}\label{F_binding_all}
\end{align}

\subsubsection{1.1.4 The work done by the pressure} 
When there is a pressure difference $\Delta P$ between the inside and outside of the MV, an additional term of energy cost $\mathcal{F}_{\rm p}=-\Delta PV$ is included. 
For the spherical cap, its volume can be calculated as $V=\pi h^2(3R-h)/3=\pi R_0^3\sqrt{\eta}(3-2\eta)/6$. 
Thus, the work done by the pressure reads as follows
\begin{align}
\mathcal{F}_{\rm p}=-\Delta PV=\frac{\pi}{6}\Delta PR_0^3\sqrt{\eta}(2\eta-3).\label{F_p_s}
\end{align}

\subsubsection{1.1.5 Total free energy of the system} 
\begin{figure*}[h]
\centering
\includegraphics[width=\linewidth]{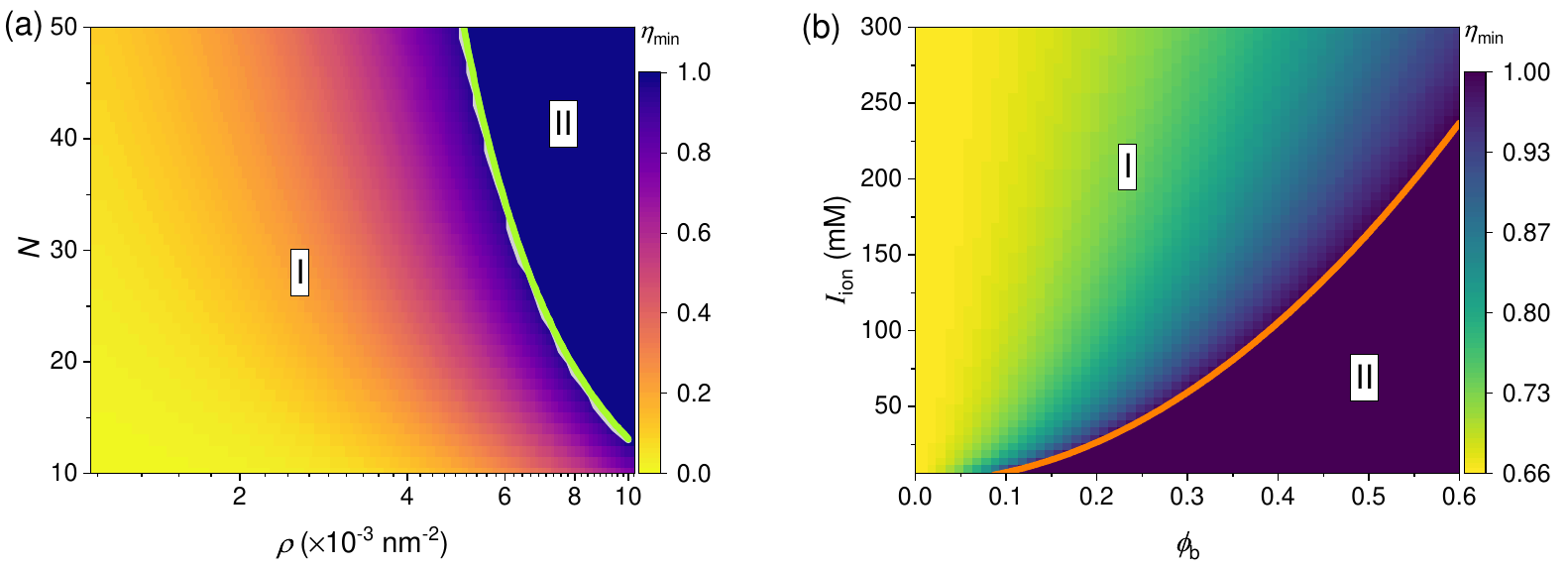}
\caption{The heat map of $\eta_{\rm min}$ on the (a) $\rho-N$ and (b) $\phi_{\rm b}-I_{\rm ion}$ spaces. 
The solid line marks the boundary of the formation of MVs, which is determined by setting Eq.~(\ref{dF/deta}) zero.
}\label{fig:Eq_phase_diagram}
\end{figure*}

Combining Eq.~(\ref{F_bending_s}), Eq.~(\ref{F_tension_s}), Eq.~(\ref{F_polymer_flat}), Eq.~(\ref{F_glycocalyx_s}), Eq.~(\ref{F_binding_all}) and Eq.~(\ref{F_p_s}) leads to the total free energy:
\begin{align}
\mathcal{F}_{\rm tot}=&N_{\rm p}k_BT\int_{R_{\rm s}}^{R_{\rm s}+L_{\rm brush}}\biggl[\frac{3}{2a^2c_{\rm p}(r)s^2(r)}+vc_{\rm p}^2(r)+\frac{\alpha_{\rm ion}^2}{2I_{\rm ion}}c_{\rm p}^2(r)\biggr]s(r)~{\rm d}r \notag\\ &+\int_\Omega\biggl[\frac{\kappa}{2}(2H-c_0)^2+\frac{k}{2}\rho_{\rm b}z^2\biggr]~{\rm d}A + \sigma\Delta A - \Delta PV \notag\\
=&\frac{9\pi R_0^3}{4\beta\xi^2\sqrt{\eta}}\biggl(\frac{3\sqrt{W}}{\xi a^2}\biggr)^{\frac{2}{3}}\biggl[\biggl(1+\frac{10N\sqrt{\eta}}{3R_0}\cdot\Bigl(\frac{Wa^2}{3\xi^2}\Bigr)^{\frac{1}{3}}\biggr)^{\frac{1}{5}}-1\biggr]+8\pi\kappa\biggl(\sqrt{\eta}-\frac{R_0c_0}{4}\biggr)^2 +\pi\sigma\eta R_0^2 \notag\\
&+\frac{\pi}{6}\Delta PR_0^3\sqrt{\eta}(2\eta-3)+\begin{cases}
\dfrac{\eta}{6}\pi k\rho_0R_0^4, \quad &0<\eta\leq (l_{\rm c}/R_0)^2 \\
\dfrac{1}{2}\pi k\rho_0l_{\rm c}^2R_0^2\biggl(1-\dfrac{l_{\rm c}}{R_0\sqrt{\eta}}\biggr) + \dfrac{1}{6\sqrt{\eta}}\pi k\rho_0R_0l_{\rm c}^3, \quad &\eta> (l_{\rm c}/R_0)^2
\end{cases} .\label{F_tot_no_mono}
\end{align}
To determine the equilibrium state of the system, we numerically compute the total free energy as a function of the relevant parameters. Minimizing this energy with respect to the shape parameter $\eta$ yields the optimal value $\eta_{\rm min}$, which characterizes the most favorable membrane morphology at equilibrium. 
Using this approach, we construct two-dimensional heat maps of $\eta_{\rm min}$ in the parameter spaces $\rho$–$N$ and $\phi_{\rm b}$–$I_{\rm ion}$, as shown in Fig.~\ref{fig:Eq_phase_diagram}. 
These phase diagrams provide a visual representation of the equilibrium membrane shape under varying glycocalyx properties and ionic conditions.
Taking a first order derivative of the total energy $\mathcal{F}_{\rm tot}$ with respect to $\eta$, one can obtain
\begin{align}
\mathcal{F}_{\rm tot}'(\eta)=\frac{{\rm d}\mathcal{F}_{\rm tot}}{{\rm d}\eta} 
=-\pi\mathcal{M}_1+\pi\mathcal{M}_2+8\pi\kappa\biggl(1-\dfrac{R_0c_0}{4\sqrt{\eta}}\biggr)+\pi R_0^2\sigma+\dfrac{\Delta P\pi R_0^3(2\eta-1)}{4\sqrt{\eta}}+\frac{\pi k\rho_0R_0l_{\rm c}^3}{6\eta^{3/2}} ,\label{dF/deta}
\end{align}
where 
\begin{align}
\mathcal{M}_1=\frac{9R_0^3}{8\beta\xi^2\eta^{3/2}}\biggl(\frac{3\sqrt{W}}{\xi a^2}\biggr)^{\frac{2}{3}}\biggl[\biggl(1+\frac{10N\sqrt{\eta}}{3R_0}\biggl(\frac{Wa^2}{3\xi^2}\biggr)^{\frac{1}{3}}\biggr)^{\frac{1}{5}}-1\biggr],\label{P1_no_mono_simple}
\end{align}
and
\begin{align}
\mathcal{M}_2=\frac{3NR_0^2}{4\beta\eta\xi^2}\biggl(\frac{3W^2}{\xi^4a^2}\biggr)^{\frac{1}{3}}\biggl[1+\frac{10N\sqrt{\eta}}{3R_0}\biggl(\frac{Wa^2}{3\xi^2}\biggr)^{\frac{1}{3}}\biggr]^{-\frac{4}{5}}.\label{P2_no_mono_simple}
\end{align}
Setting ${\rm d}\mathcal{F}_{\rm tot}/{\rm d}\eta=0$, we can obtain the analytical solution for the boundary curve that separates regime I from regime II, which is denoted by the solid line in the phase diagrams. 

Based on Eq.~(\ref{F_tot_no_mono}), the change rate of the total free energy is given by 
\begin{align}
\dot{\mathcal{F}}_{\rm tot}=&-\frac{9\pi R_0^3}{8\beta\xi^2\eta^{3/2}}\biggl(\frac{3\sqrt{W}}{\xi a^2}\biggr)^{\frac{2}{3}}\biggl[\biggl(1+\frac{10N\sqrt{\eta}}{3R_0}\biggl(\frac{Wa^2}{3\xi^2}\biggr)^{\frac{1}{3}}\biggr)^{\frac{1}{5}}-1\biggr] \dot{\eta}\notag\\
&+\frac{3N\pi R_0^2}{4\beta\eta\xi^2}\biggl(\frac{3W^2}{\xi^4a^2}\biggr)^{\frac{1}{3}}\biggl[1+\frac{10N\sqrt{\eta}}{3R_0}\biggl(\frac{Wa^2}{3\xi^2}\biggr)^{\frac{1}{3}}\biggr]^{-\frac{4}{5}} \dot{\eta}\notag\\
&+8\pi\kappa\biggl(1-\frac{R_0c_0}{4\sqrt{\eta}}\biggr)\dot{\eta}+\pi R_0^2\sigma\dot{\eta} +\frac{\pi \Delta PR_0^3(2\eta-1)}{4\sqrt{\eta}}\dot{\eta} \notag\\
&+\begin{cases}
\dfrac{1}{6}\pi k\rho_0R_0^4\dot{\eta}, \quad &0<\eta\leq (l_c/R_0)^2 \\
\dfrac{\pi k\rho_0R_0l_c^3}{6\eta^{3/2}}\dot{\eta},\quad &\eta> (l_c/R_0)^2
\end{cases}.\label{dF_tot_cap/dt_no_mono}
\end{align}

\subsection{1.2 Dissipations of the system} 
\label{Dissipations}
There are four main sources of dissipation involved in the MV formation process, including the viscous drag experienced by the formed spherical cap as it expands $\mathcal{D}_{\rm drag}$, the friction at the circle line near the detach point between the membrane and the cortex $\mathcal{D}_{\rm ct}$, and the bulk and membrane viscous dissipation, $\mathcal{D}_{\rm bulk}$ and $\mathcal{D}_{\rm mem}$. 

During the MV formation process, the protruding membrane experiences a viscous drag force as it expands. 
%We consider a Stoke's drag force ~\cite{R.A.Tyson2014}, $F_{\rm drag}$, for a spherical cap of radius $R_{\rm s}$ expanding with velocity $u=\dot{h}$, 
We modeled this using Stoke's drag force~\cite{R.A.Tyson2014}, $F_{\rm drag}$, acting on a spherical cap of radius $R_{\rm s}$ expanding with velocity $u = \dot{h}$, 
\begin{align}
F_{\rm drag}=6\pi\mu_{\rm w} uR_{\rm s}, \label{drag force}
\end{align}
where $\mu_{\rm w}$ is the dynamic viscosity of the fluid outside the cell ($\mu_{\rm w}\sim 10^{-3}~{\rm Pa\cdot s}$ for water). 
The energy dissipation function arising from Stokes drag as the membrane expands can therefore be expressed as 
\begin{align}
\mathcal{D}_{\rm drag}=\frac{1}{2}F_{\rm drag}u=3\pi \mu_{\rm w} \dot{h}^2R_{\rm s}.\label{drag dissipation_SI}
\end{align}
 
During MV formation, the membrane must detach from the underlying actin cortex, resulting energy dissipation at the membrane-cortex interface. Here, we mainly focus on the friction localized at the circular detachment line near the base of the forming vesicle. 
In the limit of small Reynolds number, the friction dissipation function is given by $\mathcal{D}_{\rm ct}=\frac{1}{2}F_{\rm friction}\dot{s}$, where the velocity $\dot{s}=R_{\rm s}\dot{\alpha}$ is the spherical cap edge shifting speed, and $L_{\rm edge}=2\pi R_{\rm s}\sin{\alpha}$ is the length of this edge. 
The friction force per unit length is given by $f_{\rm friction}=\mu_{\rm ct}\dot{s}=\mu_{\rm ct}R_{\rm s}\dot{\alpha}$. 
Then the total friction force follows as $F_{\rm friction}=f_{\rm friction}L_{\rm edge}=2\pi\mu_{\rm ct}R_{\rm s}^2\sin({\alpha})\dot{\alpha}$. Therefore, the friction dissipation function can be obtained as 
\begin{align}
\mathcal{D}_{\rm ct}=&\frac{1}{2}F_{\rm friction}\dot{s} \notag\\
=&\mu_{\rm ct}\pi R_{\rm s}^3\sin({\alpha})\dot{\alpha}^2 \notag\\
=&\mu_{\rm ct}\pi R_{\rm s}\frac{\dot{h}^2}{\sqrt{\dfrac{2h}{R_{\rm s}}-\Bigl(\dfrac{h}{R_{\rm s}}\Bigr)^2}},\label{friction dissipation_SI1}
\end{align}
where $\mu_{\rm ct}$ is the effective dynamic microviscosity of a typical order of $1~{\rm Pa\cdot s}$. 

In addition, two additional sources of dissipation arise during MV formation: bulk viscous dissipation due to the flow of intracellular fluid into the expanding spherical cap and membrane viscous dissipation resulting from the lateral flow of the membrane relative to the underlying cortex. 
These two energy dissipation contributions are given by 
\begin{align}
\mathcal{D}_{\rm bulk}=\frac{22}{5}\mu_{\rm cyt}\frac{\dot{V}^2}{\pi R_{\rm b}^3},\label{global fluid flow dissipation:PRL}
\end{align}
and
\begin{align}
\mathcal{D}_{\rm mem}= \frac{\mu_{\rm m}l_{\rm m}[{\rm d}(\Delta A)/{\rm d}t]^2}{2\pi R_{\rm b}^2},\label{global membrane flow dissipation}
\end{align}
where $\mu_{\rm cyt}$ and $\mu_{\rm m}$ are the viscosities of the cytosol and of the membrane lipid, and $l_m$ is the membrane thickness. 
Detailed derivations of Eq.~(\ref{global fluid flow dissipation:PRL}) and Eq.~(\ref{global membrane flow dissipation}) are provided in Refs.~\cite{T.M.Fisher1994,P.Sens2004,ALERT20151878}. 

Using the relations $h=R_0\sqrt{\eta}$, $R_{\rm b}=R_0\sqrt{1-\eta}$, $R_{\rm s}=R_0/(2\sqrt{\eta})$, $\Delta A=A_0-\pi R_{\rm b}^2$ and $V=\frac{\pi}{3}h^2(3R_{\rm s}-h)$, the total dissipation can then be expressed as 
\begin{align}
\mathcal{D}_{\rm tot}=&\mathcal{D}_{\rm drag}+\mathcal{D}_{\rm ct}+\mathcal{D}_{\rm bulk}+\mathcal{D}_{\rm mem} \notag\\
=&\frac{3\pi\mu_{\rm w}R_0^3}{8\eta^{3/2}}\dot{\eta}^2 + \frac{\mu_{\rm ct}\pi R_0^3}{16\eta^2\sqrt{1-\eta}}\dot{\eta}^2 + \frac{11\mu_{\rm cyt}\pi R_0^3(1-2\eta)^2}{40\eta(1-\eta)^{3/2}}\dot{\eta}^2 +\frac{\mu_{\rm m}l_{\rm m}\pi R_0^2}{2(1-\eta)}\dot{\eta}^2 . \label{Dissipation_tot_simple}
\end{align}

\subsection{1.3 Kinetic equation}
\label{kinetic_equaiton}
The summation of the change rate of the total free energy and the total energy dissipation leads to the Rayleighian $\mathcal{R}=\dot{\mathcal{F}}_{\rm tot}+\mathcal{D}_{\rm tot}$. 
Minimizing $\mathcal{R}$ with respect to $\dot{\eta}$, i.e., $\partial\mathcal{R}/\partial\dot{\eta}=0$, yields the dynamic equation governing the formation of MVs
\begin{align}
\dot{\eta}=\frac{\mathcal{M}_1-\mathcal{M}_2-8\kappa\biggl(1-\dfrac{R_0c_0}{4\sqrt{\eta}}\biggr)-R_0^2\sigma-\dfrac{\Delta PR_0^3(2\eta-1)}{4\sqrt{\eta}}}{\dfrac{3\mu_{\rm w}R_0^3}{4\eta^{3/2}}+\dfrac{\mu_{\rm ct} R_0^3}{8\eta^2\sqrt{1-\eta}}+\dfrac{\mu_{\rm m}l_{\rm m} R_0^2}{(1-\eta)}+\dfrac{11\mu_{\rm cyt} R_0^3(1-2\eta)^2}{20\eta(1-\eta)^{3/2}}} - \frac{1}{\mathcal{P}}\times\begin{cases}
\dfrac{1}{6} k\rho_0R_0^4, \quad &0<\eta\leq \Bigl(\dfrac{l_{\rm c}}{R_0}\Bigr)^2 \\
\dfrac{k\rho_0R_0l_c^3}{6\eta^{3/2}},\quad &\eta> \Bigl(\dfrac{l_{\rm c}}{R_0}\Bigr)^2
\end{cases},\label{deta_no_mono_simple}
\end{align}
which is Eq.~(\ref{deta_dt}) in the main tex. 

Equivalently, this equation can be derived using a Lagrangian description~\cite{H.Goldstein1980, P.Sens2004}, in which the kinetic equations result from balancing the ``driving force'' $f_{\rm driving}$ against the ``dissipative (friction) force'' $f_{\rm diss}$. Here, the ``driving forces'' and the ``dissipative (friction) forces'' are calculated from the variation of the total free energy and the energy
dissipated per unit time. 
Based on the Lagrangian description $\partial \mathcal{F}_{\rm tot}/\partial\eta+\partial \mathcal{D}_{\rm tot}/\partial\dot{\eta}=0$, one can also obtain exactly the same dynamical equation~(\ref{deta_no_mono_simple}). 
This dynamic equation is fairly complex and has to be solved numerically. 
The critical condition for a fully formed MV is governed by $\dot{\eta}\Big|_{\eta=1}=0$, i.e., 
\begin{align}
\frac{9R_0^3}{8\beta\xi^2}\biggl(\frac{3\sqrt{W}}{\xi a^2}\biggr)^{\frac{2}{3}}\biggl[\biggl(1+\frac{10N}{3R_0}\biggl(\frac{Wa^2}{3\xi^2}\biggr)^{\frac{1}{3}}\biggr)^{\frac{1}{5}}-1\biggr]-\frac{3NR_0^2}{4\beta\xi^2}\biggl(\frac{3W^2}{\xi^4a^2}\biggr)^{\frac{1}{3}}\biggl[1+\frac{10N}{3R_0}\biggl(\frac{Wa^2}{3\xi^2}\biggr)^{\frac{1}{3}}\biggr]^{-\frac{4}{5}} \notag\\
-8\kappa\biggl(1-\dfrac{R_0c_0}{4}\biggr)-R_0^2\sigma-\dfrac{\Delta PR_0^3}{4}-\dfrac{k\rho_0R_0l_c^3}{6}=0.\label{R0_cri}
\end{align}

\newpage
\section{2. A Stochastic Description} 
\label{Stochastic dynamics} 
\subsection{2.1 Langevin Equation}
\label{Langevin_Equation} 
When the shape parameter $\eta$ increases from $0$ to $1$, we treat this process as a thermally assisted escape from a potential wall (with $\eta=1$ being the absorbing boundary). 
Assuming that the fully formed MV will be pinched away due to the function of dynamin-related proteins, thus the condition at $\eta=1$ is absorbing. 
At a given shape parameter, we assume that thermal fluctuations or noise of the system leads to the change of membrane geometry in a stochastic way throughout the process of MV formation, such that the shape parameter $\eta$ can be treated as a random variable evolving from $0$ to $1$ with some drift due to the free energy gradient in Eq.~(\ref{F_tot_no_mono}). 
This calls for a stochastic description of the dynamics. Assuming that the membrane morphology remains axisymmetric throughout the process, its evolution can be cast as a stochastic differential equation for $\eta(t)$ in the low Reynolds number regime 
\begin{align}
-\gamma\dot{\eta}(t)+f_{\rm dirving}+f_{\rm noise}=0,\label{Langevin Equation}
\end{align}
where the first term denotes a ``dissipative force'' with $\gamma$ being an effective friction coefficient encapsulating the dissipative processes at play, the second term represents the deterministic ``driving force'', and the last term is a stochastic force describing the stochastic effects. 
The ``driving force'' is derived from the variation of an effective potential energy of the system, such that $f_{\rm driving}=\mathcal{U}'(\eta)=\partial\mathcal{U}/\partial\eta$. 
A minimal stochastic description of the dynamics may be written as 
\begin{align}
\dot{\eta}(t)=\underbrace{D_1(\eta,t)}_{\rm drift}+\underbrace{\sqrt{2D_2(\eta,t)}\zeta(t)}_{\rm random~noise},\label{simple stochastic model}
\end{align}
where $D_1$ is the drift term, $D_2$ is the ``diffusion'' coefficient (``diffusivity'' or transition rate), and $\zeta(t)$ is a Gaussian white noise with zero mean $\langle\zeta(t)\rangle=0$ and autocorrelation function given by $\langle\zeta(t)\zeta(t')\rangle=\delta(t-t')$. 
Dividing $f_{\rm driving}$ by the coefficient $\gamma$ leads to the drift coefficient: 
\begin{align}
D_1(\eta,t)=-\frac{\mathcal{U}'(\eta)}{\gamma}.\label{D_1}
\end{align}
Thus, Eq.~(\ref{simple stochastic model}) becomes
\begin{align}
\dot{\eta}(t)=-\frac{\mathcal{U}'(\eta)}{\gamma}+\sqrt{2D}\zeta(t),\label{simple stochastic model_2}
\end{align}
with a constant transition rate $D=D_2$. 
Since $\mathcal{U}'(\eta)=\mathcal{F}_{\rm tot}'(\eta)$, Eq.~(\ref{simple stochastic model_2}) can be reduced to the deterministic dynamic equation~(\ref{deta_no_mono_simple}) by assuming $\gamma=\pi\mathcal{P}$ and neglecting the noise term. 
Eq.~(\ref{simple stochastic model_2}) can be rewritten in a more general form as  
\begin{align}
\dot{\eta}(t)=\mathcal{A}(\eta)+\mathcal{B}\zeta(t), \label{simple stochastic model_3}
\end{align}
where $\mathcal{A}(\eta)=-\frac{\mathcal{U}'(\eta)}{\gamma}$ and $\mathcal{B}=\sqrt{2D}$. 

\subsection{2.2 Fokker–Planck Equation}
\label{FP_Equation}
We define $\mathbb{P}(\eta,t)$ as the probability of the membrane shape with shape parameter $\eta$ at time $t$, and it is given by $\mathbb{P}(\eta,t)=\langle\delta(\eta-\eta(t))\rangle$. 
Using $\frac{\rm d}{{\rm d}t}[\delta(\eta-\eta(t))]=-\frac{\partial}{\partial\eta}[\delta(\eta-\eta(t))]\frac{{\rm d}\eta}{{\rm d}t}$, one can obtain
\begin{align}
\frac{\partial\mathbb{P}(\eta,t)}{\partial t}=-\frac{\partial}{\partial\eta}\bigg\langle\delta(\eta-\eta(t))\frac{{\rm d}\eta}{{\rm d}t}\bigg\rangle. \label{dP/dt_1}
\end{align}
Submitting Eq.~(\ref{simple stochastic model_3}) into Eq.~(\ref{dP/dt_1}) leads to 
\begin{align}
\frac{\partial\mathbb{P}(\eta,t)}{\partial t}=&-\frac{\partial}{\partial\eta}\Bigl[\mathcal{A}(\eta)\langle\delta(\eta-\eta(t))\rangle + \mathcal{B}\langle\delta(\eta-\eta(t))\zeta(t)\rangle\Bigr] \notag\\
=&-\frac{\partial}{\partial\eta}\Bigl[\mathcal{A}(\eta)\mathbb{P}(\eta,t)+\mathcal{B}\langle\delta(\eta-\eta(t))\zeta(t)\rangle\Bigr]. \label{dP/dt_2}
\end{align}
Now our goal is to determine $\langle\delta(\eta-\eta(t))\zeta(t)\rangle$. 
Using It$\hat{\rm o}$'s lemma~\cite{C.W.Gardiner1985}, the Langevin equation can be transformed into 
\begin{align}
{\rm d}\eta=\mathcal{A}(\eta){\rm d}t+\mathcal{B}{\rm d}\mathcal{W}_t,  \label{simple stochastic model_4}
\end{align}
where ${\rm d}\mathcal{W}_t=\zeta(t){\rm d}t$. The increment of $\Delta\mathcal{W}_t$ is $\Delta\mathcal{W}_t=\mathcal{W}_{t+\Delta t}-\mathcal{W}_t\sim\mathcal{N}(0,\Delta t)$ with $\langle\Delta\mathcal{W}_t\rangle=0$ and $\langle(\Delta\mathcal{W}_t)^2\rangle=\Delta t$. 
Noting that the evolution of the probability, $\mathbb{P}(\eta,t)$, in
the Kramers-Moyal-expansion is given by~\cite{Kampen1984} 
\begin{align}
\frac{\partial\mathbb{P}(\eta,t)}{\partial t}=\sum_{n=1}^\infty\frac{(-1)^n}{n!}\frac{\partial^n}{\partial\eta^n}[D_n(\eta)\mathbb{P}(\eta,t)],  \label{simple stochastic model_5}
\end{align}
where 
\begin{align}
D_n(\eta)=\lim_{\Delta t\rightarrow{0}}\frac{\langle[\eta(t+\Delta t)-\eta(t)]^n\rangle}{\Delta t}.  \label{D_n}
\end{align} 
Using Eq.~(\ref{simple stochastic model_4}) and $\langle\Delta\mathcal{W}_t\rangle=0$ and $\langle(\Delta\mathcal{W}_t)^2\rangle=\Delta t$, 
we can obtain 
\begin{align}
\langle\Delta\eta\rangle=\mathcal{A}(\eta)\Delta t+\mathcal{B}\langle\mathcal{W}_t\rangle=\mathcal{A}(\eta)\Delta t,   \label{delta_eta}
\end{align}
and 
\begin{align}
\langle(\Delta\eta)^2\rangle=&\langle[\mathcal{A}(\eta)\Delta t+\mathcal{B}\Delta\mathcal{W}_t]^2\rangle \notag\\
=&\mathcal{A}^2(\eta)(\Delta t)^2+2\mathcal{A}(\eta)\Delta t\mathcal{B}\langle\Delta\mathcal{W}_t\rangle+\mathcal{B}^2\langle(\Delta\mathcal{W}_t)^2\rangle \notag\\
=&\mathcal{B}^2\Delta t.   \label{delta_eta_2}
\end{align}
Therefore, we have $D_1=\mathcal{A}(\eta)$ and $D_2=\mathcal{B}$. 
Keeping the Kramers-Moyal-expansion only up to the second order term results in 
\begin{align}
\frac{\partial\mathbb{P}(\eta,t)}{\partial t}=-\frac{\partial[\mathcal{A}(\eta)\mathbb{P}(\eta,t)]}{\partial\eta}+\frac{1}{2}\frac{\partial[\mathcal{B}^2\mathbb{P}(\eta,t)]}{\partial\eta^2}.  \label{simple stochastic model_6}
\end{align}
Plugging $\mathcal{A}(\eta)=-\frac{\mathcal{U}'(\eta)}{\gamma}=-\frac{1}{\gamma}\frac{\partial\mathcal{F}_{\rm tot}}{\partial\eta}$, $\mathcal{B}^2=2D$, and using the Einstein relation $D=k_BT/\gamma$, the above equation reduces to 
\begin{align}
\frac{\partial\mathbb{P}(\eta,t)}{\partial t}=D\frac{\partial}{\partial\eta}\biggl[\mathbb{P}(\eta,t)\frac{\partial}{\partial\eta}\biggl(\frac{\mathcal{F}_{\rm tot}(\eta)}{k_BT}\biggr)+\frac{\partial \mathbb{P}(\eta,t)}{\partial\eta}\biggr].  \label{simple stochastic model_7}
\end{align}

\subsection{2.3 Mean formation time}
Let $\tau(\eta_0)$ denote the mean first passage time starting from an initial state $\eta_0$. Consider a trajectory that begins at $\eta_0$ and, after a small time step $\Delta t$, reaches a new state $\eta_1 = \eta_0 + \Delta \eta$, where $\Delta \eta$ is a stochastic increment. 
The total time required to arrive at the absorbing boundary is estimated as 
\begin{align}
\tau(\eta_0)=\mathbb{E}_{\eta_0}[\Delta t+\tau(\eta_1)],  \label{T_toal_1}
\end{align}
where $\mathbb{E}_{\eta_0}[\cdot]=\langle\cdot\rangle$. 
Performing the Taylor expansion for $\tau(\eta_1)$ at $\eta_0$, one can obtain 
\begin{align}
\tau(\eta_0)=&\mathbb{E}_{\eta_0}\biggl[\Delta t+\tau(\eta_0)+\tau'(\eta_0)\Delta\eta+\frac{1}{2}\tau''(\eta_0)(\Delta\eta)^2+\mathcal{O}((\Delta\eta)^3)\biggr] \notag\\
=&\Delta t+\tau(\eta_0)+\tau'(\eta_0)\mathbb{E}_{\eta_0}[\Delta\eta]+\frac{1}{2}\tau''(\eta_0)\mathbb{E}_{\eta_0}[(\Delta\eta)^2]+\mathbb{E}_{\eta_0}[\mathcal{O}((\Delta\eta)^3)] \notag\\
=&\Delta t+\tau(\eta_0)+\tau'(\eta_0)\mathcal{A}(\eta)\Delta t+\frac{1}{2}\tau''(\eta_0)\mathcal{B}^2\Delta t.  \label{T_toal_2}
\end{align}
Then dividing the above equation by $\Delta t$ leads to the following inhomogeneous differential equation 
\begin{align}
\biggl[\mathcal{A}(\eta)\frac{\partial}{\partial\eta}+\frac{\mathcal{B}^2}{2}\frac{\partial^2}{\partial\eta^2}\biggr]\tau(\eta)=-1,  \label{T_toal_3}
\end{align}
this is called the Kolmogorov equation which has to be solved under the absorbing boundary condition that the formation process terminates as soon as the MV is fully formed $\tau(\eta=1)=0$. 
The boundary condition at $\eta=0$ is reflecting with the reflecting boundary condition $\partial_\eta\tau(\eta=0)=0$. 
If $\mathcal{A}(\eta)$ is a constant drift term, the solution of the Kolmogorov equation is given by $\tau=\frac{1}{\mathcal{A}}+\frac{\mathcal{B}^2}{2\mathcal{A}^2}(e^{-2\mathcal{A}/\mathcal{B}^2}-1)$. 
However, the drift term in our model is related to the gradient of the total free energy $\partial\mathcal{F}_{\rm tot}(\eta)/\partial\eta$, which depends on the shape parameter $\eta$. 
Using $\mathcal{A}(\eta)=-\frac{1}{\gamma}\frac{\partial\mathcal{F}_{\rm tot}}{\partial\eta}$, $\mathcal{B}^2=2D$, and $D=k_BT/\gamma$, Eq.~(\ref{T_toal_3}) can be converted to 
\begin{align}
\biggl[D\frac{\partial^2}{\partial\eta^2}-\biggl(\frac{D}{k_BT}\frac{\partial\mathcal{F}_{\rm tot}}{\partial\eta}\biggr)\frac{\partial}{\partial\eta}\biggr]\tau(\eta)=-1.   \label{T_toal_4}
\end{align}
Multiplying with an integral factor $\Gamma(\eta)=\exp(-\beta\mathcal{F}_{\rm tot})$ on both sides, the above differential equation is then solved by 
\begin{align}
\langle\tau\rangle=\frac{1}{D}\int_0^1{\rm d}\eta \int_\eta^1 {\rm d}\nu~e^{[-\mathcal{F}_{\rm tot}(\eta)+\mathcal{F}_{\rm tot}(\nu)]/k_BT},\label{mean formation time}
\end{align}
which is the mean formation time, also referred to as the mean first passage time. 
Note that in Eq.~(\ref{mean formation time}), the transition rate $D$ is assumed to be constant, consistent with the approach in Ref.~\cite{StephenMirigian2013}.

However, given that $\gamma(\eta)=\pi\mathcal{P}(\eta)$, where $\mathcal{P}(\eta)$ is a function of $\eta$, if we consider that $D=\frac{k_BT}{\gamma}$ depends on the shape parameter $\eta$, then $D$ is no longer constant. 
In such a case where the transition rate varies with the shape parameter, i.e., $D(\eta)=\frac{k_BT}{\gamma(\eta)}$, the mean first passage time takes the form
\begin{align}
\langle\tau\rangle=\int_0^1{\rm d}\eta \int_\eta^1 {\rm d}\nu~\frac{e^{[-\mathcal{F}_{\rm tot}(\eta)+\mathcal{F}_{\rm tot}(\nu)]/k_BT}}{D(\nu)}.\label{mean formation time_D}
\end{align}

The Fokker–Planck equation and the corresponding mean first passage time can also be derived using an alternative approach. 
At any given time, the probability that the shape parameter transitions from $\eta$ to a neighboring value $\eta \pm d\eta$ obeys a master equation of the form~\cite{StephenMirigian2013}: 
\begin{align}
\frac{\partial \mathbb{P}(\eta,t)}{\partial t}=D_{\eta-d\eta}\mathbb{P}(\eta-d\eta,t)-D_{\eta}'\mathbb{P}(\eta,t)-D_\eta \mathbb{P}(\eta,t)+D_{\eta+d\eta}'\mathbb{P}(\eta+d\eta,t),\label{Master equation}
\end{align}
where $D_\eta$ denotes the transition rate from a given shape parameter $\eta$ to an incrementally larger value $\eta + d\eta$, while $D'\eta$ is the rate for the reverse transition from $\eta$ to $\eta - d\eta$. 
By invoking detailed balance, $D'{\eta+d\eta}$ can be expressed in terms of $D_\eta$. In the continuous limit, this formulation leads to the Smoluchowski equation governing the stochastic dynamics~\cite{StephenMirigian2013}  
\begin{align}
\frac{\partial \mathbb{P}(\eta,t)}{\partial t}=\frac{\partial}{\partial\eta}D\biggl[\frac{\partial \mathbb{P}(\eta,t)}{\partial\eta}+\mathbb{P}(\eta,t)\frac{\partial}{\partial\eta}\biggl(\frac{\mathcal{F}_{\rm tot}(\eta)}{k_BT}\biggr)\biggr],\label{Smoluchowski equation_SI}
\end{align}
where $\mathbb{P}$ is the probability distribution of the system, $D$ is the transition rate which characterizes how rapidly the membrane edge can change its arch angle, and $k_BT$ represents the thermal energy with $k_B$ the Boltzmann constant and $T$ the temperature. 
%Also, where we have additionally assumed that the transition rate $D_\eta=D$ is constant throughout the membrane reshaping process. 
For simplicity, we assume that the transition rate $D_\eta = D$ remains constant throughout the membrane reshaping process. 
The steady-state $\partial\mathbb{P}/\partial t=0$ solution of Eq.~(\ref{Smoluchowski equation_SI}) takes the form~\cite{FuqiangSun2023} 
\begin{align}
\mathbb{P}(\eta)=\frac{q}{D}{\rm exp}\biggl(-\frac{\mathcal{F}_{\rm tot}(\eta)}{k_BT}\biggr)\int_\eta^1{\rm exp}\biggl(\frac{\mathcal{F}_{\rm tot}(\nu)}{k_BT}\biggr){\rm d}\nu,\label{probability equation}
\end{align}
in which $q$ denotes the steady-state flux of the probability from $\eta=0$ to $\eta=1$. 
Since $\mathbb{P}$ must satisfy the normalization condition, integrating both sides of Eq.~(\ref{probability equation}) leads to 
\begin{align}
\int_0^1\mathbb{P}(\eta){\rm d}\eta=\frac{q}{D}\int_0^1{\rm exp}\biggl(-\frac{\mathcal{F}_{\rm tot}(\eta)}{k_BT}\biggr)\biggl[\int_\eta^1{\rm exp}\biggl(\frac{\mathcal{F}_{\rm tot}(\nu)}{k_BT}\biggr){\rm d}\nu\biggr]{\rm d}\eta=1. \label{normalization condition}
\end{align}
Viewing the shape parameter as a stochastic variable means that the formation of MVs will occur over a distribution of times, related to the full probability distribution, $\mathbb{P}(\eta,t)$. 
We are postulating a stochastic MV formation process for which the mean formation time is of particular interest. 
The average time at which the shape parameter first equals $1$ is defined to be the mean first passage time and is given by 
\begin{align}
\langle\tau\rangle=\frac{1}{q}=&\frac{1}{D}\int_0^1{\rm exp}\biggl(-\frac{\mathcal{F}_{\rm tot}(\eta)}{k_BT}\biggr)\biggl[\int_\eta^1{\rm exp}\biggl(\frac{\mathcal{F}_{\rm tot}(\nu)}{k_BT}\biggr){\rm d}\nu\biggr]{\rm d}\eta \notag\\
=&\frac{1}{D}\int_0^1{\rm d}\eta \int_\eta^1 {\rm d}\nu~e^{[-\mathcal{F}_{\rm tot}(\eta)+\mathcal{F}_{\rm tot}(\nu)]/k_BT}.\label{first passage time}
\end{align}
Here, $\langle\tau\rangle$ represents the average time for the formation of a spherical MV. 

\newpage
\section{3. Numerical parameters}
The list of parameters used for all the numerical simulations presented in this paper is provided in Table~\ref{tab_num_params} unless otherwise specified. 
\begin{table*}[tbhp]%[!htbp]
\centering
\caption{Parameters used in our numerical estimates.}
\label{tab_num_params}
%\scalebox{0.78}{
\begin{tabular}{ccccc}
%\colrule
%\toprule
Parameter &  Meaning & Estimated range & References & Default value \\
\hline
$\kappa$ & Bending rigidity & $10\text{--} 400~k_BT$ & ~\cite{R.Lipowsky1993,J.Liu2006,J.M.Allain2004,L.Foret2014} & $20~k_BT$ \\
\hline
$\sigma$ & Membrane tension & $10^{-7}\text{--} 10^{-3}~{\rm N/m}$ & ~\cite{J.Liu2006,Z.Shi2015,R.Phillips2009,L.Foret2008} & $0.012~k_BT/{\rm nm}^2$ \\
\hline
$c_{\rm 0}$ & Spontaneous curvature & $0\text{--} 0.1~{\rm nm}^{-1}$ & ~\cite{SHURER20191757,Antonia2025} & $0~{\rm nm}^{-1}$ \\
\hline
$R_0$ & Patch radius & $20\text{--} 100~{\rm nm}$ & ~\cite{KUTTIKANDY20223674,J.Liu2006,JEPPESEN2023667,JEPPESEN2019428,Kumar2024} & $100~{\rm nm}$ \\
\hline
$\xi$ & Grafting distance & $10\text{--} 100~{\rm nm}$ & ~\cite{SHURER20191757,D.Bracha2013} & $10~{\rm nm}$ \\
\hline
$a$ & \makecell{Effective monomeric segment \\ size in the glycocalyx} & $10\text{--} 20~{\rm nm}$ & ~\cite{SHURER20191757,L.Foret2008,KUTTIKANDY20223674,J.Paturej2016,J.G.Gandhi2019,J.C.H.Kuo2018} & $10~{\rm nm}$ \\
\hline
$a_{\rm m}$ & Monomer size & $\sim0.35~{\rm nm}$ & ~\cite{SHURER20191757,KUTTIKANDY20223674,J.Paturej2016,J.G.Gandhi2019,J.C.H.Kuo2018} & $0.35~{\rm nm}$ \\
\hline  
$N$ & \makecell{Number of effective  \\ monomeric segments in a brush} & $10\text{--} 50$ & ~\cite{SHURER20191757,KUTTIKANDY20223674,J.G.Gandhi2019,J.C.H.Kuo2018} & $20$ \\
\hline
$\phi_{\rm b}$ & Degree of ionization & $0\text{--} 1$ & ~\cite{SHURER20191757,FengYuan2023} & $0.1$ \\
\hline
$I_{\rm ion}$ & Ionic concentration & $5\text{--} 1000~{\rm mM}$ & ~\cite{SHURER20191757,FengYuan2023} & $100~{\rm mM}$ \\
\hline
$\Delta P$ & Pressure difference & $0\sim 100~{\rm Pa}$ & ~\cite{R.A.Tyson2014,F.Y.Lim2012} & $2.5\times 10^{-5}~k_BT/{\rm nm^3}$ \\
\hline
$k$ & Spring constant  & $\sim 10^{-4}~{\rm N/m}$ & ~\cite{ALERT20151878,R.Alert2016} & $2.5\times10^{-2}~k_BT/{\rm nm}^2$ \\
\hline
$\rho_0$ & Density of available linkers  & $10^{14}\text{--}10^{16}~{\rm m^{-2}}$ & ~\cite{ALERT20151878,R.Alert2016} & $10^{-4}~{\rm nm}^{-2}$ \\
\hline
$l_{\rm c}$ & \makecell{Critical length of \\ membrane-cortex spring}   & $2\text{--}10~{\rm nm}$ & ~\cite{ALERT20151878,R.Alert2016} & $ 5~{\rm nm}$ \\
\hline
$l_{\rm m}$ & Membrane thickness   & $4\text{--}5~{\rm nm}$ & ~\cite{ALERT20151878,R.Alert2016} & $ 5~{\rm nm}$ \\
\hline
$\mu_{\rm w}$ & Water viscosity   & $\sim 10^{-3}~{\rm Pa\cdot s}$ & ~\cite{R.A.Tyson2014} & $0.25\times10^{-9}~k_BT\cdot{\rm s/nm^3}$ \\
\hline
$\mu_{\rm ct}$ & \makecell{Effective dynamic microviscosity \\ at contact line}  & $\sim 1~{\rm Pa\cdot s}$ & ~\cite{K.Xiao2023dynamic} & $0.25\times10^{-6}~k_BT\cdot{\rm s/nm^3}$ \\
\hline
$\mu_{\rm cyt}$ & Cytosolic viscosity   & $\sim 10^{-2}~{\rm Pa\cdot s}$ & ~\cite{A.Veksler2007,R.Shlomovitz2007PRL,B.Peleg2011,ALERT20151878,R.Alert2016} & $0.25\times10^{-8}~k_BT\cdot{\rm s/nm^3}$ \\
\hline
$\mu_{\rm m}$ &  Viscosity of membrane lipid  & $\sim 0.5\times10^{-9}~{\rm J\cdot s/m^2}$ & ~\cite{ALERT20151878,R.Alert2016,KARATEKIN20031734,D0SM01027H,V.K.Malik2022} & $2.5\times10^{-8}~k_BT\cdot{\rm s/nm^2}$ \\
\hline
$k_BT=1/\beta$ & Thermal energy & $4.1\times10^{-21}~{\rm J}$ & ~\text{--} & $4.1\times10^{-21}~{\rm J}$ \\
%\bottomrule
%\bottomrule
\end{tabular}
%}
\end{table*}

\newpage
The lists of parameter values used for all the numerical simulations
presented in this paper are provided in Table~\ref{tab_sim_params}.
\begin{table*}[tbhp]%[!htbp]
\centering
\caption{Parameter values used in our numerical simulations in all figures.}
\label{tab_sim_params}
%\scalebox{0.8}{
\begin{tabular}{cccccccc}
Parameter & Unit & \makecell{Fig.~\ref{fig:figure2_2}(a), \\ Figs.~\ref{fig:figure3}(a) and~\ref{fig:figure3}(c)}  & \makecell{Figs.~\ref{fig:figure2_2}(b)-(e) \\
and Fig.~\ref{fig:figure3}(b)} & \makecell{Fig.~\ref{fig:figure2_2}(f) \\ and Fig.~\ref{fig:figure3}(d)} & \makecell{Fig.~\ref{fig:figure2_2}(g) \\and Fig.~\ref{fig:figure3}(e)} & Fig.~\ref{fig:figure4} & Fig.~\ref{fig:figure5} \\
\hline
$\kappa$ & $k_BT$ & 20 & 20 & 20 & 20 & 20 & 20 \\
\hline
$\sigma$ & $k_BT/{\rm nm}^2$ & 0.012 & 0.012 & 0.012 & 0.012 & Varied & Varied \\
\hline
$c_0$ & ${\rm nm}^{-1}$ & 0 & 0 & 0 & 0 & Varied & Varied \\
\hline
$R_0$ & ${\rm nm}$ & Varied & Varied & 100 & 100 & Varied & Varied \\
\hline
$\xi$ & ${\rm nm}$ & Varied & Varied & Varied & 12.9 & 11.2 & 11.2 \\
\hline
$a$ & ${\rm nm}$ & 10 & 10 & 10 & 10 & 10 & 10 \\
\hline
$a_{\rm m}$ & ${\rm nm}$ & 0.35 & 0.35 & 0.35 & 0.35 & 0.35 & 0.35 \\
\hline
$N$ & 1 & Varied & Varied & Varied & 20 & 20 & 20 \\
\hline
$\phi_{\rm b}$ & 1 & Varied & Varied & 0 & Varied & 0 & 0 \\
\hline
$I_{\rm ion}$ & ${\rm mM}$ & Varied & Varied & 100 & Varied & 100 & 100 \\
\hline
$\Delta P$ & $\times 10^{-5}~k_BT/{\rm nm^3}$ & 2.5 & 2.5 & 2.5 & 2.5 & 2.5 & 2.5 \\
\hline
$k$ & $\times10^{-2}~k_BT/{\rm nm}^2$ & 2.5 & 2.5 & 2.5 & 2.5 & 2.5 & 2.5 \\
\hline
$\rho_0$ & ${\rm nm}^{-2}$ & $10^{-4}$ & $10^{-4}$ & $10^{-4}$ & $10^{-4}$ & $10^{-4}$ & $10^{-4}$ \\
\hline
$l_{\rm c}$ & ${\rm nm}$ & 5 & 5 & 5 & 5 & 5 & 5 \\
\hline
$l_{\rm m}$ & ${\rm nm}$ & 5 & 5 & 5 & 5 & 5 & 5 \\
\hline
$\mu_{\rm w}$ & $\times10^{-9}~k_BT\cdot{\rm s/nm^3}$ & 0.25 & 0.25 & 0.25 & 0.25 & 0.25 & 0.25 \\
\hline
$\mu_{\rm ct}$ & $\times10^{-6}~k_BT\cdot{\rm s/nm^3}$ & 0.25 & 0.25 & 0.25 & 0.25 & 0.25 & 0.25 \\
\hline
$\mu_{\rm cyt}$ & $\times10^{-8}~k_BT\cdot{\rm s/nm^3}$ & 0.25 & 0.25 & 0.25 & 0.25 & 0.25 & 0.25 \\
\hline
$\mu_{\rm m}$ & $\times10^{-8}~k_BT\cdot{\rm s/nm^2}$ & 2.5 & 2.5 & 2.5 & 2.5 & 2.5 & 2.5 \\
\hline
\end{tabular}
%}
\end{table*}

\newpage
\begin{figure*}[h]
\centering
\includegraphics[width=\linewidth]{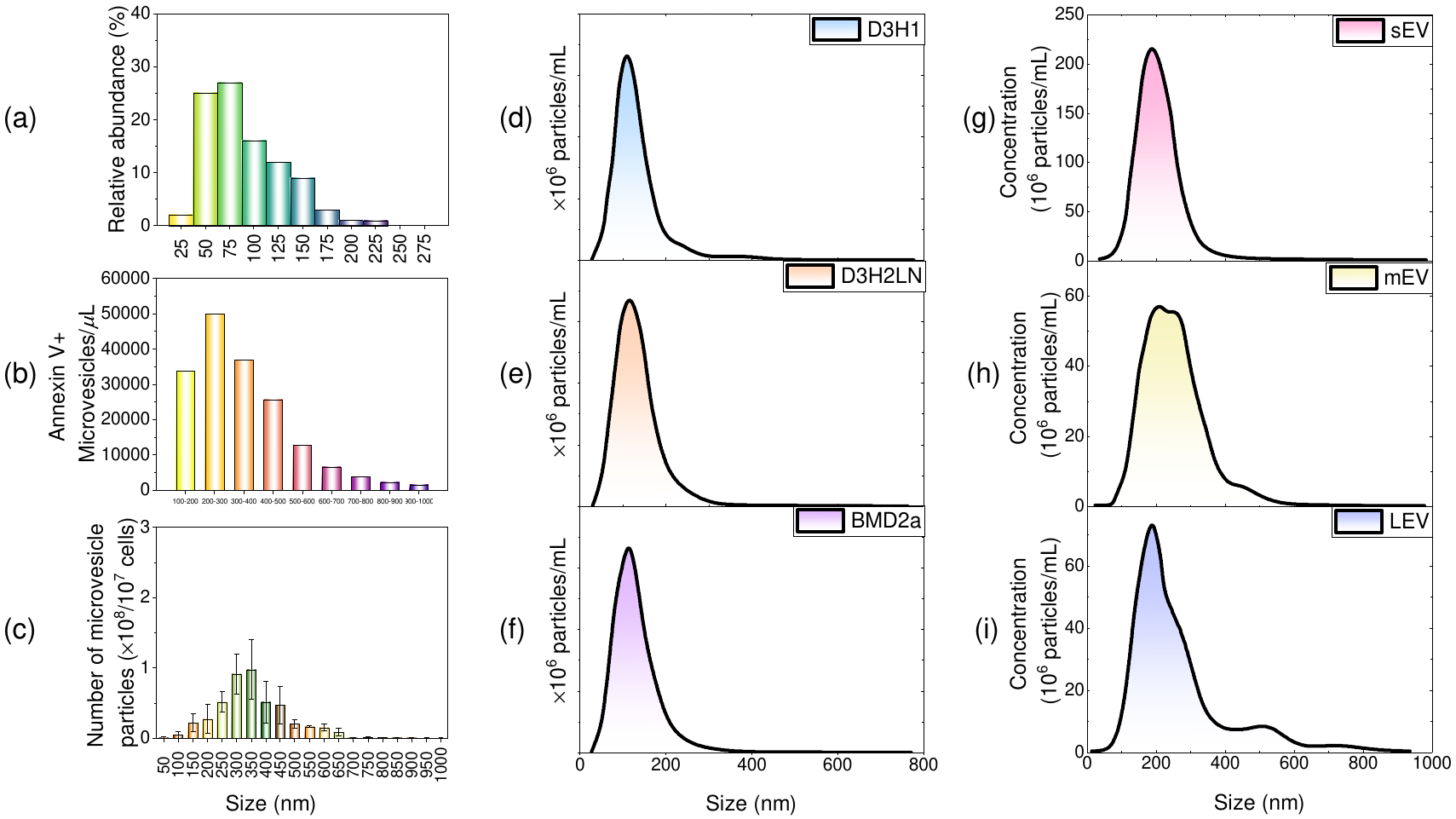}
\caption{The size distribution of MV derived from different cell types. (a) MV size distribution from GBM20/3 primary human glioblastoma cell line. Data are cited from Refs.~\cite{shao2012protein}. (b) Estimated distribution of annexin V+ MVs in human whole blood. Data are cited from Refs.~\cite{W.L.Chandler2016}. (c) Histograms of particle size distribution for MVs from MDA-231 SC human breast cancer cell lines. Data are cited from Refs.~\cite{WangTing2014}. (d)-(f) The size distribution of EVs collected from D3H1, D3H2LN, and BMD2a breast cancer cell lines. Data are cited from Refs.~\cite{Nishida-Aoki01012020}. (g)-(i) Characterization of PC3-cell-derived EVs size distribution. Data are cited from Refs.~\cite{A.Shimoda2022}.
The size distribution of MVs was determined by NTA. }\label{fig:Exp_plots}
\end{figure*}

%\bibliography{apssamp}% Produces the bibliography via BibTeX.

\iffalse
%merlin.mbs apsrev4-1.bst 2010-07-25 4.21a (PWD, AO, DPC) hacked
%Control: key (0)
%Control: author (72) initials jnrlst
%Control: editor formatted (1) identically to author
%Control: production of article title (-1) disabled
%Control: page (0) single
%Control: year (1) truncated
%Control: production of eprint (0) enabled
%
\fi

\end{document}